\documentclass[12pt]{article}
\usepackage{mathrsfs}
\makeatletter \@addtoreset{equation}{section} \makeatother
\renewcommand{\theequation}{\thesection.\arabic{equation}}
\newcommand{\ba}{\begin{array}}
\newcommand{\ea}{\end{array}}
\newcommand{\beq}{\begin{equation}}
\newcommand{\eeq}{\end{equation}}
\newcommand{\bea}{\begin{eqnarray}}
\newcommand{\eea}{\end{eqnarray}}

\def\bce{\begin{center}}
\def\ece{\end{center}}

\def\nonu{\nonumber}

\def\pa{\partial}
\def\al{\alpha}
\def\be{\beta}
\def\ga{\gamma}

\def\de{\delta}

\def\ep{\epsilon}

\def\la{\lambda}

\def\eps6{{\displaystyle \mathop{\epsilon}^{6}}{}}
\def\g6{{\displaystyle \mathop{g}^{6}}{}}

\def\nab6{{\displaystyle \mathop{\nabla}^{6}}{}}

\def\0{{\sst{(0)}}}
\def\1{{\sst{(1)}}}
\def\2{{\sst{(2)}}}
\def\3{{\sst{(3)}}}
\def\4{{\sst{(4)}}}
\def\5{{\sst{(5)}}}
\def\6{{\sst{(6)}}}
\def\7{{\sst{(7)}}}
\def\8{{\sst{(8)}}}

\def\ba{\begin{array}}
\def\ea{\end{array}}
\def\beq{\begin{equation}}
\def\eeq{\end{equation}}
\def\be{\begin{equation}}
\def\ee{\end{equation}}

\def\la{\lambda}
\def\eps{\epsilon}

\def\ba{\begin{array}}
\def\ea{\end{array}}
\def\beq{\begin{equation}}
\def\eeq{\end{equation}}
\def\be{\begin{equation}}
\def\ee{\end{equation}}

\def\la{\lambda}
\def\eps{\epsilon}

\def\eps6{{\displaystyle \mathop{\epsilon}^{6}}{}}

\def\nab6{{\displaystyle \mathop{\nabla}^{6}}{}}

\newcommand{\bean}{\begin{eqnarray*}}
\newcommand{\eean}{\end{eqnarray*}}

\usepackage{amsmath}

\usepackage{kotex}

\usepackage{amsmath}
\usepackage{graphicx}
\usepackage{amssymb}
\usepackage{amsthm}
\usepackage{grffile}
\usepackage{simplewick}
\usepackage{tikz}

\usepackage{nccmath}
\usepackage{physics}
\usepackage{wrapfig}
\usepackage{mathtools}
\usepackage{stackengine}

\usetikzlibrary{intersections,decorations.markings}

\allowdisplaybreaks

\begin{document}
\thispagestyle{empty} \addtocounter{page}{-1}
   \begin{flushright}
\end{flushright}

\vspace*{1.3cm}
  
\centerline{ \Large \bf
  The ${\cal N}=2,4$ Supersymmetric Linear $W_{\infty}[\la]$ Algebras }
\vspace*{0.5cm}
\centerline{ \Large \bf
for Generic $\lambda$ Parameter
}
\vspace*{1.5cm}
\centerline{ {\bf  Changhyun Ahn$^\dagger$},
  and {\bf Man Hea Kim$^{\ast}$}
} 
\vspace*{1.0cm} 
\centerline{\it 
$\dagger$ Department of Physics, Kyungpook National University, Taegu
41566, Korea} 
\vspace*{0.3cm}
\centerline{\it 
  $\ast$
  Department of Physics Education,  Sunchon National University,
  Sunchon 57922, Korea }
\vspace*{0.8cm} 
\centerline{\tt ahn@knu.ac.kr,
manhea.kim10000@gmail.com
} 
\vskip2cm

\centerline{\bf Abstract}
\vspace*{0.5cm}

The four different kinds of currents are given by
the multiple
$(\beta,\gamma)$ and $(b,c)$ ghost systems with a multiple product of
derivatives. We determine their complete algebra where the structure
constants depend on the deformation parameter $\la$ appearing in
the conformal weights of above fields nontrivially and depend on
the generic
spins $h_1$ and $h_2$ appearing on
the left hand sides in the (anti)commutators.
By taking the linear combinations of these currents,  
the ${\cal N}=4$ supersymmetric linear $W_{\infty}[\la]$ algebra
(and its ${\cal N}=4$ superspace description)
for generic $\lambda$ is obtained explicitly.
Moreover, we determine
the ${\cal N}=2$ supersymmetric linear $W_{\infty}[\la]$ algebra
for arbitrary $\la$.
As a by product, the $\la$ deformed
bosonic $W_{1+\infty}[\la] \times
W_{1+\infty}[\la+\frac{1}{2}]$ subalgebra
(a generalization of Pope, Romans and Shen's work  in $1990$)
is obtained.
The first factor is realized by $(b,c)$ fermionic fields
while the second factor is realized by $(\beta,\gamma)$ bosonic fields. 
The degrees of the polynomials in $\la$ for the structure constants
are given by $(h_1+h_2-2)$.
Each $w_{1+\infty}$ algebra from the celestial holography
is reproduced by taking
the vanishing limit of other deformation prameter $q$
at $\la=0$ with the contractions of the currents.

\vspace*{2cm}

\baselineskip=18pt
\newpage
\renewcommand{\theequation}
{\arabic{section}\mbox{.}\arabic{equation}}

\tableofcontents

\section{ Introduction
}

The simplest infinitely generated $W$-algebra,
which has been named as 
$w_{\infty}$ algebra \cite{Bakas},
contains generators of spin (or weight) $2,3,4, \cdots$.
The operator product expansion (OPE)
between these generators has simple form
in the sense that there are only the second and the
first order poles. The first order pole is given by
the derivative of the second order pole with different relative
coefficients.
Once the spins of the generators on the left hand side are given by
$h_1$ and $h_2$, then the spin appearing in
the second order pole on the right hand side
is equal to $(h_1+h_2-2)$.
In terms of commutator, the mode dependent structure
constant is given by $[(h_2-1)m-(h_1-1)n]$ where
$m$ is the mode for the generator of spin $h_1$ while
$n$ is the mode for the generator of spin $h_2$.
We can easily observe that this structure constant
is odd under the exchange of $h_1 \leftrightarrow h_2$
and $m\leftrightarrow n$. This extra minus sign is consistent with
the antisymmetry on the left hand side of the commutator
under this exchange.

By considering all the possible terms on the right hand side
of the commutator as well
as the central term, the $W_{\infty}$ algebra \cite{PRS1990,PRS}
has been constructed. On the right hand side of the commutator or
OPE,
the generators of
the spin $(h_1+h_2-2)$, the spin $(h_1+h_2-4)$, $\cdots$, the spin
$2$ for even $(h_1+h_2)$  or the spin $3$
for odd $(h_1+h_2)$ appear.
The structure constants and the central term are determined by using
the Jacobi identities. By rescaling the generators by
spin dependent power of a parameter $q$, the commutator has
$q$ dependence explicitly. If we take $q$ to zero limit, then
the generator of spin $(h_1+h_2-2)$ together with the
central term survive \cite{Pope1991}.
Also the structure constant becomes the one in previous paragraph.
Therefore, the centrally extended $w_{\infty}$ algebra is reproduced
under this process.
By including the spin $1$ generator in the commutator
or OPE, the $W_{1+\infty}$ algebra  \cite{PRS1990-2}
is obtained. The corresponding structure constants and the
central term are similar to the ones for $W_{\infty}$ algebra.
Moreover, the generator of the spin $2$ for even
$(h_1+h_2)$ or  the spin $1$ for odd $(h_1+h_2)$
on the right hand side appears. 

By
considering the complex fermion realization
of $W_{1+\infty}$ algebra and complex boson realization
of $W_{\infty}$ algebra,  
the ${\cal N}=2$ supersymmetric $W_{\infty}$ algebra \cite{BPRSS}
where the bosonic sector is
$W_{1+\infty} \times W_{\infty}$ is found.
By construction, there is no spin $1$ generator
from the complex bosons.
The two supersymmetry currents can be constructed
from the bilinears between the above complex bosons and
complex fermions with possible derivatives.
In particular, the right hand side of
anticommutator between the two supersymmetry
currents contains the bosonic currents of $W_{1+\infty}$ and $W_{\infty}$
as well as the central term.
All the structure constants and the central terms are determined
by Jacobi identities.

In \cite{BK},
the $U(K)$ matrix generalization of $W_{\infty}$ algebra
which is denoted by $W_{\infty}^K$ algebra 
is obtained.
From the $K$ values of complex bosons, the generators of this algebra
can be written in terms of bilinear of these complex bosons
contracted by $(K^2-1)$ generators of $SU(K)$ and $K \times K$
identity matrix with possible derivatives.
Of course, when the two generators in the commutator
are represented by this identity matrix,
then this leads to the one in \cite{PRS}.
Moreover, 
in \cite{OS}, 
the similar analysis (in the notation of
$W_{1+\infty}^L$) for  $L$ values of complex fermions
has been done and the generators are described by 
the bilinear of these complex fermions
contracted by $(L^2-1)$ generators of $SU(L)$
and the identity matrix which is $L \times L$ matrix
together with various derivatives.
Along the line of \cite{BPRSS}, the $W_{\infty}^{K,L}$ algebra
\cite{Odake}
has been obtained and the bosonic sector is given by
$W_{1+\infty}^L \times W_{\infty}^K$.
Instead of having the singlets and the adjoints of
the $SU(K)$ and $SU(L)$, the double index notation in \cite{OS}
leads to the fact that there are no antisymmetric $f$ symbol
or symmetric $d$ symbol in the commutators.
Under the condition of $K=L$, the ${\cal N}=2$ supersymmetry arises.

In \cite{BVd1,BVd2},
it is known that
the ${\cal N}=2$ supersymmetric linear $W_{\infty}[\la]$ algebra
is isomorphic to
the ${\cal N}=2$ supersymmetric linear $W_{\infty}[\frac{1}{2}-\la]$
algebra.
There exists a deformation parameter
$\la$ such that the ${\cal N}=2$ supersymmetric
$W_{\infty}[\la]$ (which is equivalent to
$W_{\infty}^{1,1}[\la]$ in later notation)
algebra can be truncated on the various subalgebras
for the particular values of $\la$.
At $\la=0$, the two fermions are identified with each other
and one of the bosons is identified with the derivative of other boson. 
Then the algebra reduces to the $W_{\infty}$ algebra.
At $\la=\frac{1}{2}$, the two bosons are identified with each other
and one of the fermions is identified with the derivative of
other fermion.
This implies the $W_{\infty}$
algebra.
As a subalgebra, the bosonic algebra contains
$W_{\infty}[\la]$ algebra and $W_{\infty}[\la+\frac{1}{2}]$ algebra.

In \cite{Ahn2203},
the multiple numbers of complex fermions and
complex bosons are introduced to find the exact correspondence
between the ${\cal N}=(2,2)$ SYK model \cite{MSW,Bulycheva,AP}
and the
${\cal N}=2$ $(\beta , \gamma)$ and $ (b , c)$ systems \cite{BVd1,BVd2}.
Because there are multiple numbers of chiral multiplets (or
Fermi multiplets) in the former, the multiple numbers of
$(\beta , \gamma)$ and $ (b , c)$ fields are crucial.
See also the previous work of 
\cite{CHR}.
Note that the fundamental OPEs between these fields
have the first order poles, compared to the ones in \cite{BPRSS}.
The generators of spin $1$ consisting of
the bilinear of $\beta$ field and $\gamma$ field with contraction and
spin $\frac{1}{2}$ consisting of $\beta$ field and $c$
field with contraction occur on the right hand sides of the
(anti)commutators.
At $\la=0$, the previous result in \cite{Ahn2202}
is reproduced. The bosonic subalgebra of the
${\cal N}=2$ supersymmetric linear $W_{\infty}^{K,K}[\la=0]$
algebra is given by
$W_{1+\infty}^K[\la=0]$ generated by fermions $(b,c)$ and
$W_{\infty}^K[\la+\frac{1}{2}=\frac{1}{2}]$
generated by bosons $(\beta,\gamma)$.
On the other hand, at $\la=\frac{1}{2}$,
the bosonic subalgebra
 of the
${\cal N}=2$ supersymmetric linear $W_{\infty}^{K,K}[\la=\frac{1}{2}]$
algebra
is given by
$W_{\infty}^K[\la=\frac{1}{2}]$ generated by fermions and
$W_{1+\infty}^K[\la+\frac{1}{2}=1]$ generated by bosons.

Recently, as noted in \cite{Ahn2208},
the structure constants obtained in \cite{Ahn2203} hold for
some constraints on the spins for the left hand sides of the
algebras with nonzero $\la$:
for general $h_1$ with the constrained $h_2$.
In \cite{Ahn2205},
some OPEs between the ${\cal N}=4$ multiplets
of low spins $h_1$  and $h_2$
are found explicitly for nonzero $\la$.
Moreover, the results in \cite{Ahn2208}
satisfy with some constraints on the spins also
\footnote{At $\la=0$, all the (anti)commutators are true.
At nonzero $\la$, under the condition of $h_1=h_2$,
all the (anti)commutators which do not contain
the spin $1$ or the spin $\frac{1}{2}$ currents
on the left hand side are valid.
If they contain one of these on the left hand sides,
the structure constants
depend on the $\la$ and cannot be written as
the known structure constants in \cite{Ahn2208}.
For $h_1\neq h_2$ at
nonzero $\la$, in general, the right hand sides
of the (anti)commutators contain the extra structures
which cannot be written in terms of the structure constants of
\cite{Ahn2208} and depend on the $\la$.
According to the results of present paper, 
all these constraints are gone and we have the explicit
(anti)commutators with complete structure constants for
any $\la, h_1$ and $h_2$.}.

In this paper, we would like to construct
the OPEs or (anti)commutators between the bifundamentals under
the $SU(2) \times SU(2)$ for {\it nonzero} deformation parameter $\la$
with {\it generic} $h_1$ and $h_2$, by reconsidering
\cite{Ahn2203,Ahn2208}. Once these are determined explicitly,
then the complete OPEs between the ${\cal N}=4$ multiplets,
related to the ${\cal N}=4$ supersymmetric linear
$W_{\infty}^{2,2}[\la]$ algebra,
are obtained and the full algebra
between the singlets and the adjoints in the $SU(K)$,
related to 
the ${\cal N}=2$ supersymmetric linear
$W_{\infty}^{K,K}[\la]$ algebra,
are determined completely.
 
In section $2$,
the complete algebra between the bifundamentals
under the $SU(2) \times SU(2)$ is obtained for any
$h_1,h_2$ and deformation parameter $\la$.
Then the ${\cal N}=4$ supersymmetric linear $W_{\infty}^{2,2}[\la]$
algebra is determined and its ${\cal N}=4$ superspace description
is presented.

In section $3$,
the complete  ${\cal N}=2$ supersymmetric linear $W_{\infty}^{K,K}[\la]$
algebra between the singlets and the adjoints
under the $SU(K)$ is found.

In section $4$, the main results of this paper are summarized.

In Appendices, all the details which do not appear in previous
sections are given \footnote{
  \label{lastfootnote} In the
  abstract, we use a simplified notation
  for $W_{\infty}[\la]$ algebra. For the ${\cal N}=4$ supersymmetric case,
  the $W_{\infty}[\la]$ stands for $W_{\infty}^{2,2}[\la]$
  and for the ${\cal N}=2$ supersymmetric case,
  the $W_{\infty}[\la]$ stands for $W_{\infty}^{K,K}[\la]$.
  The bosonic subalgebras stand for $W_{1+\infty}^K[\la] \times
  W_{1+\infty}^K[\la+\frac{1}{2}]$.
  Maybe the $W_{1+\infty}[\la]$ is better than the $W_{\infty}[\la]$
  appearing in the title but we keep the notation of \cite{BVd1,BVd2}.}.

\section{ The ${\cal N}=4$ supersymmetric linear $W_{\infty}^{2,2}[\la]$
  algebra  }

\subsection{The bifundamental currents}

The operator product expansions
of the $(\beta , \ga)$ and $(b , c)$ systems are given by \cite{CHR}
\bea
\ga^{i,\bar{a}}(z)\, \beta^{\bar{j},b}(w) =
\frac{1}{(z-w)}\, \de^{i \bar{j}}\, \de^{\bar{a} b} + \cdots\, ,
\qquad
c^{i, \bar{a}}(z) \, b^{\bar{j},b}(w) =
\frac{1}{(z-w)}\, \de^{i \bar{j}}\, \de^{\bar{a} b} + \cdots\, .
\label{fundOPE}
\eea
There exist fundamental indices $a, b $ and
antifundamental indices $\bar{a}, \bar{b}$
of $SU(2)$.
Similarly there are  fundamental indices $i, j $
and antifundamental indices $\bar{i}, \bar{j}$ of
$SU(N)$.
The $(\beta, \ga)$ fields are bosonic operators while
the $(b, c)$ fields are fermionic operators.
Note that the
conformal weights of $(\beta , \ga)$ fields are given by
$(\la,1-\la)$ while
the conformal weights of $(b , c)$ fields are given by
$(\frac{1}{2}+\la,\frac{1}{2}-\la)$, under the stress energy tensor
(\ref{Lterm}) \footnote{\label{latrans}
  Under the $\la \rightarrow \la+\frac{1}{2}$,
  the conformal weights for the former become the ones for the latter and
  under the $\la \rightarrow \la -\frac{1}{2}$,
  they for the latter become the ones for the former.
  These transformations will provide the symmetries in the bosonic
  subalgebras later.}.

The $SU(N)$ singlet currents can be determined 
by summing over the (anti)fundamental indices of $SU(N)$
\footnote{The free field theory construction we consider
  appears in the realization of
${\cal N}=4$ coset $\frac{SU(N+2)}{SU(N)}$
  model \cite{EGR}. Then the currents we consider,
  which are bilinear in the free fields,
should be invariant under the
denominator $SU(N)$. In the subsections $2.3$ and $2.4$,
the free field realization associated with
this ${\cal N}=4$ coset model, under the identification of the $\la$
with the corresponding parameter appearing in the ${\cal N}=4$ coset model
\cite{Ahn2205},
can be found.}
and are given by \cite{Ahn2205,Ahn2208} from the work of
\cite{BVd1,BVd2}
\bea
W_{F,h}^{\lambda,\bar{a}b}(z)
& = &  (-4 q)^{h-2}\sum_{i=0}^{h-1} a^{i}
(h,\lambda+\tfrac{1}{2})\,
\partial_z^{h-i-1}
((\partial_z^{i}b^{\bar{l} b})\, \delta_{\bar{l} l}\,c^{l \bar{a}})(z),
\nonu \\
W_{B,h}^{\lambda,\bar{a}b}(z)
& = & (-4 q)^{h-2}\sum_{i=0}^{h-1} a^{i}(h,\lambda)\,
\partial_z^{h-i-1}
((\partial_z^{i}\beta^{\bar{l} b})\, \delta_{\bar{l} l}\,\gamma^{l \bar{a}})(z),
\nonu \\
Q_{h+\frac{1}{2}}^{\lambda,\bar{a}b}(z)
& = & \sqrt{2}\,(-4q)^{h-1}\sum_{i=0}^{h-1} \beta^{i}(h+1,\lambda)\,
\partial_z^{h-i-1}
((\partial_z^{i}b^{\bar{l} b})\, \delta_{\bar{l} l}\,\gamma^{l \bar{a}})(z),
\nonu \\
\bar{Q}_{h+\frac{1}{2}}^{\lambda,a\bar{b}}(z)
& = & \sqrt{2}\, (-4q)^{h-1}\sum_{i=0}^{h} \alpha^{i}(h+1,\lambda)\,
\partial_z^{h-i}
((\partial_z^{i}\beta^{\bar{l} a})\,
\delta_{\bar{l} l}\,c^{l \bar{b}})(z).
\label{fourcurrents}
\eea
By counting the number of (anti)fundamental indices,
there exist four components labeled by $(\bar{a} b)=(11,12,21,22)$
in each current.
Then there are  eight bosonic currents, $W_{F,h}^{\lambda,\bar{a}b}$ and
$W_{B,h}^{\lambda,\bar{a}b}$,  for the weight
$h=1,2, \cdots $ and eight
fermionic currents, $Q_{h+\frac{1}{2}}^{\lambda,\bar{a}b}$ and $
\bar{Q}_{h+\frac{1}{2}}^{\lambda,a\bar{b}}$, 
for the weight
$h+\frac{1}{2}=\frac{3}{2}, \frac{5}{2}, \cdots$
as well as four fermionic currents 
$\bar{Q}^{\la, a \bar{b}}_{\frac{1}{2}}$
of the weight $\frac{1}{2}$ in
(\ref{fourcurrents}), which is rescaled by $\frac{1}{2}$,
compared to the one in \cite{Ahn2208}.
Note that four fermionic currents
$Q^{\la,\bar{a} b}_{ \frac{1}{2}}$ of  the weight $\frac{1}{2}$
are identically zero.

The relative coefficients appearing in (\ref{fourcurrents})
depend on the conformal weight $h$ and deformation
parameter $\la$ explicitly. They are given by the binomial
coefficients denoted by parentheses and the rising Pochhammer symbols
where $(a)_n \equiv a(a+1) \cdots (a+n-1)$ and $n$
is a nonnegative integer \cite{PRS}
as follows \cite{BVd1,BVd2}:
\bea 
 a^i(h, \la) \equiv \left(\begin{array}{c}
h-1 \\  i \\
 \end{array}\right) \, \frac{(-2\la-h+2)_{h-1-i}}{(h+i)_{h-1-i}}\, ,
 \qquad 0 \leq i \leq h-1\, ,
 \nonu \\
 \al^i(h, \la) \equiv \left(\begin{array}{c}
h-1 \\  i \\
 \end{array}\right) \, \frac{(-2\la-h+2)_{h-1-i}}{(h+i-1)_{h-1-i}}\, ,
 \qquad 0 \leq i \leq h-1\, ,
 \nonu \\
  \beta^i(h, \la) \equiv \left(\begin{array}{c}
h-2 \\  i \\
  \end{array}\right) \, \frac{(-2\la-h+2)_{h-2-i}}{(h+i)_{h-2-i}}\, ,
  \qquad 0 \leq i \leq h-2 \, .
  \label{coeff}
\eea
The $\la$ dependence appears in the numerators of (\ref{coeff}).
The degrees of the polynomials in $\la$ are given by
$(h-1-i)$, $(h-1-i)$ and $(h-2-i)$ respectively.
The corresponding ones in (\ref{fourcurrents})
are given by $(h-1)$, $(h-1)$, $(h-1)$ and $h$ respectively.

The stress energy tensor of conformal weight $2$ is given by
\cite{BVd1,BVd2,Ahn2205,Ahn2208}
\bea
L & = &
W^{\la,\bar{a} a}_{\mathrm{B},2}+
W^{\la,\bar{a} a}_{\mathrm{F},2}\, ,
\label{Lterm}
\eea 
and the corresponding central charge is 
\bea
c_{cen}= 6\,N\, (1-4\la)\, ,
\label{central}
\eea
which depends on the deformation parameter $\la$ explicitly.
The above bosonic and fermionic currents in
(\ref{fourcurrents}) are quasiprimary operators
under the stress energy tensor (\ref{Lterm}) \footnote{
\label{qsyk}
  The rank $(q_{syk}+1)$ of the interaction of the two
  dimensional ${\cal N}=(2,2)$ SYK model corresponds to
$(q_{syk}+1) =\frac{1}{2\la }$.}.

\subsection{ The OPEs or (anti)commutators between the
bifundamentals}

Let us calculate the OPEs between the currents in (\ref{fourcurrents}).
As an intermediate step, we obtain the following OPE,
from (\ref{fundOPE}), 
between the first currents without any derivatives
\footnote{The Thielemans package \cite{Thielemans} is used
  together with a mathematica \cite{mathematica} all
the times in this paper.}
\bea
((\partial_z^{i_1}b^{\bar{l} b})\, \delta_{\bar{l} l}\,c^{l \bar{a}})(z)\,
((\partial_w^{i_2}b^{\bar{l} d})\, \delta_{\bar{l} l}\,c^{l \bar{c}})(w) 
&= & \delta^{\bar{a}d}\delta^{\bar{c}b}
\,(-1)^{i_2+1}\frac{N\,i_1 ! \,j_2!}{(i_1+i_2+1)!}\,\partial_z^{i_1+i_2+1}\Bigg[\frac{1}{(z-w)}\Bigg] 
\nonumber \\ 
&+ & \delta^{\bar{a}d} \,\sum^{i_2}_{r=0} 
(-1)^r \binom{i_2}{r} \partial_z^r\Bigg[\frac{
    ((\partial_w^{i_1+i_2-r}b^{\bar{l} b})\, \delta_{\bar{l} l}\,c^{l \bar{c}})(w)}{
(z-w)} 
\Bigg] 
\nonumber \\
&- &
\delta^{\bar{c}b} \,\sum^{i_1}_{r=0} 
(-1)^r \binom{i_1}{r} \partial_w^r\Bigg[\frac{
((\partial_w^{i_1+i_2-r}b^{\bar{l} d})\, \delta_{\bar{l} l}\,c^{l \bar{a}})(w)}{
(z-w)}
\Bigg]
\nonu \\
& + & \cdots.
\label{eq:2}
\eea
The central term in (\ref{eq:2})
comes from the double contractions of this OPE.
Note that the right hand side contains
the quadratic terms in $(b,c)$ fields with various derivatives.

Then it is straightforward to calculate the following OPE
\bea
&& \partial_z^{h_1-i_1-1}\,
((\partial_z^{i_1}b^{\bar{l} b})\, \delta_{\bar{l} l}\,c^{l \bar{a}})(z)\,
\partial_w^{h_2-i_2-1}\,((\partial_w^{i_2}b^{\bar{l} d})\,
\delta_{\bar{l} l}\,c^{l \bar{c}})(w)  \nonu \\
&& =  \delta^{\bar{a}d}\delta^{\bar{c}b}
\,(-1)^{h_2}\frac{N\,i_1 ! \,i_2!\,(i_1+i_2+2)_{h_1+h_2-i_1-i_2-2}}{(h_1+h_2-1)!}\,
\partial_z^{h_1+h_2-1}\Bigg[\frac{1}{(z-w)}\Bigg] 
\nonumber
\\ 
&& +  \delta^{\bar{a}d} \,\sum^{i_2}_{r=0} 
(-1)^r \binom{i_2}{r} \partial_z^{h_1-i_1-1+r}\partial_w^{h_2-i_2-1}
\Bigg[\frac{((\partial_w^{i_1+i_2-r}b^{\bar{l} b})\, \delta_{\bar{l} l}\,
c^{l \bar{c}})(w)}{(z-w)} 
\Bigg]
\nonumber
\\
& & -
\delta^{\bar{c}b} \,\sum^{i_1}_{r=0} 
(-1)^r \binom{i_1}{r} 
\partial_z^{h_1-i_1-1}\partial_w^{h_2-i_2-1+r}
\Bigg[\frac{((\partial_w^{i_1+i_2-r}b^{\bar{l} d})\, \delta_{\bar{l} l}\,
c^{l \bar{a}})(w)}{(z-w)}
\Bigg]+\cdots.
\label{eq:3}
\eea
Let us consider the numerator appearing inside the bracket of
the second line of
(\ref{eq:3}).
We expect that we should write down it in terms of the derivative
of $W_{F,h+1}^{\lambda,\bar{a}b}$.
The number of derivative of $b$ field is taken as $s$
and the conformal weight for the derivative term is $(s+1)$.
Then we can think of all possible terms having the weight $(s+1)$
from the currents, $W_{F,1}^{\lambda,\bar{a}b}$ with $s$ derivatives,
$\cdots$, 
$W_{F,s}^{\lambda,\bar{a}b}$ with one derivative.
We introduce the undetermined coefficients appearing in these
derivative terms. They depend on the $\la$ parameter and the
number of derivative  of $b$ field $s$.
Obviously, the derivatives of $c$ field should vanish and
all the derivatives of $b$ field have the numerical values $1$.
By varying the $s$ starting from $0$ to $10$,
the above undetermined coefficients can be fixed
and the $s$ dependence can be determined explicitly.
It turns out that the following nontrivial identity
satisfies
\bea
((\partial_z^s b^{\bar{l} b})\, \delta_{\bar{l} l}\,c^{l \bar{a}})(z)
&= &
\sum_{h=0}^{s}(-1)^{s+1}\,4^{-h}\,
q^{1-h} \, Y_{W_F}(s-h,s+1,\lambda)\,\partial_z^{s-h}
W_{F,h+1}^{\lambda,\bar{a}b}(z),
\label{bcidentity}
\eea
where the $\la$ dependent coefficient
is given by
\bea
Y_{W_F}(h,s,\lambda)&\equiv &
\frac{2\,(-2\lambda-s+1)_h\,[s-1]_{h-1}}{h!\,[2s-h-1]_{h-1}}=
\frac{2\,(-2\lambda-s+1)_h\,(1-s)_{h-1}}{h!\,(1-2s+h)_{h-1}}.
\label{ywf}
\eea
The degree of this polynomial in $\la$ is $h$ and the falling
Pochhammer symbol $[a]_n \equiv a(a-1) \cdots (a-n+1)$ is used.
Moreover, the identity $[a]_n =(-1)^n\, (-a)_n$ is used.

The contribution from
$((\partial^{i_1+i_2-r}b^{\bar{l} b})\, \delta_{\bar{l} l}\,
c^{l \bar{c}})$, by using (\ref{ywf}),  can be written as 
\bea
&& \sum_{h_3=1}^{i_1+i_2-r+1}(-1)^{i_1+i_2-r+1}4^{1-h_3}\,
q^{2-h_3}\,
\frac{2\, (-i_1-i_2+r-2
\lambda)_{i_1+i_2-r-h_3+1} \,(-i_1-i_2+r)_{i_1+i_2-r-h_3}}{
  (i_1+i_2-r-h_3+1)!\,(-i_1-i_2+r-h_3)_{i_1+i_2-h_3-r}}\,
\nonu \\
&& \times \, \partial^{i_1 + i_2 - r +1- h_3}_w W_{F,h_3}^{\lambda,\bar{c}b}(w),
\label{prev}
\eea
by identifying $s =i_1+i_2-r$, $h=h_3-1$ and $\bar{a}=\bar{c}$.
After multiplying all the factors into (\ref{prev}),
we obtain
\bea
&& \delta^{\bar{a}d}\sum_{i_1=0}^{h_1-1\,}\sum_{i_2=0}^{h_2-1\,}
\sum_{r=0}^{i_2\,}
\sum_{h_3=1}^{h_1+h_2-1}
(-1)^{h_1+h_2+i_1+i_2+1}4^{h_1+h_2-h_3-3}\,
q^{h_1+h_2-h_3-2}\,
a^{i_1}(h_1,\lambda+\tfrac{1}{2})a^{i_2}(h_2,\lambda+\tfrac{1}{2})
\nonumber
\\
& & \times
\frac{2\,(-i_1-i_2+r-2\lambda)_{i_1+i_2-r-h_3+1} \,
  (-i_1-i_2+r)_{i_1+i_2-r-h_3}}{(i_1+i_2-r-h_3+1)!\,(-i_1-i_2+r-h_3)_{i_1+i_2-h_3-r}}
\binom{i_2}{r}
\nonumber
\\
&& \times\partial_z^{h_1-i_1-1+r}\partial_w^{h_2-i_2-1}
\Bigg[\frac{ \partial_w^{i_1+i_2-r-h_3+1}
W_{F,h_3}^{\lambda, \bar{c}b}(w)}{(z-w)}
\Bigg],
\label{eq:5}
\eea
where the upper bound for $h_3$, $(i_1+i_2-r+1)$, is changed into
$(h_1+h_2-1)$ because the values of $h_3 \geq (i_1+i_2-r+2)$
do not contribute due to the factorial function
$(i_1+i_2-r-h_3+1)!$ in the denominator.
This changing of upper bound in (\ref{eq:5})
is crucial to the later discussion.

By introducing a new variable $r'\equiv i_1+i_2-r$
and removing a prime notation,
the following relation satisfies
\bea
&& \delta^{\bar{a}d}\sum_{i_1=0}^{h_1-1\,}\sum_{i_2=0}^{h_2-1\,}
\sum_{r=0}^{i_1+i_2\,}
\sum_{h_3=1}^{h_1+h_2-1}
(-1)^{h_1+h_2+i_1+i_2+1}4^{h_1+h_2-h_3-3}\,
q^{h_1+h_2-h_3-2}\,
a^{i_1}(h_1,\lambda+\tfrac{1}{2})
a^{i_2}(h_2,\lambda+\tfrac{1}{2})
\nonumber
\\
&& \times
\frac{2\,\prod_{x=0}^{r-h_3}(-r-2\lambda+x)\,
(-r)_{-h_3+r}}
{(1-h_3+r)!\,(-h_3-r)_{-h_3+r}}
\binom{i_2}{i_1+i_2-r}
\nonu \\
&& \times 
\partial_z^{h_1+i_2-r-1}\partial_w^{h_2-i_2-1}
\Bigg[\frac{\partial_w^{r-h_3+1}\,
W_{F,h_3}^{\lambda, \bar{c}b}(w)}{(z-w)} 
\Bigg],
\label{eq:6}
\eea
where we intentionally add the zero value which is equal to
the summation of $r$ from $r=0$ to $r=i_1-1$ from the above
binomial.
Moreover,
we can move the derivative appearing inside of the bracket
(\ref{eq:6})
by using 
the relation
\bea
\Bigg[\frac{\partial_w^{r-h_3+1}\,
W_{F,h_3}^{\lambda, \bar{c}b}(w)}{(z-w)} 
  \Bigg] & = &
\sum_{k'=0}^{r-h_3+1} \, \binom{r-h_3+1}{r-h_3+1-k'} \partial_z^{k'}
\partial_w^{r-h_3+1-k'}
\Bigg[\frac{
W_{F,h_3}^{\lambda, \bar{c}b}(w)}{(z-w)}  \Bigg]  
\label{deriden} \\
&=&\sum_{k=h_2-i_2-1}^{h_2-h_3-i_2+r} \, \binom{r-h_3+1}{1+k-h_2+i_2}
\partial_z^{-k+h_2-h_3-i_2+r}
\partial_w^{1+k-h_2+i_2}
\Bigg[\frac{
W_{F,h_3}^{\lambda, \bar{c}b}(w)}{(z-w)}  \Bigg],
\nonu
\eea
where the new variable $k \equiv -k'+h_2-h_3-i_2+r$
which is related to the upper bound of (\ref{deriden}) is introduced.

By substituting (\ref{deriden}) into (\ref{eq:6}), the following
result satisfies
\bea
&& \delta^{\bar{a}d}
\sum_{h_3=1}^{h_1+h_2-1}
\sum_{k=0}^{h_1+h_2-h_3-1}
\sum_{i_1=0}^{h_1-1\,}\sum_{i_2=0}^{h_2-1\,}
\sum_{r=0}^{i_1+i_2\,}
(-1)^{h_1+h_2+i_1+i_2+1}4^{h_1+h_2-h_3-3}\,
a^{i_1}(h_1,\lambda+\tfrac{1}{2})
a^{i_2}(h_2,\lambda+\tfrac{1}{2})
\nonumber
\\
&& \times
q^{h_1+h_2-h_3-2}\,
\frac{2\,\prod_{x=0}^{r-h_3}(-r-2\lambda+x)\,
(-r)_{-h_3+r}}
{(1-h_3+r)!\,(-h_3-r)_{-h_3+r}}
\binom{i_2}{i_1+i_2-r} \,  \binom{r-h_3+1}{1+k-h_2+i_2}
\nonu \\
&& \times 
\partial_z^{h_1+h_2-h_3-k-1}\partial_w^{k}
\Bigg[\frac{
W_{F,h_3}^{\lambda, \bar{c}b}(w)}{(z-w)} 
\Bigg].
\label{intermediate}
\eea
Note that the upper bound and the lower bound for $k$ in
(\ref{intermediate}) are changed but
the extra terms do not contribute because of the constraint on the $k$
appearing in the binomial.

By using the relations $(-x)_n=(-1)^n (x-n+1)_n$ and $(a)_n=\frac{(a+n-1)!}{
(a-1)!}$, the following relation satisfies 
\bea
\frac{2\,
(-r)_{-h_3+r}}
     {(1-h_3+r)!\,(-h_3-r)_{-h_3+r}}=
4 \binom{r}{h_3-1}\,\frac{(2h_3-1)!}{(h_3+r)!}.     
\label{otherelation}
\eea

Therefore, the contribution from the second
term of (\ref{eq:3}) with the correct factors,
together with (\ref{otherelation}), can be described by
\bea
&& \delta^{\bar{a}d}\,
\sum_{h_3=1}^{h_1+h_2-1\,\, }\sum_{k=0}^{h_1+h_2-h_3-1}
\,
(-1)^{h_1+h_2+1}\,(4q)^{h_1+h_2-h_3-2}
(2h_3-1)!\,
S^{\,\,h_1,h_2,h_3,k}_{F,\,R}(\lambda)\,
\nonu \\
&& \times \partial_z^{h_1+h_2-h_3-k-1}\,
\partial_w^{k}\,\bigg[\frac{W_{F,h_3}^{\lambda,\bar{c}b}(w)}{(z-w)}\bigg],
\label{almostfinal}
\eea
where the $\la$ dependent coefficients in (\ref{almostfinal})
are given by
\bea
S^{\,\,h_1,h_2,h_3,k}_{F,\,R}(\lambda)&\equiv &
\sum_{i_1=0}^{h_1-1}\sum_{i_2=0}^{h_2-1}\,\sum_{r=0}^{i_1+i_2}
\Bigg[ \frac{(-1)^{i_1+i_2}}{(h_3+r)!}\,
a^{i_1}(h_1,\lambda+\tfrac{1}{2})
a^{i_2}(h_2,\lambda+\tfrac{1}{2})
\nonu \\
&\times&
\binom{r}{h_3-1}\binom{i_2}{i_1+i_2-r}\binom{1-h_3+r}{1-h_2+i_2+k}
\prod_{j=0}^{r-h_3}(-r-2\lambda+j) \Bigg].
\label{WFWFstructR}
\eea
We can do similar analysis by taking the third term of (\ref{eq:3})
\footnote{If the upper bound is less than the lower bound in the
  product notation appearing in (\ref{WFWFstructR}),
  we treat them to be $1$ according to the
  ``empty'' product. }.

Then the final OPE between
$W_{F,h_1}^{\lambda,\bar{a}b}(z)$ and $W_{F,h_2}^{\lambda, \bar{c}d}(w)$,
by considering two other contributions,
is summarized by
\bea
W_{F,h_1}^{\lambda,\bar{a}b}(z)\, W_{F,h_2}^{\lambda, \bar{c}d}(w)
&=& \delta^{\bar{a}d}\delta^{\bar{c}b}
\sum_{i_1=0}^{h_1-1}\sum_{i_2=0}^{h_2-1}(-1)^{h_1}4^{h_1+h_2-4}\,
a^{i_1}(h_1,\lambda\!+\!\tfrac{1}{2})a^{i_2}
(h_2,\lambda\!+\!\tfrac{1}{2})\frac{N\,i_1 !\,i_2 !}{(i_1+i_2+1)!}
\nonu \\
& \times & q^{h_1+h_2-4}\,
\partial_z^{h_1+h_2-1}\bigg[\frac{1}{(z-w)}\bigg]
\nonu \\
&+ & \sum_{h_3=1}^{h_1+h_2-1\,\, }\sum_{k=0}^{h_1+h_2-h_3-1}
\,
(-1)^{h_1+h_2+1}\,(4q)^{h_1+h_2-h_3-2}
(2h_3-1)!
\nonu \\
&\times &
\Bigg(\delta^{\bar{a}d}\,S^{\,\,h_1,h_2,h_3,k}_{F,\,R}(\lambda)\,
\partial_z^{h_1+h_2-h_3-k-1}\,
\partial_w^{k}\,\bigg[\frac{W_{F,h_3}^{\lambda,\bar{c}b}(w)}{(z-w)}\bigg]
\nonu \\
&- & \delta^{\bar{c}b}\,S^{\,\,h_1,h_2,h_3,k}_{F,\,L}(\lambda)\,  
\partial_z^{k}\,
\partial_w^{h_1+h_2-h_3-k-1}\,\bigg[\frac{ W_{F,h_3}^{\lambda,
    \bar{a}d}(w)}{(z-w)}\bigg]\Bigg) + \cdots,
\label{WFWF}
\eea
where other type of $\la$ dependent coefficients is
\footnote{Ther is no meaning for the assignments in the $R$ and $L$
appearing in (\ref{WFWFstructR}) and (\ref{WFWFstructL}).}
\bea
S^{\,\,h_1,h_2,h_3,k}_{F,\,L}(\lambda)&\equiv &
\sum_{i_1=0}^{h_1-1}\sum_{i_2=0}^{h_2-1}\,\sum_{r=0}^{i_1+i_2}
\Bigg[ \frac{(-1)^{i_1+i_2}}{(h_3+r)!}\,
a^{i_1}(h_1,\lambda+\tfrac{1}{2})
a^{i_2}(h_2,\lambda+\tfrac{1}{2})
\nonu \\
&\times&
\binom{r}{h_3-1}\binom{i_1}{i_1+i_2-r}\binom{1-h_3+r}{1-h_1+i_1+k}
\prod_{j=0}^{r-h_3}( -r-2\lambda+j) \Bigg].
\label{WFWFstructL}
\eea
The higest power of $\frac{1}{(z-w)}$ in the field dependent
terms of (\ref{WFWF})
is given by $(h_1+h_2-h_3)$. Then the total
conformal spin of the right hand side of this term
is given by $(h_1+h_2)$. On the other hands, the corresponding
lowest power of $\frac{1}{(z-w)}$ is $1$ when $k=h_1+h_2-h_3-1$
for $\delta^{\bar{a}d}$ and $k=0$ for $\delta^{\bar{c}b}$.
The number of derivative with respect to $w$ is given by
$(h_1+h_2-h_3-1)$.
Note that the number of powers in the derivatives
appearing in the last two lines (\ref{WFWF}) arises
in symmetric way.
Compared to the one in (\ref{WFWFstructR}), the difference in
(\ref{WFWFstructL})
appears
in some elements of binomials. Or according to the exchange of
$i_1 \leftrightarrow i_2$ and $h_1 \leftrightarrow h_2$,
the structure constant (\ref{WFWFstructR}) becomes
the structure constant (\ref{WFWFstructL}) or vice versa.
The free indices in (\ref{WFWF}) are given by $h_1$ and $h_2$ while
$i_1,i_2,h_3$ and $k$ are dummy variables.

It is straightforward to determine the commutator
corresponding to (\ref{WFWF}) by using the standard method of
conformal field theory.
According to the analysis of \cite{PRS},
the derivative terms, $\pa_z^h$ and $\pa_w^{h'}$ can
be replaced by $(-1)^h \, [m+h-1]_h$ and
$(-1)^{h'} \, [n+h'-1]_{h'}$ respectively.
It turns out that \footnote{Let us emphasize that inside of the
  summations there are various Pochhammer symbols. By using the
  mathematica \cite{mathematica},
  for example, the summation over $k$ will lead to
  some complicated combinations of special functions with
  $h_1,h_2,h_3$ and modes $m,n$ dependences. If we substitute
  the particular values for $h_1,h_2$ and $h_3$, then still
  the resulting expression is again complicated and is not written in
  terms of polynomials of $m$ and $n$. Therefore, we
  introduce a function which depends on the above variables
  $h_1,h_2$ and $h_3$ and use a
  Table and Do 
  commands inside a mathematica.
  Then we obtain the expected polynomials of $m$ $n$ for
$h_1,h_2$ and $h_3$.}
\bea
\comm{(W_{F,h_1}^{\lambda,\bar{a}b})_m}{(W_{F,h_2}^{\lambda, \bar{c}d})_n}
&=& \delta^{\bar{a}d}\delta^{\bar{c}b}\,
q^{h_1+h_2-4}\,
c_{W_F}^{h_1,h_2}(\la)\,[m+h_1-1]_{h_1+h_2-1}\,\delta_{m+n} \label{firstcomm} \\
&+ & \sum_{h_3=1}^{h_1+h_2-1 \,\, }\sum_{k=0}^{h_1+h_2-h_3-1}
\,
(-1)^{h_3}\,(4q)^{h_1+h_2-h_3-2}\,
(2h_3-1)! \nonu \\
& \times &
\Bigg[\delta^{\bar{a}d}\,S^{\,\,h_1,h_2,h_3,k}_{F,\,R}(\lambda)\, [m+h_1-1]_{h_1+h_2-h_3-k-1}\,[n+h_2-1]_{k}\,(W_{F,h_3}^{\lambda,\bar{c}b})_{m+n}
\nonu \\
&-& \delta^{\bar{c}b}\,S^{\,\,h_1,h_2,h_3,k}_{F,\,L}(\lambda)\,  [m+h_1-1]_{k}\,[n+h_2-1]_{h_1+h_2-h_3-k-1}\, (W_{F,h_3}^{\lambda,\bar{a}d})_{m+n}\Bigg],
\nonu
\eea
where mode independent central term appearing in (\ref{firstcomm})
is given by
\bea
c_{W_F}^{h_1,h_2}(\la)=  N \sum_{i_1=0}^{h_1-1}\sum_{i_2=0}^{h_2-1}
(-1)^{h_2-1}\,4^{h_1+h_2-4}
a^{i_1}(h_1,\lambda+\tfrac{1}{2})a^{i_2}(h_2,\lambda+\tfrac{1}{2})
\frac{i_1 ! i_2 !}{(i_1+i_2+1)!}.
\label{cwf}
\eea

The OPE (\ref{WFWF}) or commutator (\ref{firstcomm})
is one of the main result of this paper.
In Appendix $A$ we will present other remaining OPEs or
(anti)commutators. We have found the complete algebra
between the currents in (\ref{fourcurrents})
for generic deformation parameter $\la$ for any
conformal weights $h_1$ and $h_2$ where $h_1, h_2 = 1, 2,
3, \cdots$.
At $\la=0$, obviously, the previous results \cite{Ahn2208}
mare reproduced.

\subsection{  The ${\cal N}=4$ supersymmetric linear $W_{\infty}^{2,2}[\la]$
  algebra }

The ${\cal N}=4$ multiplet can be described by
\cite{Schoutens,BCG,AK1509}
the following ${\cal N}=4$ super field
\bea
\bold{\Phi}^{(h)}(Z)
    =\Phi^{(h)}_0(z)
    +\theta^i\,\Phi^{(h),i}_\frac{1}{2}(z)
    +\theta^{4-ij}\,\Phi^{(h),ij}_1(z)
    +\theta^{4-i}\,\Phi^{(h),i}_\frac{3}{2}(z)
     +\theta^{4-0}\,\Phi^{(h)}_2(z),
\label{bigPhi}
\eea
where $SO(4)$ indices $i, j=1,2, \cdots, 4$.
The conventions for the ${\cal N}=4$ superspace coordinates and the left
covariant spinor derivatives are taken from \cite{Schoutens,AK1509}.
The Grassmann coordinate $\theta^i$ has a weight $-\frac{1}{2}$
\footnote{The coordinates of ${\cal N}=4$  superspace 
can be written in terms of 
$(Z, \overline{Z})$ where 
$Z=(z, \theta^i)$, $\overline{Z} =(\bar{z}, \bar{\theta}^i)$
and the $SO(4)$-vector index $i$ is given by $i=1, \cdots, 4$. 
The left covariant spinor derivative 
is defined by 
$ D^i = \theta^i \frac{\pa}{\pa z } +  \frac{\pa}{\pa {\theta^i}}$
with the anticommutators 
$
\{ D^i, D^j \} = 2 \delta^{ij} \frac{\pa}{\pa z}$.
The notation
$\theta^{4-0}$ stands for 
$\theta^1 \,\theta^2 \,\theta^3 \,\theta^4$. 
The complement $4-i$ is defined such that
$\theta^1 \,\theta^2 \,\theta^3 \, \theta^{4} = \theta^{4-i}\,
\theta^{i}$ \cite{Schoutens,AK1509}.
}.

The lowest component of (\ref{bigPhi}) having a weight $h$
is described by \cite{Ahn2208}
\bea
\Phi_0^{(h)} & = &\frac{q^{2-h}(-4)^{h-2}}{(2h-1)}\,
\Bigg[ -(h-2\la)\,
  W^{\la,\bar{a} a }_{\mathrm{F},h}+
  (h-1+2\la) \, W^{\la,\bar{a} a}_{\mathrm{B},h} \Bigg],
\label{lowest}
\eea
where the bosonic currents are given by (\ref{fourcurrents}) and there
are summation over $SU(2)$ indices.

It is known that the next lowest components in (\ref{bigPhi})
can be determined by using the four supersymmetry currents of
${\cal N}=4$ stress energy tensor and the above lowest component
(\ref{lowest}).
By computing the first order poles of these OPEs, we obtain
\cite{Ahn2208} with an extra minus sign
\bea
\Phi^{(h),1}_{\frac{1}{2}}
&
=& - q^{1-h}\, (-4)^{h-3}\,
\Bigg[-\frac{1}{2}\,(
Q^{\la,11}_{h+\frac{1}{2}}
+i\sqrt{2}\,Q^{\la,12}_{h+\frac{1}{2}}
+2i \sqrt{2}\,Q^{\la,21}_{h+\frac{1}{2}}
-2\,Q^{\la,22}_{h+\frac{1}{2}}
\nonu \\
& + & 2\,\bar{Q}^{\la,11}_{h+\frac{1}{2}}
+2i \sqrt{2}\, \bar{Q}^{\la,12}_{h+\frac{1}{2}}
+i\sqrt{2}\,\bar{Q}^{\la,21}_{h+\frac{1}{2}}
-\bar{Q}^{\la,22}_{h+\frac{1}{2}}
)\,\Bigg],
\nonu\\
\Phi^{(h),2}_{\frac{1}{2}}
&
=& -  q^{1-h}\, (-4)^{h-3}\,\Bigg[
\frac{i}{2}\,(
Q^{\la,11}_{h+\frac{1}{2}}
+2i\sqrt{2} \,Q^{\la,21}_{h+\frac{1}{2}}
-2 \,Q^{\la,22}_{\frac{3}{2}}
+2\, \bar{Q}^{\la,11}_{h+\frac{1}{2}}
+2i\sqrt{2} \, \bar{Q}^{\la,12}_{h+\frac{1}{2}}
-\bar{Q}^{\la,22}_{h+\frac{1}{2}}
)\, \Bigg],
\nonu\\
\Phi^{(h),3}_{\frac{1}{2}}
&
=& -  q^{1-h}\, (-4)^{h-3}\,\Bigg[
\frac{i}{2}\,(
Q^{\la,11}_{h+\frac{1}{2}}
+i\sqrt{2} \,Q^{\la,12}_{h+\frac{1}{2}}
-2\,Q^{\la,22}_{h+\frac{1}{2}}
+2\, \bar{Q}^{\la,11}_{h+\frac{1}{2}}
+i \sqrt{2} \, \bar{Q}^{\la,21}_{h+\frac{1}{2}}
-\bar{Q}^{\la,22}_{h+\frac{1}{2}}
)\, \Bigg],
\nonu\\
\Phi^{(h),4}_{\frac{1}{2}}
&
=& -  q^{1-h}\, (-4)^{h-3}\,\Bigg[
\frac{1}{2}\,Q^{\la,11}_{h+\frac{1}{2}}
+Q^{\la,22}_{h+\frac{1}{2}}
-\bar{Q}^{\la,11}_{h+\frac{1}{2}}
-\frac{1}{2}\,\bar{Q}^{\la,22}_{h+\frac{1}{2}} \Bigg],
\label{nextlowest}
\eea
where the fermionic currents are given in (\ref{fourcurrents}).

The following six components of (\ref{bigPhi})
can be determined by calculating the OPEs between
the previous four supersymmetry currents
and the components appearing in (\ref{nextlowest}) and reading off the
first order poles \cite{Ahn2208}
 with an extra minus sign
\bea
\Phi^{(h),12}_{1}
&
=&   q^{1-h}\, (-4)^{h-3}\,\Bigg[
2i\,W^{\la,11}_{\mathrm{B},h+1}
-\sqrt{2}\,W^{\la,12}_{\mathrm{B},h+1}
-2i\,\,W^{\la,22}_{\mathrm{B},h+1}
\nonu \\
& + & 2i\,W^{\la,11}_{\mathrm{F},h+1}
-2\sqrt{2}\,W^{\la,12}_{\mathrm{F},h+1}
-2i\,W^{\la,22}_{\mathrm{F},h+1}\, \Bigg],
\nonu\\
\Phi^{(h),13}_{1}
&
=& 
  q^{1-h}\, (-4)^{h-3}\,\Bigg[
-2i\,W^{\la,11}_{\mathrm{B},h+1}
+4\sqrt{2}\,W^{\la,21}_{\mathrm{B},h+1}
+2i\,\,W^{\la,22}_{\mathrm{B},h+1}
\nonu \\
& - & 2i\,W^{\la,11}_{\mathrm{F},h+1}
+2\sqrt{2}\,W^{\la,21}_{\mathrm{F},h+1}
+2i\,W^{\la,22}_{\mathrm{F}+h+1}\, \Bigg],
\nonu\\
\Phi^{(h),14}_{1}
&
=&   q^{1-h}\, (-4)^{h-3}\,\Bigg[
2\,W^{\la,11}_{\mathrm{B},h+1}
+i\sqrt{2}\,W^{\la,12}_{\mathrm{B},h+1}
+4i\sqrt{2}\,\,W^{\la,21}_{\mathrm{B},h+1}
-2\,W^{\la,22}_{\mathrm{B},h+1}
\nonu \\
& - & 2\,W^{\la,11}_{\mathrm{F},h+1}
-2i\sqrt{2}\,W^{\la,12}_{\mathrm{F},h+1}
 -  2i\sqrt{2}\,W^{\la,21}_{\mathrm{F},h+1}
+2\,W^{\la,22}_{\mathrm{F},h+1}\, \Bigg],
\nonu\\
\Phi^{(h),23}_{1}
&
=&    q^{1-h}\, (-4)^{h-3}\,\Bigg[
-2\,W^{\la,11}_{\mathrm{B},h+1}
-i\sqrt{2}\,W^{\la,12}_{\mathrm{B},h+1}
-4i\sqrt{2}\,\,W^{\la,21}_{\mathrm{B},h+1}
+2\,W^{\la,22}_{\mathrm{B},h+1}
\nonu \\
& - & 2\,W^{\la,11}_{\mathrm{F},h+1}
-2i\sqrt{2}\,W^{\la,12}_{\mathrm{F},h+1}
-  2i\sqrt{2}\,W^{\la,21}_{\mathrm{F},h+1}
+2\,W^{\la,22}_{\mathrm{F},h+1}\, \Bigg],
\nonu\\
\Phi^{(h),24}_{1}
&
=&   q^{1-h}\, (-4)^{h-3}\,\Bigg[
-2i\,W^{\la,11}_{\mathrm{B},h+1}
+4\sqrt{2}\,W^{\la,21}_{\mathrm{B},h+1}
+2i\,\,W^{\la,22}_{\mathrm{B},h+1}
\nonu \\
& + & 2i\,W^{\la,11}_{\mathrm{F},h+1}
-2\sqrt{2}\,W^{\la,21}_{\mathrm{F},h+1}
-2i\,W^{\la,22}_{\mathrm{F},h+1}\, \Bigg],
\nonu\\
\Phi^{(h),34}_{1}
&
=&    q^{1-h}\, (-4)^{h-3}\,\Bigg[
-2i\,W^{\la,11}_{\mathrm{B},h+1}
+\sqrt{2}\,W^{\la,12}_{\mathrm{B},h+1}
+2i\,\,W^{\la,22}_{\mathrm{B},h+1}
\nonu \\
& + & 2i\,W^{\la,11}_{\mathrm{F},h+1}
-2\sqrt{2}\,W^{\la,12}_{\mathrm{F},h+1}
-2i\,W^{\la,22}_{\mathrm{F},h+1}\, \Bigg],
\label{sixcomp}
\eea
where the bosonic currents are given by (\ref{fourcurrents}).

The next components \cite{Ahn2208}
with an extra minus sign
can be determined by
calculating the OPEs between the supersymmetry currents
and the previous components (\ref{sixcomp})
\bea
\tilde{\Phi}^{(h),1}_{\frac{3}{2}}
&
\equiv &
\Phi^{(h),1}_{\frac{3}{2}} -\frac{1}{(2h+1)}\, (1-4\la)\,
\pa \,\Phi^{(h),1}_{\frac{1}{2}}
\nonu \\
&=&   q^{-h}\, (-4)^{h-3}\,\Bigg[
  -\frac{1}{2}\,(
Q^{\la,11}_{h+\frac{3}{2}}
+i\sqrt{2}\,Q^{\la,12}_{h+\frac{3}{2}}
+2i\sqrt{2}\,Q^{\la,21}_{h+\frac{3}{2}}
-2\,Q^{\la,22}_{h+\frac{3}{2}}
\nonu \\
& - & 2\,\bar{Q}^{\la,11}_{h+\frac{3}{2}}
-2i\sqrt{2}\,\bar{Q}^{\la,12}_{h+\frac{3}{2}}
-i\sqrt{2}\,\bar{Q}^{\la,21}_{h+\frac{3}{2}}
+\bar{Q}^{\la,22}_{h+\frac{3}{2}}
)\, \Bigg],
\nonu\\
\tilde{\Phi}^{(h),2}_{\frac{3}{2}}
&
\equiv &
\Phi^{(h),2}_{\frac{3}{2}} -\frac{1}{(2h+1)}\, (1-4\la)\,
\pa \,\Phi^{(h),2}_{\frac{1}{2}}
\nonu \\
&
=&    q^{-h}\, (-4)^{h-3}\,\Bigg[
\frac{i}{2}\,(\,
Q^{\la,11}_{h+\frac{3}{2}}
+2i\sqrt{2}\,Q^{\la,21}_{h+\frac{3}{2}}
-2\,Q^{\la,22}_{h+\frac{3}{2}}
-2\,\bar{Q}^{\la,11}_{h+\frac{3}{2}}
-2i\sqrt{2}\,\bar{Q}^{\la,12}_{h+\frac{3}{2}}
+\bar{Q}^{\la,22}_{h+\frac{3}{2}}
)\, \Bigg],
\nonu\\
\tilde{\Phi}^{(h),3}_{\frac{3}{2}}
&
\equiv &
\Phi^{(h),3}_{\frac{3}{2}} -\frac{1}{(2h+1)}\, (1-4\la)\,
\pa \,\Phi^{(h),3}_{\frac{1}{2}}
\nonu \\
&
=&   q^{-h}\, (-4)^{h-3}\,\Bigg[
\frac{i}{2}\,(
Q^{\la,11}_{h+\frac{3}{2}}
+i\sqrt{2}\,Q^{\la,12}_{h+\frac{3}{2}}
-2\,Q^{\la,22}_{h+\frac{3}{2}}
-2\,\bar{Q}^{\la,11}_{h+\frac{3}{2}}
-i\sqrt{2}\,\bar{Q}^{\la,21}_{h+\frac{3}{2}}
+\bar{Q}^{\la,22}_{h+\frac{3}{2}}
)\, \Bigg],
\nonu\\
\tilde{\Phi}^{(h),4}_{\frac{3}{2}}
&
\equiv &
\Phi^{(h),4}_{\frac{3}{2}} -\frac{1}{(2h+1)}\, (1-4\la)\,
\pa \,\Phi^{(h),4}_{\frac{1}{2}}
\nonu \\
&
=&    q^{-h}\, (-4)^{h-3}\,\Bigg[
\frac{1}{2}\,
(
Q^{\la,11}_{h+\frac{3}{2}}
+2\,Q^{\la,22}_{h+\frac{3}{2}}
+2\,\bar{Q}^{\la,11}_{h+\frac{3}{2}}
+\bar{Q}^{\la,22}_{h+\frac{3}{2}}
)\, \Bigg],
\label{3half}
\eea
which are quasiprimary under the stress energy tensor (\ref{Lterm}).
The eight fermionic currents for fixed $h$ in (\ref{fourcurrents})
are distributed into (\ref{nextlowest}) and (\ref{3half})
where $h$ is replaced by $(h-1)$.

The last component
\cite{Ahn2208}
 with an extra minus sign
of (\ref{bigPhi}) obtained by
the OPE between the supersymmetry currents and the components
in (\ref{3half}), which is quasiprimary
under the stress energy tensor (\ref{Lterm}) is 
\bea
\tilde{\Phi}^{(h)}_{2}
&
\equiv &
\Phi^{(h)}_{2} -\frac{1}{(2h+1)}\, (1-4\la)\,
\pa^2 \,\Phi^{(h)}_{0}
=
 -  q^{-h}\, (-4)^{h-3}\,\Bigg[
2\,(
W^{\la,\bar{a} a}_{\mathrm{B},h+2}
+W^{\la, \bar{a} a}_{\mathrm{F},h+2}
) \Bigg].
\label{lastcomp}
 \eea
Again, the bosonic currents are given by (\ref{fourcurrents}).
The eight bosonic currents for fixed $h$ in (\ref{fourcurrents})
are distributed into (\ref{lowest}), (\ref{sixcomp})
where $h$ is replaced by $(h-1)$ and
(\ref{lastcomp}) where $h$  is replaced by $(h-2)$.

It is straightforward, from
 (\ref{firstcomm})
and (\ref{otherbifun}) or (\ref{WFWF}) and (\ref{remainingbi}),
 to calculate all the (anti)commutators
 or OPEs
 between the ${\cal N}=4$ multiplet explicitly.
 For example, the commutator between the lowest components
 is described by \footnote{For simplicity, we do not include
 $q$ dependence explicitly. Or we consider $q=1$ case.}
\bea
&& \comm{(\Phi^{(h_1)}_{0})_m}{(\Phi^{(h_2)}_{0})_n}
=  C_{0,0}^{h_1,h_2}(\la)\,
[m+h_1-1]_{h_1+h_2-1}\,\delta_{m+n}
\nonu \\
&&+ \sum_{h_3=1}^{h_1+h_2-1\,\,}\sum_{k=0}^{h_1+h_2-h_3-1}
\frac{(-1)^{h_1+h_2+1}\,4^{2(h_1+h_2-h_3)-4}}{(2h_1-1)(2h_2-1)}
(2h_3-1)! \nonu \\
& & \times  \Bigg( \bigg[
(h_1-2\lambda)(h_2-2\lambda)S^{\,\,h_1,h_2,h_3,k}_{F,\,R}
-(h_1-1+2\lambda)(h_2-1+2\lambda)S^{\,\,h_1,h_2,h_3,k}_{B,\,R}
\bigg] \nonu \\
&& \times  [m+h_1-1]_{h_1+h_2-h_3-k-1}[n+h_2-1]_k
\nonu \\
&& - 
\bigg[
(h_1-2\lambda)(h_2-2\lambda)S^{\,\,h_1,h_2,h_3,k}_{F,\,L}
-(h_1-1+2\lambda)(h_2-1+2\lambda)S^{\,\,h_1,h_2,h_3,k}_{B,\,L}
\bigg]
\nonu \\
& & \times  [m+h_1-1]_k [n+h_2-1]_{h_1+h_2-h_3-k-1}\,\,
\Bigg)(\Phi^{(h_3)}_0)_{m+n}
\nonu \\
& & + \sum_{h_3=-1}^{h_1+h_2-3}\sum_{k=0}^{h_1+h_2-h_3-3}
\frac{(-1)^{h_1+h_2}\,4^{2(h_1+h_2-h_3)-5}}{2(2h_1-1)(2h_2-1)}
\,
(2h_3+2)!
\nonu \\
& & \times \Bigg(\bigg[
  (h_1-2\lambda)(h_2-2\lambda)(h_3+1+2\lambda)
  S^{\,\,h_1,h_2,h_3+2,k}_{F,\,R}
\nonu \\
&& +  (h_1-1+2\lambda)(h_2-1+2\lambda)(h_3+2-2\lambda)
S^{\,\,h_1,h_2,h_3+2,k}_{B,\,R}\bigg]
\nonu \\
& & \times
[m+h_1-1]_{h_1+h_2-h_3-k-3}[n+h_2-1]_k
\nonu \\
& & - 
\bigg[
  (h_1-2\lambda)(h_2-2\lambda)(h_3+1+2\lambda)
  S^{\,\,h_1,h_2,h_3+2,k}_{F,\,L}\nonu \\
&& + 
  (h_1-1+2\lambda)(h_2-1+2\lambda)(h_3+2-2\lambda)
  S^{\,\,h_1,h_2,h_3+2,k}_{B,\,L}
\bigg] \nonu \\
&& \times  [m+h_1-1]_k [n+h_2-1]_{h_1+h_2-h_3-k-3}
\,\,\Bigg) (\tilde{\Phi}^{(h_3)}_2)_{m+n},
\label{zerozerocomm}
\eea
where the mode independent central term is given by
\footnote{\label{commtoope} We obtain the corresponding OPE as follows.  
  The central term has the additional sign factor
  $(-1)^{h_1+h_2-1}$ from the falling Pochhammer symbol.
  Moreover, the derivative term
  $\pa_z^{h_1+h_2-1} \, \Big[\frac{1}{(z-w)}\Big]$  should be multiplied.
  For the two falling Pochhammer symbols in (\ref{zerozerocomm}),
  we should transform as
  $(-1)^{h_1+h_2-h_3-1}\, \pa_z^{h_1+h_2-h_3-k-1} \, 
  \, \pa_w^{k}$ where the power of minus sign
  comes from the upper bound of $k$ in the summation
  and the corresponding operator
  appears as $\frac{\Phi_0^{(h_3)}(w)}{(z-w)}$
  on the right hand side. Other terms can be done similarly.}
\bea
C_{0,0}^{h_1,h_2}(\la) &= &
\frac{2\,N\,(-1)^{h_1+1}\,4^{2(h_1+h_2-4)}\,}{(2h_1-1)(2h_2-1)}\,
\sum_{i_1=0}^{h_1-1}\sum_{i_2=0}^{h_2-1}
\frac{i_1 !\,i_2 !}{(i_1+i_2+1)!} \nonu
\\
&
\times &
\Bigg[(h_1-2\lambda)(h_2-2\lambda)
  a^{i_1}(h_1,\lambda+\tfrac{1}{2})a^{i_2}
  (h_2,\lambda+\tfrac{1}{2})\nonu \\
  & - &
  (h_1-1+2\lambda)(h_2-1+2\lambda)a^{i_1}(h_1,\lambda)a^{i_2}
(h_2,\lambda)
\Bigg].
\label{C00}
\eea
We can easily check that
if we interchange between $h_1 \leftrightarrow h_2$
and $m \leftrightarrow n$,
the right hand side of (\ref{zerozerocomm}) becomes
the original one with an extra minus sign, by using the
definitions in (\ref{STRUCT}) together with (\ref{Sidentity}).
The combination of $(h_1+h_2)$
appears and the $h_1$ dependent factor and its $h_2$ dependent factor
occur simultaneously.
The corresponding OPE for (\ref{zerozerocomm})
can be found in (\ref{n4sope}).
From (\ref{lowest}) and (\ref{lastcomp}),
the commutators between the first component and the last component
consist of the ones between the bosonic currents in (\ref{fourcurrents}).
This implies that the bosonic subalgebras of the
${\cal N}=4$ supersymmetric linear $W_{\infty}[\la]$ algebra arises
from the linear combination of (\ref{lowest}) and (\ref{lastcomp}).

The remaining $14$ (anti)commutators are presented in
(\ref{n4relations}).
The complete algebra
between the ${\cal N}=4$ multiplet in (\ref{bigPhi})
for generic deformation parameter $\la$ for any
conformal weights $h_1$ and $h_2$ where $h_1, h_2 = 1, 2,
3, \cdots$ is summarized by (\ref{zerozerocomm}) and
(\ref{n4relations}).

\subsection{  The ${\cal N}=4$ supersymmetric OPE in ${\cal N}=4$
superspace}

By using the ${\cal N}=4$ supersymmetry, th above
algebra  (\ref{zerozerocomm}) and
(\ref{n4relations}) can be written in terms of the following
${\cal N}=4$ super OPE
\footnote{
\label{JJfootnote}
  The ${\cal N}=4$ super OPE between the stress energy tensor is
  \bea
{\bf J}(Z_{1})\,{\bf J}(Z_{2})  & = & 
-\frac{\theta_{12}^{4-0}}{z_{12}^{2}}\,
N
+\frac{\theta_{12}^{4-i}}{z_{12}}\,
D^i\, {\bf J}
(Z_{2})
+\frac{\theta_{12}^{4-0}}{z_{12}}\,
2\, \pa\, {\bf J}
(Z_{2}) + \cdots.
\nonu
\eea
The ${\cal N}=4$ super OPE between the ${\cal N}=4$ stress energy
tensor and the ${\cal N}=4$ multiplet is presented in (\ref{jphi}).}
\bea
&& \bold{\Phi}^{(h_1)}(Z_1)\,\bold{\Phi}^{(h_2)}(Z_2)
=
\frac{1}{z_{12}^{h_1+h_2}}\,C_{0,0}^{h_1,h_2}(\lambda)\,(h_1+h_2-1)!
\nonu \\
&& +
\frac{\theta^{4-0}}{z_{12}^{h_1+h_2+2}}\,\bigg[(-1)^{h_1+h_2}C_{0,2}^{h_2,h_1}(\lambda)+\frac{(1-4\lambda)}{(2h_1+1)}C_{0,0}^{h_1,h_2}(\lambda)\bigg](h_1+h_2+1)!
+
\sum_{p=1}^{h_1+h_2-1\,\,}\frac{1}{z_{12}^p}
\nonu \\
&& \times 
\Bigg[ \sum_{h_3=1}^{h_1+h_2-p} f^{\,\,h_1,h_2,h_3,p}_1(\lambda)\,
  \partial^{h_1+h_2-h_3-p} \bold{\Phi}^{(h_3)}
  - \frac{1}{256}\,  f^{\,\,h_1,h_2,1,p}_2(\lambda)\,
\pa^{h_1+h_2-p} \, {\bf J}
\nonu \\
&& +\sum_{h_3=2}^{h_1+h_2-p}
f^{h_1,h_2,h_3,p}_2(\lambda)\,\partial^{h_1+h_2-h_3-p}
\bigg(D^{4-0}-\frac{(1-4\lambda)}{(2h_3-3)}\partial^{2}\bigg)
\bold{\Phi}^{(h_3-2)}
  \Bigg](Z_2)\nonu \\
&&
+ \sum_{p=1}^{h_1+h_2\,\,} \frac{\theta^{i}}{z_{12}^p} \Bigg[
  \sum_{h_3=0}^{h_1+h_2-p}
  f^{h_1,h_2,h_3,p}_3(\lambda)\,\partial^{h_1+h_2-h_3-p}\, D^i \nonu \\
  && +
 \sum_{h_3=0}^{h_1+h_2-p-1} f^{h_1,h_2,h_3,p}_4(\lambda)\,
\partial^{h_1+h_2-h_3-p-1}\,\bigg( D^{4-i}
+\frac{(1-4\lambda)}{(2h_3+1)}\,\partial D^{i}\bigg)
\Bigg]\bold{\Phi}^{(h_3)}(Z_2)
\nonu \\
& &
+
\sum_{p=1}^{h_1+h_2\,\,}\frac{\theta^{4-ij}}{z_{12}^p}\,\,
\Bigg[ \sum_{h_3=0}^{h_1+h_2-p}
  f^{h_1,h_2,h_3,p}_5(\lambda)\,\,\partial^{h_1+h_2-h_3-p}
  D^{4-ij}
  \nonu \\
  &&
+\sum_{h_3=0}^{h_1+h_2-p} f^{h_1,h_2,h_3,p}_6(\lambda)\,\partial^{h_1+h_2-h_3-p} D^{ij}
\Bigg]\bold{\Phi}^{(h_3)}(Z_2)
\nonu \\
&&
+ \sum_{p=1}^{h_1+h_2+1\,\,}
\frac{\theta^{4-i}}{z_{12}^p}\,
\,\Bigg[ \sum_{h_3=0}^{h_1+h_2-p} f^{h_1,h_2,h_3,p}_7(\lambda)\,
\,\partial^{h_1+h_2-h_3-p}
  \bigg( D^{4-i}+\frac{(1-4\la)}{(2h_3+1)}\,\pa D^i \bigg)
  \nonu
\\
&  &+  \sum_{h_3=0}^{h_1+h_2-p+1} f^{h_1,h_2,h_3,p}_8(\lambda)\,
\partial^{h_1+h_2-h_3-p+1}
  \partial D^{i}\
\Bigg]\bold{\Phi}^{(h_3)}(Z_2)
\nonu \\
&& 
+ \sum_{p=1}^{h_1+h_2+1\,\,}
\frac{\theta^{4-0}}{z_{12}^p}
\Bigg[ \sum_{h_3=2}^{h_1+h_2-p+2} f^{h_1,h_2,h_3,p}_9(\lambda)\,
  \partial^{h_1+h_2-h_3-p+2}
\bigg(D^{4-0}-\frac{(1-4\lambda)}{(2h_3-3)}\partial^2 \bigg) \bold{\Phi}^{(h_3-2)}
\nonu \\
&& - \frac{1}{256}\,  f^{\,\,h_1,h_2,1,p}_9(\lambda)\,
\pa^{h_1+h_2-p+2} \, {\bf J}+
\sum_{h_3=1}^{h_1+h_2-p+2} f^{h_1,h_2,h_3,p}_{10}(\lambda)\,
\partial^{h_1+h_2-h_3-p+2}\bold{\Phi}^{(h_3)}
\Bigg](Z_2) \nonu \\
&& + \cdots.
\label{SUPER}
\eea
The central term comes from the two contributions
in (\ref{n4sope}).
Depending on the number of Grassmann coordinates,
each component OPE in (\ref{n4sope}) is located 
at the right place.
The conformal weight of $(h_1+h_2)$
on the left hand side is satisfied also on the right hand side.
Note that there is a relation $\bold{\Phi}^{(0)} \equiv \frac{1}{64}\,
\bold{J}$.
The $\bold{J}$ dependent terms
arise on the right hand side of (\ref{SUPER}).
The various structure constants are given by (\ref{bigC})
and (\ref{10f})
\footnote{If the upper bound is less than the lower bound in the
  summation notation appearing in (\ref{SUPER}),
  we treat them to be $0$ according to the
  ``empty'' sum. }.
The upper bound for $h_3$ summation comes from the
vanishing of the derivatives.
Similarly the upper bound for $p$ summation comes from the
fact that the conformal weight is given by $(h_1+h_2)$.
There are $(1-4\la)$ factors in (\ref{SUPER}) where
the previous central term (\ref{central}) associated with
the stress energy tensor vanishes at $\la=\frac{1}{4}$.
The result of (\ref{SUPER}) is a generalization of \cite{Ahn2205}
where the $h_1+h_2 \leq 4$ cases are obtained.
At $\la=0$, the previous results in \cite{Ahn2208} are reproduced.

Therefore, the above expression (\ref{SUPER})
is equivalent to the component results in (\ref{zerozerocomm}) and
(\ref{n4relations}).

\section{ The ${\cal N}=2$ supersymmetric linear $W_{\infty}^{K,K}[\la]$
algebra  }

\subsection{ The singlet currents and the adjoint currents}

By multiplying the generators of $SU(K)$ $t^{\hat{A}}_{a \bar{b}}$
and $\de_{a \bar{b}}$
into
(\ref{fourcurrents}) \footnote{In this section, the previous
$SU(2)$ is replaced by $SU(K)$.}, we construct 
the following singlets and adjoint currents \cite{Ahn2202}
\bea
W_{\mathrm{F},h} \!& \equiv
\! & W^{\bar{a} b}_{\mathrm{F},h}\, \delta_{b
\bar{a}} \ , \qquad
W^{\hat{A}}_{\mathrm{F},h} \equiv
W^{\bar{a} b}_{\mathrm{F},h}\, t^{\hat{A}}_{b \bar{a}}\ ,
\qquad
W_{\mathrm{B},h} \!  \equiv \!  W^{\bar{a} b}_{\mathrm{B},h}\,
\delta_{b \bar{a}} \ , \qquad
W^{\hat{A}}_{\mathrm{B},h}  \equiv 
W^{\bar{a} b}_{\mathrm{B},h}\, t^{\hat{A}}_{b \bar{a}} \ ,
\label{singletadjoint}
\\
Q_{h+\frac{1}{2}} \! & \equiv
& \!  Q^{\bar{a} b}_{h+\frac{1}{2}}\, \delta_{b
  \bar{a}} \ , \qquad
Q^{\hat{A}}_{h+\frac{1}{2}} \equiv
Q^{\bar{a} b}_{h+\frac{1}{2}}\, t^{\hat{A}}_{b \bar{a}}\ ,
\qquad
\bar{Q}_{h+\frac{1}{2}} \! \equiv
\!  \bar{Q}^{a \bar{b} }_{h+\frac{1}{2}}\, \delta_{a
\bar{b}} \ , \qquad
\bar{Q}^{\hat{A}}_{h+\frac{1}{2}} \equiv
\bar{Q}^{a \bar{b} }_{h+\frac{1}{2}}\, t^{\hat{A}}_{a \bar{b}}\ ,
\nonu
\eea
where $\hat{A}  =   1, 2, \cdots, (K^2-1)$.
It is straightforward to calculate the OPEs between
the singlets and the adjoints by using the OPEs appearing in
section $2$.

\subsection{ The OPEs or (anti)commutators between the
singlets and the adjoints}

The OPE between the singlets, from (\ref{singletadjoint})
and (\ref{WFWF}),  can be summarized by
\bea
&& W_{F,h_1}^{\lambda}(z) \, W_{F,h_2}^{\lambda}(w) 
=
K\,c_{W_F}^{h_1,h_2}(\la)\,q^{h_1+h_2-4}
(-1)^{h_1+h_2+1}\,\partial_z^{h_1+h_2-1}\Bigg[\frac{1}{(z-w)}\Bigg]
\nonu \\
&& + \sum_{h_3=1}^{h_1+h_2-1\,\, }\sum_{k=0}^{h_1+h_2-h_3-1}
\,
(-1)^{h_1+h_2+1}\,(4q)^{h_1+h_2-h_3-2}\,(2h_3-1)! \nonu \\
&& \times 
\big(S^{\,\,h_1,h_2,h_3,k}_{F,\,R}\,\partial_z^{h_1+h_2-h_3-k-1}\partial_w^k
-S^{\,\,h_1,h_2,h_3,k}_{F,\,L}\, \partial_z^k\,
\partial_w^{h_1+h_2-h_3-k-1}\big)
\Bigg[\frac{W_{F,h_3}^{\lambda}(w)}{(z-w)}\Bigg]+\cdots.    
\label{one}
\eea
Due to the contractions between the group indices,
the field dependent term appears as a common factor.
The relations (\ref{cwf}), (\ref{WFWFstructR}) and (\ref{WFWFstructL})
are assumed.

With one adjoint index, the following OPE satisfies
\bea
&& W_{F,h_1}^{\lambda}(z) \, W_{F,h_2}^{\lambda,\hat{A}}(w) 
=
\sum_{h_3=1}^{h_1+h_2-1\,\, }\sum_{k=0}^{h_1+h_2-h_3-1}
\,
(-1)^{h_1+h_2+1}\,(4q)^{h_1+h_2-h_3-2}\,(2h_3-1)! \nonu \\
&& \times
\big(S^{\,\,h_1,h_2,h_3,k}_{F,\,R}\,\partial_z^{h_1+h_2-h_3-k-1}\partial_w^k
-S^{\,\,h_1,h_2,h_3,k}_{F,\,L}\, \partial_z^k\,\partial_w^{h_1+h_2-h_3-k-1}
\big)\Bigg[\frac{W_{F,h_3}^{\lambda,\hat{A}}(w)}{(z-w)}\Bigg]+\cdots.    
\label{two}
\eea
There is no central term and
the expression (\ref{two}) is the same as the one without a central
term in (\ref{one}) except the adjoint index appearing on the current.

With two adjoint indices, we obtain
\bea
&& W_{Fh_1}^{\lambda,\hat{A}}(z) \, W_{F,h_2}^{\lambda,\hat{B}}(w) 
=
\delta^{\hat{A}\hat{B}}\,c_{W_F}^{h_1,h_2}(\la)\,q^{h_1+h_2-4}\,
(-1)^{h_1+h_2+1}\,
\partial_z^{h_1+h_2-1}\Bigg[\frac{1}{(z-w)}\Bigg] \nonu \\
&& + \frac{1}{2}\sum_{h_3=1}^{h_1+h_2-1\,\, }\sum_{k=0}^{h_1+h_2-h_3-1}
\,
(-1)^{h_1+h_2+1}\,(4q)^{h_1+h_2-h_3-2}\,(2h_3-1)! \nonu \\
&& \times 
\Bigg(
  \big(S^{\,\,h_1,h_2,h_3,k}_{F,\,R}\, \partial_z^{h_1+h_2-h_3-k-1} \partial_w^k
  -S^{\,\,h_1,h_2,h_3,k}_{F,\,L}\,
  \partial_z^k\, \partial_w^{h_1+h_2-h_3-k-1}\,\big)
\nonu \\
&& \times  \bigg[\frac{ (
\frac{2}{K}\delta^{\hat{A}\hat{B}}\,\,W_{F,h_3}^{\lambda}
+d^{\hat{A}\hat{B}\hat{C}}\,W_{F,h_3}^{\lambda,\hat{C}})(w)
}{(z-w)}
\bigg]
\nonu \\
&& +
\big(S^{\,\,h_1,h_2,h_3,k}_{F,\,R}\, \partial_z^{h_1+h_2-h_3-k-1} \partial_w^k
+S^{\,\,h_1,h_2,h_3,k}_{F,\,L}\,  \partial_z^k\,
\partial_w^{h_1+h_2-h_3-k-1}\,\big)
\nonu \\
&& \times  \bigg[\frac{ i\,f^{\hat{A}\hat{B}\hat{C}}\,
    W_{F,h_3}^{\lambda,\hat{C}}(w)}{(z-w)}
 \bigg] \Bigg)
+ \cdots.
\label{three}
\eea
On the right hand side of (\ref{three}), there are singlets and adjoints.
Because we are dealing with $K \geq 3$, there exists
a nontrivial symmetric $d$ symbol.

We present some OPEs between the singlets where $h_1+h_2 \leq 6$
corresponding to (\ref{one})
as follows:
\bea
W_{F,1}^{\lambda}(z)\,W_{F,1}^{\lambda}(w)
&= & \frac{1}{(z-w)^2}\,\frac{N\,K}{16 \, q^2}+\cdots\,,
\nonu \\
W_{F,1}^{\lambda}(z)\,W_{F,2}^{\lambda}(w)
&= & \frac{1}{(z-w)^3}\,\frac{N\,K\,\lambda}{2\, q}
+\frac{1}{(z-w)^2}\,W_{F,1}^{\lambda}(w)+\cdots\,,
\nonu \\
W_{F,1}^{\lambda}(z)\,W_{F,3}^{\lambda}(w)
&=&
\frac{1}{(z-w)^4}\,2\,N\,K\,\lambda^2
+\frac{1}{(z-w)^3}\,8 \, q \, \lambda\,W_{F,1}^{\lambda}(w)
\nonu \\
&+& \frac{1}{(z-w)^2}\,\Bigg[
2\,W_{F,2}^{\lambda}
-4\, q\, \lambda\,\partial_w W_{F,1}^{\lambda}
\Bigg](w)+\cdots,
\nonu \\
W_{F,2}^{\lambda}(z)\,W_{F,2}^{\lambda}(w)
&=&
\frac{1}{(z-w)^4}\,\frac{N\,K\,(1-12\lambda^2)}{2}
+\frac{1}{(z-w)^2}\,2\,W_{F,2}^{\lambda}(w)
\nonu
\\
& + & \frac{1}{(z-w)}\,\partial_w W_{F,2}^{\lambda}(w) +\cdots,
\nonu \\
W_{F,2}^{\lambda}(z)\,W_{F,3}^{\lambda}(w)
&=&
\frac{1}{(z-w)^5}\,8\, q\,N\,K\,\lambda(1-4\lambda^2)
\nonu \\
& + & \frac{1}{(z-w)^4}\, 16\, q^2\,
(1-4\lambda^2)\,W_{F,1}^{\lambda}(w)
\nonu \\
&
+& \frac{1}{(z-w)^2}\,3\, W_{F,3}^{\lambda}(w)
+\frac{1}{(z-w)}\,\partial_w W_{F,3}^{\lambda}(w)+\cdots,
\nonu \\
W_{F,3}^{\lambda}(z)\,W_{F,3}^{\lambda}(w)
&=&
\frac{1}{(z-w)^6}\,\frac{32 \, q^2\, N\,K}{3}(1-15\lambda^2+20\lambda^4)
\nonu \\
&
+& \frac{1}{(z-w)^4}\,64 \, q^2\,(1-\lambda^2)\,
W_{F,2}^{\lambda}(w)
\nonu \\
& + & \frac{1}{(z-w)^3}\,32\, q^2\,(1-\lambda^2)
\,\partial_w W_{F,2}^{\lambda}(w)
\nonu \\
&
+& \frac{1}{(z-w)^2}
\Bigg[
  4\,W_{F,4}^{\lambda}+\frac{48\,q^2}{5}
(1-\lambda^2)\,\partial_w^2 W_{F,2}^{\lambda}
\Bigg](w)
\nonu \\
&
+& \frac{1}{(z-w)}
\Bigg[
  2\,\partial W_{F,4}^{\lambda}+
  \frac{32\, q^2}{15}(1-\lambda^2)\,\partial_w^3 W_{F,2}^{\lambda}
\Bigg](w)
+\cdots.
\label{six}
\eea
Compared to the ones in \cite{Pope1991},
the above OPEs (\ref{six}) becomes the corresponding expressions
appearing in \cite{Pope1991} with some typo (in the last expression
of $(3.38)$ in \cite{Pope1991}, the numerical value $\frac{1}{6}$
should be $1$)
when we take $\la \rightarrow 0$
limit with $q=\frac{1}{4}$.
The $N K$ corresponds to their central term.
In other words, the $W_{1+\infty}^K[\la=0]$ algebra is realized.
We also add the third and the last ones in (\ref{six}).
We present other OPEs in Appendix $D$ corresponding to (\ref{two})
and (\ref{three}) \footnote{
\label{twofour}
  At $\la =\pm \frac{1}{2}$, we can easily
  see that the last three OPEs in (\ref{six}) and the next OPE
  below 
  will lead to the ones in $(3.21)$ of \cite{Pope1991}
  which are some OPEs in the $W_{\infty}$ algebra.
  For convenience, we present the following OPE
\bea
&& W_{F,2}^{\lambda}(z)\,W_{F,4}^{\lambda}(w)
 = 
\frac{1}{(z-w)^6}\,32\, q^2\,N\,K\,\lambda^2(1-4\lambda^2)
 +
\frac{1}{(z-w)^5}\,128\, q^3\,\lambda(1-4\lambda^2)\,
W_{F,1}^{\lambda}(w)
\nonu \\
&& +  \frac{1}{(z-w)^4}
\Bigg[
\frac{48\, q^2}{5}(7-12\lambda^2)\,W_{F,2}^{\lambda}
 -64\, q^3 \,\lambda(1-4\lambda^2)\,\partial_w W_{F,1}^{\lambda}
\Bigg](w)
 + \frac{1}{(z-w)^2}\,4\,W_{F,4}^{\lambda}(w)
 \nonu \\
&& +\frac{1}{(z-w)}\,\partial_w W_{F,4}^{\lambda}(w)
+\cdots.
\nonu
\eea
In this OPE, at $\la =\pm \frac{1}{2}$,
the spin $1$ current does not appear due to the $(1-4\la^2)$ factor.
This also happens for the OPE between the spin $2$ and the spin $3$
of (\ref{six}). We can check that this OPE reduces to the one in
\cite{Pope1991} at $\la=0$ with $q=\frac{1}{4}$.
Therefore, in our notation, the $W_{\infty}^K[\la =\pm \frac{1}{2}]$ algebra
is obtained. We have checked that
the structure constant in (\ref{largeP})
satisfies $P_F^{h_1,h_2,1}(m,n, \pm \frac{1}{2})=0$
for $h_1, h_2 > 1$. This implies that there is a decoupling of
$W_{F,1}^{\la}(w)$ at $\la = \pm \frac{1}{2}$ in the full OPEs.}.

In (\ref{26}) where the three commutators (\ref{one}), (\ref{two})
and (\ref{three}) are also added for convenience,
the complete (anti)commutators for the
${\cal N}=2$ supersymmetric $W_{\infty}^{K,K}[\la]$ algebra
are presented.

In particular, at $\la=0$,
the above three OPEs, (\ref{one}), (\ref{two}) and
(\ref{three}), or the first three commutators appearing in (\ref{26}),
lead to the one in \cite{OS}.
In this sense, these for nonzero $\la$ are the $\la$ deformed
$W_{1+\infty}$ algebra (or $\hat{SU}(K)_{k=N}$  algebra) which is denoted
as $W_{1+\infty}^{K}[\la]$ algebra in the footnote \ref{lastfootnote}
of the introduction.
By taking the contractions of the currents  with vanishing
$q$ limit for $\la=0$, as noted by \cite{Ahn2111},
the $w_{1+\infty}$ algebra (which is the extension of
$w_{\infty}$ algebra \cite{Bakas}) is reproduced from 
(\ref{one}).
In the OPE language, there are only second and first order poles
in (\ref{six}) 
under this limit by ignoring the central terms
\footnote{Inside Thielemans package \cite{Thielemans},
  the command {\tt SetOPEOptions[OPEMethod, ClassicalOPEs];}
  provides no double contractions and this leads to the vanishing
central terms in the OPEs.}.
For $h_1=1$, there exists only second order pole.

We can analyze the similar OPEs or (anti)commutators corresponding to
the bosonic subalgebra coming from the $(\beta,\gamma)$ fields.
By comparing (\ref{firstcomm}) with the first relation of
(\ref{otherbifun}) together with (\ref{Sidentity}),
the three OPEs corresponding to (\ref{one}), (\ref{two}) and
(\ref{three}) can be obtained by changing $\la \rightarrow \la -
\frac{1}{2}$ in the structure constants and the central terms
\footnote{We observe that at $\la =\frac{1}{2}$, the
  corresponding $W_{1+\infty}^K[\la +\frac{1}{2}=1]$
  algebra can be reproduced. Similarly,
  the $W_{\infty}^K[\la +\frac{1}{2}=\frac{1}{2}]$ algebra
  is obtained  at $\la =0$.
  As in the footnote \ref{twofour},
  the structure constant in (\ref{largeP})
  where $S^{\,\,h_1,h_2,h_3,k}_{F,\,R}(\lambda)$
  is replaced by $S^{\,\,h_1,h_2,h_3,k}_{B,\,R}(\lambda)$
(and $S^{\,\,h_1,h_2,h_3,k}_{F,\,L}(\lambda)$
  is replaced by $S^{\,\,h_1,h_2,h_3,k}_{B,\,L}(\lambda)$)  
satisfies $P_B^{h_1,h_2,1}(m,n,0)=0$
for $h_1, h_2 > 1$. There is a decoupling of
$W_{B,1}^{\la}(w)$ at $\la=0$ in the full OPEs.}.
There exists an extra minus sign in the central term.
These for nonzero $\la$ are the $\la$ deformed
$W_{1+\infty}$ algebra  which is denoted
as $W_{1+\infty}^{K}[\la+\frac{1}{2}]$
algebra in the footnote \ref{lastfootnote}
of the introduction.

Therefore, the bosonic subalgebra of
${\cal N}=2$ supersymmetric linear
$W_{\infty}[\la]$ algebra is described by
$W_{1+\infty}^K[\la] \times W_{1+\infty}^K[\la+\frac{1}{2}]$.
Note that the transformations described in the footnote
\ref{latrans} can be applied here.
The point is that the $(b,c)$ fields realization is
applied in the first factor while
the $(\beta,\gamma)$ fields realization is applied in the
second factor.
We realize that the $W_{1+\infty}$ algebra
is realized by both fermionic and bosonic constructions. 
The deformation parameter $\la$ appears in our construction and
plays the role in each bosonic subalgebra with a little different form.
The first one is described by $\la$ while
the second one is described by $(\la +\frac{1}{2})$.
Therefore, we have found a matrix generalization of \cite{BVd1,BVd2}
with explicit forms for the structure constants
\footnote{
\label{ttrans}
  As done in \cite{BVd2} where the $\la$
  is replaced by $(\frac{1}{2}-\la)$,
  for the transformation (with multiple numbers of each free field) of
  \bea
  c^{l\bar{a}} \mapsto
\epsilon_1\, \delta^{l}_{\,\,\,\bar{l}}\,\delta^{\bar{a}}_{\,\,\,b}\,
  \beta^{\bar{l} b}\,,\qquad
\beta^{\bar{l}a} \mapsto
-\mathrm{i}\,
\epsilon_2\, \delta^{\bar{l}}_{\,\,\,l}\,\delta^{a}_{\,\,\,\bar{b}}\,
c^{l\bar{b}}
\qquad
\gamma^{l \bar{a}} \mapsto
-\epsilon_1\, \delta^{l}_{\,\,\,\bar{l}}\,\delta^{\bar{a}}_{\,\,\,b}\,
b^{\bar{l} b}\,,\qquad 
   b^{\bar{l}a} \mapsto 
   \mathrm{i}\,
\epsilon_2\, \delta^{\bar{l}}_{\,\,\,l}\,\delta^{a}_{\,\,\,\bar{b}}\,
\gamma^{l \bar{b}},
\label{trans1}
\eea
where $\ep_1$ and $\ep_2$ are real anticommuting parameters,
the corresponding currents are transformed as
follows
\bea
 W_{F,h}^{\lambda,\bar{a}b}(z) &\longmapsto& 
   (-1)^{h}\,\mathrm{i}\,(\epsilon_1 \epsilon_2)\,W_{B,h}^{\frac{1}{2}-\lambda,\,b\bar{a}}(z),
   \qquad
     W_{B,h}^{\lambda,\bar{a}b}(z) \longmapsto 
   (-1)^{h}\,\mathrm{i}\,(\epsilon_1 \epsilon_2)\,W_{F,h}^{\frac{1}{2}-\lambda,\,b\bar{a}}(z),
   \nonu \\
        Q_{h+\frac{1}{2}}^{\lambda,\bar{a}b}(z) &\longmapsto& 
   (-1)^{h+1}\,\mathrm{i}\,(\epsilon_1 \epsilon_2)\, Q_{h+\frac{1}{2}}^{\frac{1}{2}-\lambda,\,b\bar{a}}(z),
   \qquad
    \bar{Q}_{h+\frac{1}{2}}^{\lambda,a\bar{b}}(z) \longmapsto 
   (-1)^{h+1}\,\mathrm{i}\,(\epsilon_1 \epsilon_2)\, \bar{Q}_{h+\frac{1}{2}}^{\,\,\frac{1}{2}-\lambda,\,\bar{b}a}(z).
\label{trans}
\eea
Note that the group indices are reversed after transformation.
We can make the above transformations (\ref{trans})
into the previous OPE (\ref{WFWF}) and it turns out that
we obtain the singular terms of the
OPE of $(-1)^{h_1}W_{B,h_1}^{\frac{1}{2}-\la,b\bar{a}}(z)\,
(-1)^{h_2+1}Q_{h_2+\frac{1}{2}}^{\frac{1}{2}-\la,d\bar{c}}(w)$.
On the other hand, we can extract this OPE from the first equation
of (\ref{remainingbi}). The singular terms are exactly the same as
the previous one and the central term has opposite sign due to the
different statistics in (\ref{trans1}).
Similarly,
the singular terms for the OPEs
$(-1)^{h_1}W_{B,h_1}^{\frac{1}{2}-\la,b\bar{a}}
(z)\,(-1)^{h_2+1}Q_{h_2+\frac{1}{2}}^{\frac{1}{2}-\la,d\bar{c}}(w)$,
$(-1)^{h_1}W_{B,h_1}^{\frac{1}{2}-\la,b\bar{a}}(z)\,
(-1)^{h_2+1}\bar{Q}_{h_2+\frac{1}{2}}^{\frac{1}{2}-\la,\bar{d}c}(w)$
and $(-1)^{h_1+1}Q_{h_1+\frac{1}{2}}^{\frac{1}{2}-\la,\bar{a}b}(z)\,
(-1)^{h_2+1}\bar{Q}_{h_1+\frac{1}{2}}^{\frac{1}{2}-\la,c\bar{d}}(w)$
can be determined from the second, the third, and
the last in (\ref{remainingbi}) by using the
transformations (\ref{trans}) respectively.
Or we can extract them from the fourth, the fifth and the last OPEs
in (\ref{remainingbi}). The singular terms
are the same. For the last one,
the central term has opposite sign as before (\ref{trans1}).
This analysis implies that
the first, the fourth and the fifth of (\ref{remainingbi})
are redundant because they can be determined from
(\ref{WFWF}), the second and the third of (\ref{remainingbi})
up to the central term.
Recall that the enveloping algebra of $OSp(1,2)$
in an associative algebra between the wedge generators
is factored over the ideal in \cite{BVd2}.
The quadratic Casimir operator of $OSp(1,2)$
is given by $\la(\la -\frac{1}{2})$ which is invariant under the
above transformation $\la \rightarrow \frac{1}{2} -\la$.
The linear combination described by the Casimir is proportional to
the identity operator. The multiplication properties of this algebra
depends on $\la$ via the Casimir operator value $\la(\la-\frac{1}{2})$.
We thank the referee
for raising this point.}.




\section{ Conclusions and outlook}

In this paper, we have found the OPEs or (anti)commutators
between the bifundamentals in (\ref{WFWF}) and (\ref{remainingbi})
(or (\ref{firstcomm}) and (\ref{otherbifun})).
In deriving these,
it was crucial to
prove (\ref{bcidentity}) and (\ref{otheridentity}).
Based on these results,
the ${\cal N}=4$ supersymmetric $W_{\infty}^{2,2}[\la]$ algebra
is obtained by (\ref{zerozerocomm}) and (\ref{n4relations})
(or (\ref{SUPER})).
Finally,
the ${\cal N}=2$ supersymmetric $W_{\infty}^{K,K}[\la]$ algebra
is determined by (\ref{26}).
In arriving at these two algebras, the above
 OPEs or (anti)commutators
 between the bifundamentals are heavily used.
 One of the main results in this paper, as mentioned in the
 abstract, is a generalization of $W_{1+\infty}$ algebra
 found by \cite{PRS1990-2} by the presence of the
 deformation parameter $\la$.
 Originally, this parameter appears in the conformal weights
 in the $(\beta,\gamma)$ and $(b, c)$ ghost systems (See also the footnote
 \ref{latrans})
 and plays the role of the rank of the interaction in the
 two dimensional ${\cal N}=(2,2)$ SYK model.
 See also the footnote \ref{qsyk}.

It is an open problem
to describe the present results in the
${\cal N}=2$ superspace starting from the commuting $B$
field and an anticommuting $C$ field \cite{BVd1,BVd2}.
It is known that
the bosonic $W_{\infty}$ algebra with $\la=0$ can be obtained from
the bosonic $W_{1+\infty}$ algebra with $\la =0$ by decoupling the
spin $1$ current. It is an open problem to observe
whether that kind of truncation for the nonzero $\la$,
by introducing the new currents of
spins from the old currents of spins via (non)linear
combinations, appears or not. 
It would be interesting to find how the large $(N,k)$ 't Hooft like limit
in the ${\cal N}=4$ coset model appears in the present result
for any $h_1$ and $h_2$ \cite{Ahn2205}.

Recently,
the wedge subalgebra of $w_{1+\infty}$ algebra \cite{Bakas}
has been identified with the symmetries on the celestial sphere
\cite{Strominger,Stromingerprl} in the context of celestial conformal
field theory \cite{PPR}. See also relevant works in
\cite{FSTZ,GHPS,HPS,Jiang,MRSV,BHS,MNS,DK} \footnote{
  Moreover, in \cite{CPS}, the chiral algebra \cite{BLLPRv} from the
  ${\cal N}=2$ gauge theories in four dimensions is studied
  in the context of  twisted holography. See also the relevant
  work in \cite{BHS-1}.
We thank the referee for pointing this out.
  In particular, the small ${\cal N}=4$
  superconformal algebra \cite{Ademolloetal1,Ademolloetal2}
  is reproduced in \cite{BLLPRv}. On the other hand,
  when $\la =0$, the $(A.1)$ of \cite{Ahn2205} (which is the
  component results corresponding to
  the expression in the footnote \ref{JJfootnote}) contains
  the equations of $(2,1)$, $(2.2)$ and $(2.3)$ of \cite{AGK},
  as a subalgebra,
  which describe the above small ${\cal N}=4$ superconformal
  algebra. In their construction \cite{BLLPRv}, the conformal weights
  for $(\beta, \ga)$ and $(b,c)$
  are given by $\frac{1}{2}$, $\frac{1}{2}$, $1$ and $0$ respectively.
  This implies the case of $\la = \frac{1}{2}$.
  Then by taking the symmetry between $\la$ and $(\frac{1}{2}-\la)$
  described in the footnote \ref{ttrans}, we arrive at the previous
  $\la=0$ case. It would be interesting to see whether their
  construction provides the extension of small ${\cal N}=4$ superconformal
algebra or not.}.
The ${\cal N}=1$ supersymmetric $w_{1+\infty}$ algebra plays the important
role in the corresponding ${\cal N}=1$ supersymmetric Einstein-Yang-Mills
theory \cite{Ahn2111,Ahn2202}.
It would be interesting to find whether the results of this paper
provide the corresponding celestial conformal field theories or not,
by taking the proper limits on the two parameters $q$ and $\la$.
More explicitly, can we
see the mode dependent structure constants in the (anti)commutators
from the celestial conformal field theories?


\vspace{.7cm}

\centerline{\bf Acknowledgments}

CA thanks M. Pate for general discussions on the celestial holography. 
This work was supported by the National
Research Foundation of Korea(NRF) grant funded by the
Korea government(MSIT) 
(No. 2023R1A2C1003750).

\newpage

\appendix

\renewcommand{\theequation}{\Alph{section}\mbox{.}\arabic{equation}}

\section{ The remaining OPEs or (anti)commutators between the
bifundamentals}

The other three identities in the context of (\ref{bcidentity})
can be summarized by
\bea
((\partial_z^s \beta^{\bar{l} b})\, \delta_{\bar{l} l}\,\gamma^{l \bar{a}})(z)
&=&
\sum_{h=0}^{s}(-1)^{s+1}\,4^{-h}\,
q^{1-h}\, Y_{W_B}(s-h,s+1,\lambda)\,\partial_z^{s-h}
W_{B,h+1}^{\lambda,\bar{a}b}(z),
\nonu \\
((\partial_z^s b^{\bar{l} b})\, \delta_{\bar{l} l}\,\gamma^{l \bar{a}})(z)
&=&
\sum_{h=0}^{s}(-1)^{s}\,2^{-\frac{1}{2}}\,4^{-h}\,
q^{-h}\,
Y_{Q}(s-h,s+1,\lambda)\,\partial_z^{s-h}
Q_{h+\frac{3}{2}}^{\lambda,\bar{a}b}(z),
\nonu
\\
((\partial_z^s \beta^{\bar{l} b})\, \delta_{\bar{l} l}\,c^{l \bar{a}})(z)
&=&
\sum_{h=0}^{s}(-1)^{s+1}\,2^{\frac{3}{2}}\,4^{-h}\,
q^{1-h}\,
Y_{\bar{Q}}(s-h,s+1,\lambda)\,
\partial_z^{s-h}
\bar{Q}_{h+\frac{1}{2}}^{\lambda,b\bar{a}}(z).
\label{otheridentity}
\eea
The left hand sides of (\ref{otheridentity})
appear in the last three bifundamentals (\ref{fourcurrents})
without derivatives.
Here the coefficients appearing on the right hand sides of
(\ref{otheridentity}) are
\bea
Y_{W_B}(h,s,\lambda)& \equiv &
\frac{2\,(-2\lambda-s+2)_h\,[s-1]_{h-1}}{h!\,[2s-h-1]_{h-1}},
\nonu    \\
Y_{Q}(h,s,\lambda)& \equiv&
\frac{(-2\lambda-s+1)_h\,[s-1]_{h}}{h!\,[2s-h]_{h}},
\nonu    \\
Y_{\bar{Q}}(h,s,\lambda)& \equiv&
\frac{(-2\lambda-s+2)_h\,[s-1]_{h}}{h!\,[2s-h-2]_{h}}.
\label{remY}
\eea
In the similar expressions to (\ref{eq:2}),
the identities in (\ref{otheridentity})
enable us to express the right hand sides of the corresponding
OPEs by using the original bifundamentals (\ref{fourcurrents})
with possible
derivatives.

The remaining OPEs between the bifundamentals,
by taking the similar analysis done in
the subsection $2.2$ together with
(\ref{otheridentity}) and (\ref{remY}),  are 
\bea
W_{B,h_1}^{\lambda,\bar{a}b}(z)\, W_{B,h_2}^{\lambda, \bar{c}d}(w)
&=& \delta^{\bar{a}d}\delta^{\bar{c}b}
\sum_{i_1=0}^{h_1-1}\sum_{i_2=0}^{h_2-1}
(-1)^{h_1+1}(4q)^{h_1+h_2-4}\,
a^{i_1}\big(h_1,\lambda\big)a^{i_2}\big(h_2,\lambda\big)\frac{N\,i_1 !\,i_2 !}{(i_1+i_2+1)!}\nonu \\
& \times & \partial_z^{h_1+h_2-1}\,\bigg[\frac{1}{(z-w)}\bigg]
\nonu \\
&+ & \sum_{h_3=1}^{h_1+h_2-1 \,\, }\sum_{k=0}^{h_1+h_2-h_3-1}\,
(-1)^{h_1+h_2+1}\,(4q)^{h_1+h_2-h_3-2}(2h_3-1)! \nonu \\
& \times &
\Bigg(\delta^{\bar{a}d}\,S^{\,\,h_1,h_2,h_3,k}_{B,\,R}(\lambda)\, \partial_z^{h_1+h_2-h_3-k-1}\,\partial_w^{k}\,\bigg[\frac{ W_{B,h_3}^{\lambda,\bar{c}b}(w)}{
    (z-w)}\bigg] \nonu
\\
&- & \delta^{\bar{c}b}\,S^{\,\,h_1,h_2,h_3,k}_{B,\,L}(\lambda)\,  \partial_z^{k}\,\partial_w^{h_1+h_2-h_3-k-1}\,\bigg[\frac{ W_{B,h_3}^{\lambda,\bar{a}d}(w)}{(z-w)}\bigg] \Bigg)+\cdots,
\nonu \\
W_{F,h_1}^{\lambda,\bar{a}b}(z)\, Q_{h_2+\frac{1}{2}}^{\lambda, \bar{c}d}(w)
&=& \delta^{\bar{a}d}
\sum_{h_3=1}^{h_1+h_2-1\,\,  }\sum_{k=0}^{h_1+h_2-h_3-1}
(-1)^{h_1+h_2+1}\,(4q)^{h_1+h_2-h_3-2}\,(2h_3)!\, \nonu
\\
&\times & 
T_F^{\,\,h_1,h_2,h_3,k}(\lambda)\,\,\partial_z^{h_1+h_2-h_3-k-1}\,\partial_w^{k}\,\bigg[\frac{ Q_{h_3+\frac{1}{2}}^{\lambda, \bar{c}b}(w)}{(z-w)}\bigg]
+\cdots,
\nonu \\
W_{F,h_1}^{\lambda,\bar{a}b}(z)\, \bar{Q}_{h_2+\frac{1}{2}}^{\lambda, c\bar{d}}(w)
&=& \delta^{\bar{d}b}
\sum_{h_3=0}^{h_1+h_2-1 \,\, }\sum_{k=0}^{h_1+h_2-h_3-1}
(-1)^{h_1+h_2+1}\,(4q)^{h_1+h_2-h_3-2}\,(2h_3)! \nonu \\
&\times & 
\bar{T}_F^{\,\,h_1,h_2,h_3,k}(\lambda)\,\,\partial_z^{k}\,\partial_w^{h_1+h_2-h_3-k-1}\,\bigg[\frac{ \bar{Q}_{h_3+\frac{1}{2}}^{\lambda, c\bar{a}}(w)}{(z-w)}
  \bigg]+\cdots,
\nonu \\
W_{B,h_1}^{\lambda,\bar{a}b}(z)\, Q_{h_2+\frac{1}{2}}^{\lambda, \bar{c}d}(w)
&=& \delta^{\bar{c}b}
\sum_{h_3=1}^{h_1+h_2-1 \,\, }\sum_{k=0}^{h_1+h_2-h_3-1}
(-1)^{h_1+h_2}\,(4q)^{h_1+h_2-h_3-2}\,(2h_3)!\nonu \\
&\times& 
T_B^{\,\,h_1,h_2,h_3,k}(\lambda)\,\,\partial_z^{k}\,\partial_w^{h_1+h_2-h_3-k-1}\,\bigg[\frac{ Q_{h_3+\frac{1}{2}}^{\lambda, \bar{a}d}(w)}{(z-w)}\bigg]+\cdots,
\nonu \\
W_{B,h_1}^{\lambda,\bar{a}b}(z)\, \bar{Q}_{h_2+\frac{1}{2}}^{\lambda, c\bar{d}}(w)
&=& \delta^{\bar{a}c}
\sum_{h_3=1}^{h_1+h_2-1 \,\, }\sum_{k=0}^{h_1+h_2-h_3-1}
(-1)^{h_1+h_2}\,(4q)^{h_1+h_2-h_3-2}\,(2h_3)!\nonu \\
&\times& 
\bar{T}_B^{\,\,h_1,h_2,h_3,k}(\lambda)\,\,\partial_z^{h_1+h_2-h_3-k-1}\,\partial_w^{k}\,\bigg[\frac{\bar{Q}_{h_3+\frac{1}{2}}^{\lambda,b\bar{d}}(w)}{(z-w)}
  \bigg]+\cdots,
\nonu \\
Q_{h_1+\frac{1}{2}}^{\lambda,\bar{a}b}(z)\, \bar{Q}_{h_2+\frac{1}{2}}^{\lambda, c\bar{d}}(w)
&=& \delta^{\bar{a}c}\delta^{\bar{d}b}
\sum_{i_1=0}^{h_1-1}\sum_{i_2=0}^{h_2}2(-1)^{h_1+1}(4q)^{h_1+h_2-2}
\,\beta^{i_1}(h_1\!+\!1,\lambda)
\alpha^{i_2}(h_2\!+\!1,\lambda)\nonu \\
& \times & \frac{N\,i_1 !\,i_2 !}{(i_1+i_2+1)!}
\, \partial_z^{h_1+h_2}\bigg[\frac{1}{(z-w)}\bigg]\nonu \\
&+ & \sum_{h_3=1}^{h_1+h_2 \,\, }\sum_{k=0}^{h_1+h_2-h_3}
\,
2\,(-1)^{h_1+h_2+1}\,(4q)^{h_1+h_2-h_3}\,(2h_3-1)! \nonu \\
&\times&
\Bigg( \delta^{\bar{a}c}\,U_F^{\,\,h_1,h_2,h_3,k}(\lambda)\, 
\partial_z^{h_1+h_2-h_3-k}\,\partial_w^{k}\,\bigg[\frac{  W_{F,h_3}^{\lambda,\bar{d}b}(w)}{(z-w)}
\bigg]
\nonu
\\
&+ & \delta^{\bar{d}b}\,U_B^{\,\,h_1,h_2,h_3,k}(\lambda)\, 
\partial_z^{k}\,\partial_w^{h_1+h_2-h_3-k}\,\bigg[\frac{ W_{B,h_3}^{\lambda,\bar{a}c}(w)}{(z-w)} \bigg] \Bigg) +\cdots.
\label{remainingbi}
\eea
We observe that the powers of derivatives in the second(fourth) OPE
appear in the third(fifth) OPE under the exchange of
$z \leftrightarrow w$.
Compared to the ones for $\la=0$, the algebra
given by (\ref{WFWF}) and (\ref{remainingbi})
has the summation over the dummy variable $k$
appearing in the structure constants and derivatives.
In general, the highest order poles occur
when the bosonic currents have the spin $1$
or the fermionic currents have the spin $\frac{3}{2}$
(or $\frac{1}{2}$) and the lowest order pole is
given by the first order pole. Between them,
all the other structures of poles can be read off
from the explicit structure constants and the derivatives.

The structure constants appearing in (\ref{remainingbi})
as well as the ones in section $2$
are given by
\bea
S^{\,\,h_1,h_2,h_3,k}_{F,\,R}(\lambda)&=&
\sum_{i_1=0}^{h_1-1}\sum_{i_2=0}^{h_2-1}\,\sum_{r=0}^{i_1+i_2}
\Bigg[\frac{(-1)^{i_1+i_2}}{(h_3+r)!}\,
a^{i_1}(h_1,\lambda+\tfrac{1}{2})
a^{i_2}(h_2,\lambda+\tfrac{1}{2})\nonu \\
&\times&
\binom{r}{h_3-1}\binom{i_2}{i_1+i_2-r}\binom{1-h_3+r}{1-
h_2 +i_2+k}
\prod_{j=0}^{r-h_3}(-r-2\lambda+j)\Bigg],
\nonu \\
S^{\,\,h_1,h_2,h_3,k}_{F,\,L}(\lambda)&=&
\sum_{i_1=0}^{h_1-1}\sum_{i_2=0}^{h_2-1}\,\sum_{r=0}^{i_1+i_2}
\Bigg[\frac{(-1)^{i_1+i_2}}{(h_3+r)!}\,
a^{i_1}(h_1,\lambda+\tfrac{1}{2})
a^{i_2}(h_2,\lambda+\tfrac{1}{2})
\nonu \\
&\times&
\binom{r}{h_3-1}\binom{i_1}{i_1+i_2-r}
\binom{1-h_3+r}{1-h_1+i_1+k}
\prod_{j=0}^{r-h_3}( -r-2\lambda+j)\Bigg],
\nonu \\
S^{\,\,h_1,h_2,h_3,k}_{B,\,R}(\lambda)&=&
\sum_{i_1=0}^{h_1-1}\sum_{i_2=0}^{h_2-1}\,\sum_{r=0}^{i_1+i_2}
\Bigg[\frac{(-1)^{i_1+i_2}}{(h_3+r)!}
\,a^{i_1}(h_1,\lambda\big)a^{i_2}(h_2,\lambda\big)
\nonu \\
&\times&
\binom{r}{h_3-1}
\binom{i_2}{i_1+i_2-r}\binom{1-h_3+r}{1-h_2+i_2+k}
\prod_{j=0}^{r-h_3}(1-r-2\lambda+j)\Bigg],
\nonu \\
S^{\,\,h_1,h_2,h_3,k}_{B,\,L}(\lambda)&=&
\sum_{i_1=0}^{h_1-1}\sum_{i_2=0}^{h_2-1}\,\sum_{r=0}^{i_1+i_2}
\Bigg[\frac{(-1)^{i_1+i_2}}{(h_3+r)!}
\,a^{i_1}(h_1,\lambda)a^{i_2}(h_2,\lambda)
\nonu \\
&\times&
\binom{r}{h_3-1}
\binom{i_1}{i_1+i_2-r}\binom{1-h_3+r}{1-h_1+i_1+k}
\prod_{j=0}^{r-h_3} (1-r-2\lambda+j)\Bigg],
\nonu \\
T_{F}^{\,\,h_1,h_2,h_3,k}(\lambda)&=&
\sum_{i_1=0}^{h_1-1}\sum_{i_2=0}^{h_2-1}\,\sum_{r=0}^{i_1+i_2}
\Bigg[\frac{(-1)^{i_1+i_2}}{(h_3+r+1)!}\,
a^{i_1}(h_1,\lambda+\tfrac{1}{2})\beta^{i_2}(h_2+1,\lambda)
\nonu \\
&\times&
\binom{r}{h_3-1}\binom{i_2}{i_1+i_2-r}\binom{1-h_3+r}{1-h_2+
i_2+k}
\prod_{j=0}^{r-h_3}(-r-2\lambda+j)
\Bigg],
\nonu \\
\bar{T}_{F}^{\,\,h_1,h_2,h_3,k}(h_1,\lambda)&=&
\sum_{i_1=0}^{h_1-1}\sum_{i_2=0}^{h_2}\,\sum_{r=0}^{i_1+i_2}
\Bigg[\frac{(-1)^{i_1+i_2}}{(h_3+r)!}\,
a^{i_1}(h_1,\lambda+\tfrac{1}{2})\alpha^{i_2}(h_2+1,\lambda)
\nonu \\
&\times&
\binom{r}{h_3}\binom{i_1}{i_1+i_2-r}\binom{-h_3+r}{1-h_1+
i_1+k}
\prod_{j=0}^{r-h_3-1}( 1-r-2\lambda+j)\Bigg],
\nonu \\
T_{B}^{\,\,h_1,h_2,h_3,k}(\lambda)&=&
\sum_{i_1=0}^{h_1-1}\sum_{i_2=0}^{h_2-1}\,\sum_{r=0}^{i_1+i_2}
\Bigg[\frac{(-1)^{i_1+i_2}}{(h_3+r+1)!}\,
a^{i_1}(h_1,\lambda)
\beta^{i_2}(h_2+1,\lambda)
\nonu \\
&\times&
\binom{r}{h_3-1}\binom{i_1}{i_1+i_2-r}\binom{1-h_3+r}{
  1-h_1+i_1+k}
\prod_{j=0}^{r-h_3}(-r-2\lambda+j)
\Bigg],
\nonu \\
\bar{T}_{B}^{\,\,h_1,h_2,h_3,k}(\lambda)&=&
\sum_{i_1=0}^{h_1-1}\sum_{i_2=0}^{h_2}\,\sum_{r=0}^{i_1+i_2}
\Bigg[\frac{(-1)^{i_1+i_2}}{(h_3+r)!}\,
a^{i_1}(h_1,\lambda)\alpha^{i_2}(h_2+1,\lambda)
\nonu \\
&\times&
\binom{r}{h_3}\binom{i_2}{i_1+i_2-r}\binom{-h_3+r}{-
h_2+i_2+k}
\prod_{j=0}^{r-h_3-1}(1-r-2\lambda+j)\Bigg],
\nonu \\
U_F^{\,\,h_1,h_2,h_3,k}(\lambda)&=&
\sum_{i_1=0}^{h_1-1}\sum_{i_2=0}^{h_2}\sum_{r=0}^{i_1+i_2}
\Bigg[\frac{(-1)^{i_1+i_2}}{(h_3+r)!}\,
\beta^{i_1}(h_1+1,\lambda)\alpha^{i_2}(h_2+1,\lambda)
\nonu \\
&\times&
\binom{r}{h_3-1}
\binom{i_2}{i_1+i_2-r}\binom{1-h_3+r}{-h_2+
i_2+k}
\prod_{j=0}^{r-h_3}(-r-2\lambda+j )\Bigg],
\nonu \\
U_B^{\,\,h_1,h_2,h_3,k}(\lambda)&=& 
\sum_{i_1=0}^{h_1-1}\sum_{i_2=0}^{h_2}\sum_{r=0}^{i_1+i_2}
\Bigg[\frac{(-1)^{i_1+i_2}}{(h_3+r)!}\,
\beta^{i_1}(h_1+1,\lambda)\alpha^{i_2}(h_2+1,\lambda)
\nonu \\
&\times&
\binom{r}{h_3-1}
\binom{i_1}{i_1+i_2-r}\binom{1-h_3+r}{1-h_1+i_1+k}
\prod_{j=0}^{r-h_3}( 1-r-2\lambda+j)
\Bigg].
\label{STRUCT}
\eea
The $k$ dependence appears in one of the binomials and the
dummy variables $i_1, i_2$ and $r$ appear in various places.
The $\la$ dependence appears in the numerators.

Note that there are some relations, from the symmetries
of bosonic subalgebra,
\bea
S^{\,\,h_1,h_2,h_3,k}_{B,\,R}(\lambda)=
S^{\,\,h_1,h_2,h_3,k}_{F,\,R}(\lambda-\tfrac{1}{2}), \qquad
S^{\,\,h_1,h_2,h_3,k}_{B,\,L}(\lambda)=S^{\,\,h_1,h_2,h_3,k}_{F,\,L}(\lambda-
\tfrac{1}{2}).
\label{Sidentity}
\eea
This can be understood by changing the conformal weights
$(\frac{1}{2}+\la, \frac{1}{2}-\la)$ 
for $(b,c)$ fields
into $(\frac{1}{2}+\la-\frac{1}{2},\frac{1}{2}-\la+\frac{1}{2})$
which are the ones for $(\beta,\gamma)$ fields.
See also the footnote \ref{latrans}.
Then the OPE or commutator between the currents
$W_{B,h}^{\lambda,\bar{a}b}$ can be obtained also from the
one between the currents $W_{F,h}^{\lambda,\bar{a}b}$ with the
relation (\ref{Sidentity}) up to a central term
\footnote{Sometimes we do not write down the structure constants
with argument $\la$ for the space of lines.
They do depend on the $\la$ except the relations in
(\ref{zerostruct}).}.

The (anti)commutators
corresponding to (\ref{remainingbi}), by following
the description around (\ref{firstcomm}), can be described by
\bea
\comm{(W_{B,h_1}^{\lambda,\bar{a}b})_m}{(W_{B,h_2}^{\lambda, \bar{c}d})_n}
&=& \delta^{\bar{a}d}\delta^{\bar{c}b}\,c_{W_B}^{h_1,h_2}(\la)
\, q^{h_1+h_2-4} \, [m+h_1-1]_{h_1+h_2-1}\,\delta_{m+n} \nonu \\
&+ & \sum_{h_3=1}^{h_1+h_2-1 \,\, }\sum_{k=0}^{h_1+h_2-h_3-1}
\,
(-1)^{h_3}\,(4q)^{h_1+h_2-h_3-2}\,(2h_3-1)! \nonu \\
&\times&
\Bigg[\delta^{\bar{a}d}S^{\,\,h_1,h_2,h_3,k}_{B,\,R}
[m+h_1-1]_{h_1+h_2-h_3-k-1}[n+h_2-1]_{k}(W_{B,h_3}^{\lambda,\bar{c}b})_{m+n}
\nonu \\
&-& \delta^{\bar{c}b}S^{\,\,h_1,h_2,h_3,k}_{B,\,L}
[m+h_1-1]_{k}[n+h_2-1]_{h_1+h_2-h_3-k-1} (W_{B,h_3}^{\lambda,\bar{a}d})_{m+n}
\Bigg],
\nonu \\
\comm{(W_{F,h_1}^{\lambda,\bar{a}b})_m}{(Q_{h_2+\frac{1}{2}}^{\lambda, \bar{c}d})_r}
&=& \delta^{\bar{a}d}
\sum_{h_3=1}^{h_1+h_2-1 \,\, }\sum_{k=0}^{h_1+h_2-h_3-1}
(-1)^{h_3}\,(4q)^{h_1+h_2-h_3-2}\,(2h_3)! \nonu \\
&\times& 
T_F^{\,\,h_1,h_2,h_3,k}\,\,[m+h_1-1]_{h_1+h_2-h_3-k-1}\,[r+h_2-\tfrac{1}{2}]_k\,
(Q_{h_3+\frac{1}{2}}^{\lambda, \bar{c}b})_{m+r},
\nonu \\
\comm{(W_{F,h_1}^{\lambda,\bar{a}b})_m}{(\bar{Q}_{h_2+\frac{1}{2}}^{\lambda, c\bar{d}})_r}
&= & \delta^{\bar{d}b}
\sum_{h_3=0}^{h_1+h_2-1 \,\, }\sum_{k=0}^{h_1+h_2-h_3-1}
(-1)^{h_3}\,(4q)^{h_1+h_2-h_3-2}\,(2h_3)! \nonu \\
&\times& 
\bar{T}_F^{\,\,h_1,h_2,h_3,k}\,\,[m+h_1-1]_{k}\,[r+h_2-\tfrac{1}{2}]_{h_1+h_2-h_3-k-1}
\,(\bar{Q}_{h_3+\frac{1}{2}}^{\lambda, c\bar{a}})_{m+r},
\nonu \\
\comm{(W_{B,h_1}^{\lambda,\bar{a}b})_m}{(Q_{h_2+\frac{1}{2}}^{\lambda, \bar{c}d})_r}
&=& \delta^{\bar{c}b}
\sum_{h_3=1}^{h_1+h_2-1 \,\, }\sum_{k=0}^{h_1+h_2-h_3-1}
(-1)^{h_3+1}\,(4q)^{h_1+h_2-h_3-2}\,(2h_3)! \nonu \\
&\times &
T_B^{h_1,h_2,h_3,k}[m+h_1-1]_{k}
[r+h_2-\tfrac{1}{2}]_{h_1+h_2-h_3-k-1}
(Q_{h_3+\frac{1}{2}}^{\lambda, \bar{a}d}(w))_{m+r},
\nonu \\
\comm{(W_{B,h_1}^{\lambda,\bar{a}b})_m}{(\bar{Q}_{h_2+\frac{1}{2}}^{\lambda, c\bar{d}})_r}
&=& \delta^{\bar{a}c}
\sum_{h_3=0}^{h_1+h_2-1 \,\, }\sum_{k=0}^{h_1+h_2-h_3-1}
(-1)^{h_3+1}\,(4q)^{h_1+h_2-h_3-2}\,(2h_3)!\nonu \\
&\times& 
\bar{T}_B^{\,\,h_1,h_2,h_3,k}\,\,[m+h_1-1]_{h_1+h_2-h_3-k-1}\,[r+h_2-\tfrac{1}{2}]_{k}\,(\bar{Q}_{h_3+\frac{1}{2}}^{\lambda,b\bar{d}})_{m+r},
\nonu \\
\acomm{(Q_{h_1+\frac{1}{2}}^{\lambda, \bar{a}b})_r}{(\bar{Q}_{h_2+\frac{1}{2}}^{\lambda, c\bar{d}})_s}
&=& \delta^{\bar{a}c}\delta^{\bar{d}b}\,
q^{h_1+h_2-2}\,
c_{Q\bar{Q}}^{h_1,h_2}(\la)
\,
[r+h_1-\tfrac{1}{2}]_{h_1+h_2}\,\delta_{r+s} \nonu \\
&+ & \sum_{h_3=1}^{h_1+h_2 \,\, }\sum_{k=0}^{h_1+h_2-h_3}
2\,(-1)^{h_3+1}\,(4q)^{h_1+h_2-h_3}\,(2h_3-1)! \nonu \\
&\times&
\Bigg[\delta^{\bar{a}c}\,U_F^{\,\,h_1,h_2,h_3,k}\, 
[r+h_1-\tfrac{1}{2}]_{h1+h_2-h_3-k}\,[s+h_2-\tfrac{1}{2}]_{k}\,(W_{F,h_3}^{\lambda,\bar{d}b})_{r+s}
\nonu \\
&+ & \delta^{\bar{d}b}U_B^{\,\,h_1,h_2,h_3,k} 
      [r+h_1-\tfrac{1}{2}]_{k}[s+h_2-\tfrac{1}{2}]_{h1+h_2-h_3-k}
      (W_{B,h_3}^{\lambda,\bar{a}c})_{r+s}\Bigg].
\label{otherbifun}
\eea
The $\la$ dependences for the structure constants
are ignored to save the space but they do depend on the $\la$
as in (\ref{STRUCT}).

The central terms appearing in (\ref{otherbifun}) are given by
\bea
c_{W_B}^{h_1,h_2}(\la) & \equiv & N\,\sum_{i_1=0}^{h_1-1}\sum_{i_2=0}^{h_2-1}
(-1)^{h_2}\,4^{h_1+h_2-4}\,
a^{i_1}(h_1,\lambda)a^{i_2}(h_2,\lambda)
\frac{i_1 !\,i_2 !}{(i_1+i_2+1)!},
\label{remcentral}
\\
c_{Q\bar{Q}}^{h_1,h_2}(\la) & \equiv & N\,\sum_{i_1=0}^{h_1-1}\sum_{i_2=0}^{h_2}
2\,(-1)^{h_2+1}\,4^{h_1+h_2-2}
\,\beta^{i_1}(h_1+1,\lambda)
\alpha^{i_2}(h_2+1,\lambda)\frac{i_1 !\,i_2 !}{(i_1+i_2+1)!}.
\nonu
\eea

Then the complete algebra between the
bifundamentals is given by either (\ref{firstcomm})
and (\ref{otherbifun}) or (\ref{WFWF}) and (\ref{remainingbi}).
The central terms are characterized by
(\ref{cwf}) and (\ref{remcentral}) while mode independent
structure constants are given by (\ref{STRUCT}).

There are nontrivial relations between the structure constants
(\ref{STRUCT}) at $\la=0$
and the ones in \cite{Ahn2208} as follows:
\bea
& &   \sum_{k=0}^{h_1+h_2-h_3-1}
(-1)^{h_1+h_2}\,4^{h_1+h_2-h_3-2}(2h_3-1)!
S^{\,\,h_1,h_2,h_3,k}_{F,\,R}(0)
[m+h_1-1]_{h_1+h_2-h_3-k-1}[n+h_2-1]_k
\nonu \\
&& = \frac{1}{2}\,p_F^{h_1,h_2,h_1+h_2-h_3-2}(m,n),
\nonu \\
& & \sum_{k=0}^{h_1+h_2-h_3-1}
(-1)^{h_3+1}4^{h_1+h_2-h_3-2}(2h_3-1)!
S^{\,\,h_1,h_2,h_3,k}_{F,\,L}(0)
[m+h_1-1]_{k}[n+h_2-1]_{h_1+h_2-h_3-k-1}
\nonu \\
&& = \frac{1}{2}\,p_F^{h_1,h_2,h_1+h_2-h_3-2}(m,n),
\nonu \\
&& \sum_{k=0}^{h_1+h_2-h_3-1}
(-1)^{h_1+h_2}4^{h_1+h_2-h_3-2}(2h_3-1)!
S^{\,\,h_1,h_2,h_3,k}_{B,\,R}(0)
[m+h_1-1]_{h_1+h_2-h_3-k-1}[n+h_2-1]_k
\nonu \\
&& = \frac{1}{2}\,p_B^{h_1,h_2,h_1+h_2-h_3-2}(m,n),
\nonu 
\\
&& \sum_{k=0}^{h_1+h_2-h_3-1}
(-1)^{h_3+1}\,4^{h_1+h_2-h_3-2}(2h_3-1)!
S^{\,\,h_1,h_2,h_3,k}_{B,\,L}(0)
[m+h_1-1]_{k}[n+h_2-1]_{h_1+h_2-h_3-k-1}
\nonu \\
&& =\frac{1}{2}\,p_B^{h_1,h_2,h_1+h_2-h_3-2}(m,n),
\nonu
\\
&& \sum_{k=0}^{h_1+h_2-h_3-1}
(-1)^{h_3}\,4^{h_1+h_2-h_3-2}\,(2h_3)!\,
T^{\,\,h_1,h_2,h_3,k}_{F}(0)\,
[m+h_1-1]_{h_1+h_2-h_3-k-1}[r+h_2-\tfrac{1}{2}]_k
\nonu \\
&& =q_F^{h_1,h_2+\frac{1}{2},h_1+h_2-h_3-2}(m,r),
\nonu
\\
&& \sum_{k=0}^{h_1+h_2-h_3-1}
(-1)^{h_1+h_2}\,4^{h_1+h_2-h_3-2}\,(2h_3)!\,
\bar{T}^{\,\,h_1,h_2,h_3,k}_{F}(0)\,
[m+h_1-1]_{k}[r+h_2-\tfrac{1}{2}]_{h_1+h_2-h_3-k-1}
\nonu \\
&& =q_F^{h_1,h_2+\frac{1}{2},h_1+h_2-h_3-2}(m,r),
\nonu 
\\
&& \sum_{k=0}^{h_1+h_2-h_3-1}
(-1)^{h_3+1}\,4^{h_1+h_2-h_3-2}\,(2h_3)!\,
T^{\,\,h_1,h_2,h_3,k}_{B}(0)\,
[m+h_1-1]_{k}[r+h_2-\tfrac{1}{2}]_{h_1+h_2-h_3-k-1}
\nonu \\
&& =q_B^{h_1,h_2+\frac{1}{2},h_1+h_2-h_3-2}(m,r),
\nonu
\\
&& \sum_{k=0}^{h_1+h_2-h_3-1}
(-1)^{h_1+h_2+1}4^{h_1+h_2-h_3-2}(2h_3)!
\bar{T}^{\,\,h_1,h_2,h_3,k}_{B}(0)
[m+h_1-1]_{h_1+h_2-h_3-k-1}[r+h_2-\tfrac{1}{2}]_{k}
\nonu \\
&& =q_B^{h_1,h_2+\frac{1}{2},h_1+h_2-h_3-2}(m,r),
\nonu
\\
&& \sum_{k=0}^{h_1+h_2-h_3}
(-1)^{h_3+1}\,4^{h_1+h_2-h_3}\,(2h_3-1)!\,
U^{\,\,h_1,h_2,h_3,k}_{F}(0)\,
[r+h_1-\tfrac{1}{2}]_{h_1+h_2-h_3-k}[s+h_2-\tfrac{1}{2}]_{k}
\nonu \\
&& =\frac{1}{2}\, o_F^{h_1+\frac{1}{2},h_2+\frac{1}{2},h_1+h_2-h_3}(r,s),
\nonu
\\
&& \sum_{k=0}^{h_1+h_2-h_3}
(-1)^{h_3+1}\,4^{h_1+h_2-h_3}\,(2h_3-1)!\,
U^{\,\,h_1,h_2,h_3,k}_{B}(0)\,
[r+h_1-\tfrac{1}{2}]_{k}[s+h_2-\tfrac{1}{2}]_{h_1+h_2-h_3-k}
\nonu \\
&& =\frac{1}{2}\, o_B^{h_1+\frac{1}{2},h_2+\frac{1}{2},h_1+h_2-h_3}(r,s).
\label{zerostruct}
\eea
We obtain these relations by using 
(\ref{firstcomm})
and (\ref{otherbifun}) at $\la =0$. The six
structure constants
appearing in (\ref{zerostruct}), $p_F^{h_1,h_2,h_3}(m,n)$,
$p_B^{h_1,h_2,h_3}(m,n)$, $q_F^{h_1,h_2,h_3}(m,r)$,
$q_B^{h_1,h_2,h_3}(m,r)$, $o_F^{h_1,h_2,h_3}(r,s)$ and
$o_B^{h_1,h_2,h_3}(r,s)$
are given in \cite{Ahn2208} explicitly.
That is, each structure constant appearing on the left hand side
has the dummy variable $k$ and 
for generic $h_1,h_2$ and $h_3$, only after this is summed over
$k$ together with mode and $k$ dependent factors,
the previous structure constant appearing on the
right hand side is reproduced.

\section{  The remaining
  ${\cal N}=4$ supersymmetric linear $W_{\infty}^{2,2}[\la]$
  algebra }

The total number of (anti)commutators
between the ${\cal N}=4$ multiplet is given by $15$.
One of them is given by (\ref{zerozerocomm}).
The $14$ remaining (anti)commutators are given by
\bea
&& \comm{(\Phi^{(h_1)}_{0})_m}{(\Phi^{(h_2),i}_{\frac{1}{2}})_r}
=\sum_{h_3=0}^{h_1+h_2-1\,\,}\sum_{k=0}^{h_1+h_2-h_3-1}
\frac{2\,(-1)^{h_1+h_2+1}\,4^{2(h_1+h_2-h_3)-5}}{(2h_1-1)}
(2h_3)! \nonu \\
&& \times \Bigg(\bigg[
(h_1-2\lambda)T^{\,\,h_1,h_2,h_3,k}_{F}
+(h_1-1+2\lambda)\bar{T}^{\,\,h_1,h_2,h_3,k}_{B}
\bigg]\nonu \\
&& \times[m+h_1-1]_{h_1+h_2-h_3-k-1}[r+h_2-\tfrac{1}{2}]_k
\nonu \\
&&+
\bigg[
(h_1-2\lambda)\bar{T}^{\,\,h_1,h_2,h_3,k}_{F}
+(h_1-1+2\lambda)T^{\,\,h_1,h_2,h_3,k}_{B}
\bigg]
\nonu \\
&& \times[m+h_1-1]_k [r+h_2-\tfrac{1}{2}]_{h_1+h_2-h_3-k-1}\,\,
\Bigg)(\Phi^{(h_3),i}_\frac{1}{2})_{m+r}
\nonu \\
&& +\sum_{h_3=0}^{h_1+h_2-2\,\,}\sum_{k=0}^{h_1+h_2-h_3-2}
\frac{2\,(-1)^{h_1+h_2+1}\,4^{2(h_1+h_2-h_3)-6}}{(2h_1-1)}
\,(2h_3+2)! \nonu \\
&& \times\Bigg(\bigg[
(h_1-2\lambda)T^{\,\,h_1,h_2,h_3+1,k}_{F}
-(h_1-1+2\lambda)
\bar{T}^{\,\,h_1,h_2,h_3+1,k}_{B}\bigg] \nonu \\
&& \times
[m+h_1-1]_{h_1+h_2-h_3-k-2}[r+h_2-\tfrac{1}{2}]_k
\nonu \\
&&-
\bigg[
(h_1-2\lambda)\bar{T}^{\,\,h_1,h_2,h_3+1,k}_{F}
-(h_1-1+2\lambda)T^{\,\,h_1,h_2,h_3+1,k}_{B}
\bigg] \nonu \\
&& \times[m+h_1-1]_k [r+h_2-\tfrac{1}{2}]_{h_1+h_2-h_3-k-2}
\,\,\Bigg)(\tilde{\Phi}^{(h_3),i}_\frac{3}{2})_{m+r},
\nonu \\
&& \comm{(\Phi^{(h_1)}_{0})_m}{(\Phi^{(h_2),ij}_{1})_n} 
 =\sum_{h_3=0}^{h_1+h_2\,\,}\sum_{k=0}^{h_1+h_2-h_3-1}
\frac{2\,(-1)^{h_1+h_2}\,4^{2(h_1+h_2-h_3)-5}}{(2h_1-1)}
(2h_3+1)! \nonu \\
&& \times \Bigg(\bigg[
(h_1-2\lambda)S^{\,\,h_1,h_2+1,h_3+1,k}_{F,\,R}
-(h_1-1+2\lambda)S^{\,\,h_1,h_2+1,h_3+1,k}_{B,\,R}
\bigg] \nonu \\
&& \times[m+h_1-1]_{h_1+h_2-h_3-k-1}[n+h_2]_k
\nonu \\
&&-
\bigg[
(h_1-2\lambda)S^{\,\,h_1,h_2+1,h_3+1,k}_{F,\,L}
-(h_1-1+2\lambda)S^{\,\,h_1,h_2+1,h_3+1,k}_{B,\,L}
\bigg]
\nonu \\
&& \times[m+h_1-1]_k [n+h_2]_{h_1+h_2-h_3-k-1}\,\,
\Bigg)(\Phi^{(h_3),ij}_1)_{m+n}
\nonu \\
&& +\sum_{h_3=0}^{h_1+h_2\,\,}\sum_{k=0}^{h_1+h_2-h_3-1}
\frac{2\,(-1)^{h_1+h_2}\,4^{2(h_1+h_2-h_3)-5}}{(2h_1-1)}
\,(2h_3+1)! \nonu \\
&& \times \Bigg(\bigg[
(h_1-2\lambda)S^{\,\,h_1,h_2+1,h_3+1,k}_{F,\,R}
+(h_1-1+2\lambda)
S^{\,\,h_1,h_2+1,h_3+1,k}_{B,\,R}\bigg] \nonu \\
&& \times
[m+h_1-1]_{h_1+h_2-h_3-k-1}[n+h_2]_k
\nonu \\
&& -
\bigg[
(h_1-2\lambda)S^{\,\,h_1,h_2+1,h_3+1,k}_{F,\,L}
+(h_1-1+2\lambda)S^{\,\,h_1,h_2+1,h_3+1,k}_{B,\,L}
\bigg] \nonu \\
&& \times[m+h_1-1]_k [n+h_2]_{h_1+h_2-h_3-k-1}
\,\,\Bigg)(\tilde{\Phi}^{(h_3),ij}_1)_{m+n},
\nonu \\
&& \comm{(\Phi^{(h_1)}_{0})_m}{(\tilde{\Phi}^{(h_2),i}_{\frac{3}{2}})_r}
=\sum_{h_3=0}^{h_1+h_2\,\,}\sum_{k=0}^{h_1+h_2-h_3}
\frac{2\,(-1)^{h_1+h_2}\,4^{2(h_1+h_2-h_3)-4}}{(2h_1-1)}
(2h_3)! \nonu \\
&& \times \Bigg(\bigg[
(h_1-2\lambda)T^{\,\,h_1,h_2+1,h_3,k}_{F}
-(h_1-1+2\lambda)\bar{T}^{\,\,h_1,h_2+1,h_3,k}_{B}
\bigg] \nonu \\
&& \times[m+h_1-1]_{h_1+h_2-h_3-k}[r+h_2+\tfrac{1}{2}]_k
\nonu \\
&& -
\bigg[
(h_1-2\lambda)\bar{T}^{\,\,h_1,h_2+1,h_3,k}_{F}
-(h_1-1+2\lambda)T^{\,\,h_1,h_2+1,h_3,k}_{B}
\bigg]
\nonu \\
&& \times[m+h_1-1]_k [r+h_2+\tfrac{1}{2}]_{h_1+h_2-h_3-k}\,\,
\Bigg)(\Phi^{(h_3),i}_\frac{1}{2})_{m+r}
\nonu \\
&& +\sum_{h_3=0}^{h_1+h_2-1\,\,}\sum_{k=0}^{h_1+h_2-h_3-1}
\frac{2\,(-1)^{h_1+h_2}\,4^{2(h_1+h_2-h_3)-5}}{(2h_1-1)}
\,(2h_3+2)! \nonu \\
&& \times \Bigg(\bigg[
(h_1-2\lambda)T^{\,\,h_1,h_2+1,h_3+1,k}_{F}
+(h_1-1+2\lambda)
\bar{T}^{\,\,h_1,h_2+1,h_3+1,k}_{B}\bigg] \nonu \\
&& \times
[m+h_1-1]_{h_1+h_2-h_3-k-1}[r+h_2+\tfrac{1}{2}]_k
\nonu \\
&& +
\bigg[
(h_1-2\lambda)\bar{T}^{\,\,h_1,h_2+1,h_3+1,k}_{F}
+(h_1-1+2\lambda)T^{\,\,h_1,h_2+1,h_3+1,k}_{B}
\bigg] \nonu \\
& & \times[m+h_1-1]_k [r+h_2+\tfrac{1}{2}]_{h_1+h_2-h_3-k-1}
\,\,\Bigg)(\tilde{\Phi}^{(h_3),i}_\frac{3}{2})_{m+r},
\nonu \\
&& \comm{(\Phi^{(h_1)}_{0})_m}{(\tilde{\Phi}^{(h_2)}_{2})_n}
=C_{0,2}^{h_1,h_2}(\lambda)\,[m+h_1-1]_{h_1+h_2+1}\,\delta_{m+n}
\nonu
\\
&& +\sum_{h_3=1}^{h_1+h_2+1\,\,}\sum_{k=0}^{h_1+h_2-h_3+1}
\frac{2(-1)^{h_1+h_2}\,4^{2(h_1+h_2-h_3)-3}}{(2h_1-1)}
(2h_3-1)! \nonu \\
&& \times \Bigg( \bigg[
(h_1-2\lambda)S^{\,\,h_1,h_2+2,h_3,k}_{F,\,R}
+(h_1-1+2\lambda)S^{\,\,h_1,h_2+2,h_3,k}_{B,\,R}
\bigg] \nonu \\
&& \times[m+h_1-1]_{h_1+h_2-h_3-k+1}[n+h_2+1]_k
\nonu \\
&&-
\bigg[
(h_1-2\lambda)S^{\,\,h_1,h_2+2,h_3,k}_{F,\,L}
+(h_1-1+2\lambda)S^{\,\,h_1,h_2+2,h_3,k}_{B,\,L}
\bigg]
\nonu \\
&& \times[m+h_1-1]_k [n+h_2+1]_{h_1+h_2-h_3-k+1}\,\,
\Bigg)(\Phi^{(h_3)}_0)_{m+n}
\nonu \\
&& +\sum_{h_3=-1}^{h_1+h_2-1\,\,}\sum_{k=0}^{h_1+h_2-h_3-1}
\frac{(-1)^{h_1+h_2+1}\,4^{2(h_1+h_2-h_3)-4}}{(2h_1-1)}
\,(2h_3+2)! \nonu \\
&& \times \Bigg(\bigg[
(h_1-2\lambda)(h_3+1+2\lambda)S^{\,\,h_1,h_2+2,h_3+2,k}_{F,\,R}
\nonu \\
& & -(h_1-1+2\lambda)(h_3+2-2\lambda)
S^{\,\,h_1,h_2+2,h_3+2,k}_{B,\,R}\bigg] \nonu \\
&& \times
[m+h_1-1]_{h_1+h_2-h_3-k-1}[n+h_2+1]_k
\nonu \\
&& -
\bigg[
  (h_1-2\lambda)(h_3+1+2\lambda)S^{\,\,h_1,h_2+2,h_3+2,k}_{F,\,L}
  \nonu \\
&&
-(h_1-1+2\lambda)(h_3+2-2\lambda)S^{\,\,h_1,h_2+2,h_3+2,k}_{B,\,L}
\bigg] \nonu \\
&& \times[m+h_1-1]_k [n+h_2+1]_{h_1+h_2-h_3-k-1}
\,\,\Bigg)(\tilde{\Phi}^{(h_3)}_2)_{m+n},
\nonu \\
&& \acomm{(\Phi^{(h_1),i}_{\frac{1}{2}})_r}{(\Phi^{(h_2),j}_{\frac{1}{2}})_s}
=
\delta^{ij}\,C_{\frac{1}{2},\frac{1}{2}}^{h_1,h_2}(\lambda)\,
      [r+h_1-\tfrac{1}{2}]_{h_1+h_2}\,\delta_{r+s}
      \nonu \\
&& +
\delta^{ij}
\sum_{h_3=1}^{h_1+h_2\,\,}\sum_{k=0}^{h_1+h_2-h_3}
(-1)^{h_1+h_2}\,4^{2(h_1+h_2-h_3)-4}\,
(2h_3-1)!
\nonu \\
&& \times\Bigg(
\Big(U^{\,\,h_2,h_1,h_3,k}_{B}-U^{\,\,h_1,h_2,h_3,k}_{F}\Big)
[r+h_1-\tfrac{1}{2}]_{h_1+h_2-h_3-k}
[s+h_2-\tfrac{1}{2}]_{k}
\nonu \\
&&+\Big(U^{\,\,h_1,h_2,h_3,k}_{B}-U^{\,\,h_2,h_1,h_3,k}_{F}\Big)
[r+h_1-\tfrac{1}{2}]_{k}
[s+h_2-\tfrac{1}{2}]_{h_1+h_2-h_3-k}
\,\Bigg)(\Phi^{(h_3)}_{0})_{r+s}
\nonu \\
&& 
+\delta^{ij}
\sum_{h_3=-1}^{h_1+h_2-2\,\,}\sum_{k=0}^{h_1+h_2-h_3-2}
2\,(-1)^{h_1+h_2}\,4^{2(h_1+h_2-h_3)-6}\,
(2h_3+2)!
\nonu \\
&& \times\Bigg(
\Big((h_3+1+2\lambda)U^{\,\,h_1,h_2,h_3+2,k}_{F}
+(h_3+2-2\lambda)U^{\,\,h_2,h_1,h_3+2,k}_{B}\Big)
\nonu \\
&& \times
[r+h_1-\tfrac{1}{2}]_{h_1+h_2-h_3-k-2}
[s+h_2-\tfrac{1}{2}]_{k} \nonu
\\
&& +
\Big(
(h_3+1+2\lambda)U^{\,\,h_2,h_1,h_3+2,k}_{F}
+(h_3+2-2\lambda)U^{\,\,h_1,h_2,h_3+2,k}_{B}
\Big)
\nonu \\
&& \times
[r+h_1-\tfrac{1}{2}]_{k}
[s+h_2-\tfrac{1}{2}]_{h_1+h_2-h_3-k-2} 
\,\Bigg)(\tilde{\Phi}^{(h_3)}_{2})_{r+s}
\nonu \\
&& +
\sum_{h_3=0}^{h_1+h_2-1\,\,}\sum_{k=0}^{h_1+h_2-h_3-1}
(-1)^{h_1+h_2}\,4^{2(h_1+h_2-h_3)-5}\,(2h_3+1)!\,
\Bigg[
\nonu \\
&&
\times\Bigg(
\Big(U^{\,\,h_1,h_2,h_3+1,k}_{F}+U^{\,\,h_2,h_1,h_3+1,k}_{B}\Big)
[r+h_1-\tfrac{1}{2}]_{h_1+h_2-h_3-k-1}
[s+h_2-\tfrac{1}{2}]_{k}
\nonu \\
&&-
\Big(U^{\,\,h_2,h_1,h_3+1,k}_{F}+U^{\,\,h_1,h_2,h_3+1,k}_{B}\Big)
[r+h_1-\tfrac{1}{2}]_{k}
[s+h_2-\tfrac{1}{2}]_{h_1+h_2-h_3-k-1}
\Bigg)(\Phi^{(h_3),ij}_{1})_{r+s}
\nonu \\
&&
+\Bigg(
\Big(U^{\,\,h_1,h_2,h_3+1,k}_{F}-U^{\,\,h_2,h_1,h_3+1,k}_{B}\Big)
[r+h_1-\tfrac{1}{2}]_{h_1+h_2-h_3-k-1}
[s+h_2-\tfrac{1}{2}]_{k}
\nonu \\
&&
-\Big(U^{\,\,h_2,h_1,h_3+1,k}_{F}-U^{\,\,h_1,h_2,h_3+1,k}_{B}\Big)
[r+h_1-\tfrac{1}{2}]_{k}
[s+h_2-\tfrac{1}{2}]_{h_1+h_2-h_3-k-1}
 \Bigg) (\tilde{\Phi}^{(h_3),ij}_{1})_{r+s}\Bigg],
\nonu \\
&& \comm{(\Phi^{(h_1),i}_{\frac{1}{2}})_r}{(\Phi^{(h_2),jk}_{1})_m}
=\delta^{ij}\sum_{h_3=0}^{h_1+h_2\,\,}\sum_{t=0}^{h_1+h_2-h_3}
(-1)^{h_1+h_2+1}\,4^{2(h_1+h_2-h_3)-4}\,(2h_3)!
\nonu \\
&& \times\Bigg(
\Big(
T^{\,\,h_2+1,h_1,h_3,t}_{F}-\bar{T}^{\,\,h_2+1,h_1,h_3,t}_{B}
\Big)[r+h_1-\tfrac{1}{2}]_t [m+h_2]_{h_1+h_2-h_3-t}
\nonu \\
&&+
\Big(
T^{\,\,h_2+1,h_1,h_3,t}_{B}-\bar{T}^{\,\,h_2+1,h_1,h_3,t}_{F}
\Big)[r+h_1-\tfrac{1}{2}]_{h_1+h_2-h_3-t} [m+h_2]_{t}
\Bigg)(\Phi^{(h_3),k}_{\frac{1}{2}})_{r+m}
\nonu \\
&& +
\delta^{ij}\sum_{h_3=0}^{h_1+h_2-1\,\,}\sum_{t=0}^{h_1+h_2-h_3-1}
(-1)^{h_1+h_2+1}\,4^{2(h_1+h_2-h_3)-5}\,(2h_3+2)!
\nonu \\
&& \times\Bigg(
\Big(
T^{\,\,h_2+1,h_1,h_3+1,t}_{F}+\bar{T}^{\,\,h_2+1,h_1,h_3+1,t}_{B}
\Big)[r+h_1-\tfrac{1}{2}]_t [m+h_2]_{h_1+h_2-h_3-t-1}
\nonu \\
&&+
\Big(
T^{\,\,h_2+1,h_1,h_3+1,t}_{B}+\bar{T}^{\,\,h_2+1,h_1,h_3+1,t}_{F}
\Big)[r+h_1-\tfrac{1}{2}]_{h_1+h_2-h_3-t-1} [m+h_2]_{t}
\Bigg)(\tilde{\Phi}^{(h_3),k}_{\frac{3}{2}})_{r+m}
\nonu \\
&& -\delta^{ik}\Bigg[\,\, j\leftrightarrow k\,\,\Bigg]
\nonu \\
&&
+\epsilon^{ijkl}\sum_{h_3=0}^{h_1+h_2\,\,}\sum_{t=0}^{h_1+h_2-h_3}
(-1)^{h_1+h_2+1}\,4^{2(h_1+h_2-h_3)-4}\,(2h_3)!
\nonu \\
&& \times 
\Bigg(
\Big(
T^{\,\,h_2+1,h_1,h_3,t}_{F}+\bar{T}^{\,\,h_2+1,h_1,h_3,t}_{B}
\Big)[r+h_1-\tfrac{1}{2}]_t [m+h_2]_{h_1+h_2-h_3-t}
\nonu \\
&&-
\Big(
T^{\,\,h_2+1,h_1,h_3,t}_{B}+\bar{T}^{\,\,h_2+1,h_1,h_3,t}_{F}
\Big)[r+h_1-\tfrac{1}{2}]_{h_1+h_2-h_3-t} [m+h_2]_{t}
\Bigg)(\Phi^{(h_3),l}_{\frac{1}{2}})_{r+m}
\nonu \\
&&
+\epsilon^{ijkl}
\sum_{h_3=0}^{h_1+h_2-1\,\,}\sum_{t=0}^{h_1+h_2-h_3-1}
(-1)^{h_1+h_2+1}\,4^{2(h_1+h_2-h_3)-5}\,(2h_3+2)!
\nonu \\
&& \times 
\Bigg(
\Big(
T^{\,\,h_2+1,h_1,h_3+1,t}_{F}-\bar{T}^{\,\,h_2+1,h_1,h_3+1,t}_{B}
\Big)[r+h_1-\tfrac{1}{2}]_t [m+h_2]_{h_1+h_2-h_3-t-1}
\nonu \\
&&-
\Big(
T^{\,\,h_2+1,h_1,h_3+1,t}_{B}-\bar{T}^{\,\,h_2+1,h_1,h_3+1,t}_{F}
\Big)[r+h_1-\tfrac{1}{2}]_{h_1+h_2-h_3-t-1} [m+h_2]_{t}
\Bigg)(\tilde{\Phi}^{(h_3),l}_{\frac{3}{2}})_{r+m},
\nonu \\
&&
\acomm{(\Phi^{(h_1),i}_{\frac{1}{2}})_r}{(\tilde{\Phi}^{(h_2),j}_{\frac{3}{2}})_s}
=
\delta^{ij}\,C_{\frac{1}{2},\frac{3}{2}}^{h_1,h_2}(\lambda)\,
      [r+h_1-\tfrac{1}{2}]_{h_1+h_2+1}\,\delta_{r+s}
\nonu \\     
&& +
\delta^{ij}
\sum_{h_3=1}^{h_1+h_2+1\,\,}\sum_{k=0}^{h_1+h_2-h_3+1}
(-1)^{h_1+h_2+1}\,4^{2(h_1+h_2-h_3)-3}\,
(2h_3-1)!
\nonu \\
&& \times\Bigg[
\Big(U^{\,\,h_1,h_2+1,h_3,k}_{F}+U^{\,\,h_2+1,h_1,h_3,k}_{B}\Big)
[r+h_1-\tfrac{1}{2}]_{h_1+h_2-h_3-k+1}
[s+h_2+\tfrac{1}{2}]_{k}
\nonu \\
&&
-\Big(U^{\,\,h_1,h_2+1,h_3,k}_{B}+U^{\,\,h_2+1,h_1,h_3,k}_{F}\Big)
[r+h_1-\tfrac{1}{2}]_{k}
[s+h_2+\tfrac{1}{2}]_{h_1+h_2-h_3-k+1}
\,\Bigg](\Phi^{(h_3)}_{0})_{r+s}
\nonu \\
&&
+\delta^{ij}
\sum_{h_3=-1}^{h_1+h_2-1\,\,}\sum_{k=0}^{h_1+h_2-h_3-1}
2\,(-1)^{h_1+h_2}\,4^{2(h_1+h_2-h_3)-5}\,
(2h_3+2)!
\nonu \\
&& \times\Bigg[
\Big((h_3+1+2\lambda)U^{\,\,h_1,h_2+1,h_3+2,k}_{F}
-(h_3+2-2\lambda)U^{\,\,h_2+1,h_1,h_3+2,k}_{B}\Big)
\nonu \\
&& \times
[r+h_1-\tfrac{1}{2}]_{h_1+h_2-h_3-k-1}
[s+h_2+\tfrac{1}{2}]_{k} 
\nonu \\
&&+
\Big(
(h_3+2-2\lambda)U^{\,\,h_1,h_2+1,h_3+2,k}_{B}
-(h_3+1+2\lambda)U^{\,\,h_2+1,h_1,h_3+2,k}_{F}
\Big)
\nonu \\
&& \times
[r+h_1-\tfrac{1}{2}]_{k}
[s+h_2+\tfrac{1}{2}]_{h_1+h_2-h_3-k-1} 
\,\Bigg](\tilde{\Phi}^{(h_3)}_{2})_{r+s}
\nonu \\
&& +
\sum_{h_3=0}^{h_1+h_2\,\,}\sum_{k=0}^{h_1+h_2-h_3}
(-1)^{h_1+h_2}\,4^{2(h_1+h_2-h_3)-4}\,(2h_3+1)!\,
\Bigg[
\nonu \\
&&
\times\Bigg(
\Big(U^{\,\,h_1,h_2+1,h_3+1,k}_{F}-U^{\,\,h_2+1,h_1,h_3+1,k}_{B}\Big)
[r+h_1-\tfrac{1}{2}]_{h_1+h_2-h_3-k}
[s+h_2+\tfrac{1}{2}]_{k}
\nonu \\
&&
+\Big(U^{\,\,h_2+1,h_1,h_3+1,k}_{F}-U^{\,\,h_1,h_2+1,h_3+1,k}_{B}\Big)
[r+h_1-\tfrac{1}{2}]_{k}
[s+h_2+\tfrac{1}{2}]_{h_1+h_2-h_3-k}
 \Bigg) (\Phi^{(h_3),ij}_{1})_{r+s}
\nonu \\
&& +\Bigg(
\Big(U^{\,\,h_1,h_2+1,h_3+1,k}_{F}+U^{\,\,h_2+1,h_1,h_3+1,k}_{B}\Big)
[r+h_1-\tfrac{1}{2}]_{h_1+h_2-h_3-k}
[s+h_2+\tfrac{1}{2}]_{k}
\nonu \\
&&
+\Big(U^{\,\,h_2+1,h_1,h_3+1,k}_{F}+U^{\,\,h_1,h_2+1,h_3+1,k}_{B}\Big)
[r+h_1-\tfrac{1}{2}]_{k}
[s+h_2+\tfrac{1}{2}]_{h_1+h_2-h_3-k}
\Bigg)(\tilde{\Phi}^{(h_3),ij}_{1})_{r+s}
\,\Bigg],
\nonu \\
&& \comm{(\Phi^{(h_1),i}_{\frac{1}{2}})_r}{(\tilde{\Phi}^{(h_2)}_{2})_m}
=
\sum_{h_3=0}^{h_1+h_2+1\,\,}\sum_{k=0}^{h_1+h_2-h_3+1}
(-1)^{h_1+h_2+1}\,4^{2(h_1+h_2-h_3)-3}\,(2h_3)!
\nonu \\
&&
\times\Bigg[
\Big(T^{\,\,h_2+2,h_1,h_3,k}_{F}-\bar{T}^{\,\,h_2+2,h_1,h_3,k}_{B}\Big)
[r+h_1-\tfrac{1}{2}]_{k}
[m+h_2+1]_{h_1+h_2-h_3-k+1}
\nonu \\
&&
+\Big(\bar{T}^{\,\,h_2+2,h_1,h_3,k}_{F}-T^{\,\,h_2+2,h_1,h_3,k}_{B}\Big)
[r+h_1-\tfrac{1}{2}]_{h_1+h_2-h_3-k+1}
[m+h_2+1]_{k}
 \Bigg] (\Phi^{(h_3),i}_{\frac{1}{2}})_{r+m}
\nonu \\
&& +
\sum_{h_3=0}^{h_1+h_2\,\,}\sum_{k=0}^{h_1+h_2-h_3}
(-1)^{h_1+h_2+1}\,4^{2(h_1+h_2-h_3)-4}\,(2h_3+2)!
\nonu \\
&&
\times\Bigg[
\Big(T^{\,\,h_2+2,h_1,h_3+1,k}_{F}+\bar{T}^{\,\,h_2+2,h_1,h_3+1,k}_{B}\Big)
[r+h_1-\tfrac{1}{2}]_{k}
[m+h_2+1]_{h_1+h_2-h_3-k}
\nonu \\
&&
-\Big(\bar{T}^{\,\,h_2+2,h_1,h_3+1,k}_{F}+T^{\,\,h_2+2,h_1,h_3+1,k}_{B}\Big)
[r+h_1-\tfrac{1}{2}]_{h_1+h_2-h_3-k}
[m+h_2+1]_{k}
 \Bigg] (\tilde{\Phi}^{(h_3),i}_{\frac{3}{2}})_{r+m},
\nonu \\
&& \comm{(\Phi^{(h_1),ij}_{1})_m}{(\Phi^{(h_2),kl}_{1})_n}
=\Big(\delta^{ik}\delta^{jl}-\delta^{il}\delta^{jk}\Big)
\,C_{1,1,\,-}^{h_1,h_2}(\lambda)\,[m+h_1]_{h_1+h_2+1}\,\delta_{m+n}
\nonu \\
&&+
\epsilon^{ijkl}\,\,\,C_{1,1,\,+}^{h_1,h_2}(\lambda)\,[m+h_1]_{h_1+h_2+1}\,\delta_{m+n}
\nonu
\\
&&
+\Big(\delta^{ik}\delta^{jl}-\delta^{il}\delta^{jk}\Big)
\sum_{h_3=1}^{h_1+h_2+1\,\,}
\sum_{t=0}^{h_1+h_2-h_3+1}
(-1)^{h_1+h_2}\,4^{2(h_1+h_2-h_3)-3}\,(2h_3-1)!
\nonu \\
&& \times\Bigg[ 
\Big(
S^{\,\,h_1+1,h_2+1,h_3,t}_{F,\,R}
-S^{\,\,h_1+1,h_2+1,h_3,t}_{B,\,R}
\Big)
[m+h_1]_{h_1+h_2-h_3-t+1}[n+h_2,t]
\nonu \\
&& -
\Big(
S^{\,\,h_1+1,h_2+1,h_3,t}_{F,\,L}
-S^{\,\,h_1+1,h_2+1,h_3,t}_{B,\,L}
\Big)
[m+h_1]_{t}[n+h_2]_{h_1+h_2-h_3-t+1}
\Bigg](\Phi^{(h_3)}_0)_{m+n}
\nonu \\
&&
+\Big(\delta^{ik}\delta^{jl}-\delta^{il}\delta^{jk}\Big)
\sum_{h_3=-1}^{h_1+h_2-1\,\,}
\sum_{t=0}^{h_1+h_2-h_3-1}
(-1)^{h_1+h_2+1}\,2^{-1}\,4^{2(h_1+h_2-h_3)-4}\,(2h_3+2)! \nonu \\
&& \times \Bigg[
\Big(
(h_3+1+2\lambda)S^{\,\,h_1+1,h_2+1,h_3+2,t}_{F,\,R}
+(h_3+2-2\lambda)S^{\,\,h_1+1,h_2+1,h_3+2,t}_{B,\,R}
\Big)
\nonu \\
&& \times[m+h_1]_{h_1+h_2-h_3-t-1}[n+h_2]_t \nonu \\
&&-
\Big(
(h_3+1+2\lambda)S^{\,\,h_1+1,h_2+1,h_3+2,t}_{F,\,L}
+(h_3+2-2\lambda)S^{\,\,h_1+1,h_2+1,h_3+2,t}_{B,\,L}
\Big)
\nonu \\
&& \times[m+h_1]_{t}[n+h_2]_{h_1+h_2-h_3-t-1}\,
\Bigg](\tilde{\Phi}^{(h_3)}_2)_{m+n}
\nonu \\
&& +
\epsilon^{ijkl}
\sum_{h_3=0}^{h_1+h_2+1\,\,}
\sum_{t=0}^{h_1+h_2-h_3+1}
(-1)^{h_1+h_2}\,4^{2(h_1+h_2-h_3)-3}\,(2h_3-1)! \nonu \\
&& \times \Bigg[
\Big(
S^{\,\,h_1+1,h_2+1,h_3,t}_{F,\,R}
+S^{\,\,h_1+1,h_2+1,h_3,t}_{B,\,R}
\Big)
[m+h_1]_{h_1+h_2-h_3-t+1}[n+h_2]_t \nonu \\
&&-
\Big(
S^{\,\,h_1+1,h_2+1,h_3,t}_{F,\,L}
+S^{\,\,h_1+1,h_2+1,h_3,t}_{B,\,L}
\Big)
[m+h_1]_{t}[n+h_2]_{h_1+h_2-h_3-t+1}\,
\Bigg](\Phi^{(h_3)}_0)_{m+n}.
\nonu \\
&& +\epsilon^{ijkl}
\sum_{h_3=-1}^{h_1+h_2-1\,\,}
\sum_{t=0}^{h_1+h_2-h_3-1}
(-1)^{h_1+h_2+1}\,2^{-1}\,4^{2(h_1+h_2-h_3)-4}\,(2h_3+2)!
\nonu \\
&& \times \Bigg[
\Big(
(h_3+1+2\lambda)S^{\,\,h_1+1,h_2+1,h_3+2,t}_{F,\,R}
-(h_3+2-2\lambda)S^{\,\,h_1+1,h_2+1,h_3+2,t}_{B,\,R}\Big)
\nonu \\
&& \times[m+h_1]_{h_1+h_2-h_3-t-1}[n+h_2]_t
\nonu \\
&&-
\Big(
(h_3+1+2\lambda)S^{\,\,h_1+1,h_2+1,h_3+2,t}_{F,\,L}
-(h_3+2-2\lambda)S^{\,\,h_1+1,h_2+1,h_3+2,t}_{B,\,L}
\Big)
\nonu \\
&& \times[m+h_1]_{t}[n+h_2]_{h_1+h_2-h_3-t-1}\,
\Bigg](\tilde{\Phi}^{(h_3)}_2)_{m+n}
\nonu \\
&& +
\delta^{ik}\sum_{h_3=0}^{h_1+h_2\,\,}\sum_{t=0}^{h_1+h_2-h_3}
(-1)^{h_1+h_2+1}\,4^{2(h_1+h_2-h_3)-4}\,(2h_3+1)!\,\Bigg[
  \nonu \\
&& \times \Bigg(
\Big(
S^{\,\,h_1+1,h_2+1,h_3+1,t}_{F,\,R}
+S^{\,\,h_1+1,h_2+1,h_3+1,t}_{B,\,R}
\Big)
[m+h_1]_{h_1+h_2-h_3-t}[n+h_2]_t \nonu \\
&&+
\Big(
S^{\,\,h_1+1,h_2+1,h_3+1,t}_{F,\,L}
+S^{\,\,h_1+1,h_2+1,h_3+1,t}_{B,\,L}
\Big)
[m+h_1]_{t}[n+h_2]_{h_1+h_2-h_3-t}\,
\Bigg)(\Phi^{(h_3),jl}_1)_{m+n}
\nonu \\
&& +
\Bigg(
\Big(
S^{\,\,h_1+1,h_2+1,h_3+1,t}_{F,\,R}
-S^{\,\,h_1+1,h_2+1,h_3+1,t}_{B,\,R}
\Big)
[m+h_1]_{h_1+h_2-h_3-t}[n+h_2]_t\nonu \\
&&+
\Big(
S^{\,\,h_1+1,h_2+1,h_3+1,t}_{F,\,L}
-S^{\,\,h_1+1,h_2+1,h_3+1,t}_{B,\,L}
\Big)
[m+h_1]_{t}[n+h_2]_{h_1+h_2-h_3-t}\,
\Bigg)(\tilde{\Phi}^{(h_3),jl}_1)_{m+n}\,\Bigg]
\nonu
\\
&&
-\delta^{il}\Bigg[\,\, j\leftrightarrow k\,\,\Bigg]
-\delta^{jk}\Bigg[\,\, i\leftrightarrow j\,\,\Bigg]
+\delta^{jl}\Bigg[\,\, i\leftrightarrow j\,,\,\,k\leftrightarrow l\,\,
  \Bigg],
\nonu \\
&& \comm{(\Phi^{(h_1),ij}_{1})_m}{(\tilde{\Phi}^{(h_2),k}_{\frac{3}{2}})_r}
=\delta^{ik}
\sum_{h_3=0}^{h_1+h_2+1\,\,}
\sum_{t=0}^{h_1+h_2-h_3+1}
(-1)^{h_1+h_2+1}\,4^{2(h_1+h_2+1-h_3)-3}\,(2h_3)!
\nonu \\
&& \times
\Bigg[
\Big(
T^{\,\,h_1+1,h_2+1,h_3,t}_{F}+\bar{T}^{\,\,h_1+1,h_2+1,h_3,t}_{B}
\Big)[m+h_1]_{h_1+h_2-h_3-t+1}[r+h_2+\tfrac{1}{2}]_{t}
\nonu \\
&&+
\Big(
T^{\,\,h_1+1,h_2+1,h_3,t}_{B}+\bar{T}^{\,\,h_1+1,h_2+1,h_3,t}_{F}
\Big)[m+h_1]_{t}[r+h_2+\tfrac{1}{2}]_{h_1+h_2-h_3-t+1}
\Bigg](\Phi^{(h_3),j}_{\frac{1}{2}})_{m+r}
\nonu \\
&& +
\delta^{ik}
\sum_{h_3=0}^{h_1+h_2\,\,}
\sum_{t=0}^{h_1+h_2-h_3}
(-1)^{h_1+h_2+1}\,4^{2(h_1+h_2+1-h_3)-4}\,(2h_3+2)!
\nonu \\
&& \times
\Bigg[
\Big(
T^{\,\,h_1+1,h_2+1,h_3+1,t}_{F}-\bar{T}^{\,\,h_1+1,h_2+1,h_3+1,t}_{B}
\Big)[m+h_1]_{h_1+h_2-h_3-t}[r+h_2+\tfrac{1}{2}]_{t}
\nonu \\
&&
+
\Big(
T^{\,\,h_1+1,h_2+1,h_3+1,t}_{B}-\bar{T}^{\,\,h_1+1,h_2+1,h_3+1,t}_{F}
\Big)[m+h_1]_{t}[r+h_2+\tfrac{1}{2}]_{h_1+h_2-h_3-t}
\Bigg](\tilde{\Phi}^{(h_3),j}_{\frac{3}{2}})_{m+r}
\nonu
\\
&& -\delta^{jk}\Bigg[\,\, i\leftrightarrow j\,\,\Bigg]
\nonu \\
&&
+\epsilon^{ijkl}
\sum_{h_3=0}^{h_1+h_2+1\,\,}
\sum_{t=0}^{h_1+h_2-h_3+1}
(-1)^{h_1+h_2+1}\,4^{2(h_1+h_2+1-h_3)-3}\,(2h_3)!
\nonu \\
&& \times
\Bigg[
\Big(
T^{\,\,h_1+1,h_2+1,h_3,t}_{F}-\bar{T}^{\,\,h_1+1,h_2+1,h_3,t}_{B}
\Big)[m+h_1]_{h_1+h_2-h_3-t+1}[r+h_2+\tfrac{1}{2}]_{t}
\nonu \\
&&
-
\Big(
T^{\,\,h_1+1,h_2+1,h_3,t}_{B}-\bar{T}^{\,\,h_1+1,h_2+1,h_3,t}_{F}
\Big)[m+h_1]_{t}[r+h_2+\tfrac{1}{2}]_{h_1+h_2-h_3-t+1}
\Bigg](\Phi^{(h_3),l}_{\frac{1}{2}})_{m+r}
\nonu \\
&&
+
\epsilon^{ijkl}
\sum_{h_3=0}^{h_1+h_2\,\,}
\sum_{t=0}^{h_1+h_2-h_3}
(-1)^{h_1+h_2+1}\,4^{2(h_1+h_2+1-h_3)-4}\,(2h_3+2)!
\nonu \\
&& \times
\Bigg[
\Big(
T^{\,\,h_1+1,h_2+1,h_3+1,t}_{F}+\bar{T}^{\,\,h_1+1,h_2+1,h_3+1,t}_{B}
\Big)[m+h_1]_{h_1+h_2-h_3-t}[r+h_2+\tfrac{1}{2}]_{t}
\nonu \\
&&
-
\Big(
T^{\,\,h_1+1,h_2+1,h_3+1,t}_{B}+\bar{T}^{\,\,h_1+1,h_2+1,h_3+1,t}_{F}
\Big)[m+h_1]_{t}[r+h_2+\tfrac{1}{2}]_{h_1+h_2-h_3-t}
\Bigg](\tilde{\Phi}^{(h_3),l}_{\frac{3}{2}})_{m+r},
\nonu \\
&& \comm{(\Phi^{(h_1),ij}_{1})_m}{(\tilde{\Phi}^{(h_2)}_{2})_n}
=
\sum_{h_3=0}^{h_1+h_2+1\,\,}
\sum_{k=0}^{h_1+h_2-h_3+1}
(-1)^{h_1+h_2+1}\,4^{2(h_1+h_2-h_3)-3}\,(2h_3+1)!\,\nonu \\
&& \times \Bigg[
 \Bigg(
\Big(
S^{\,\,h_1+1,h_2+2,h_3+1,k}_{F,\,R}
+S^{\,\,h_1+1,h_2+2,h_3+1,k}_{B,\,R}
\Big)
[m+h_1]_{h_1+h_2-h_3-k+1}[n+h_2+1]_k \nonu \\
&&-
\Big(
S^{\,\,h_1+1,h_2+2,h_3+1,k}_{F,\,L}
+S^{\,\,h_1+1,h_2+2,h_3+1,k}_{B,\,L}\Big)
[m+h_1]_{k}[n+h_2+1]_{h_1+h_2-h_3-k+1}\,
\Bigg)(\Phi^{(h_3),ij}_1)_{m+n}
\nonu \\
&& +
 \Bigg(
\Big(
S^{\,\,h_1+1,h_2+2,h_3+1,k}_{F,\,R}
-S^{\,\,h_1+1,h_2+2,h_3+1,k}_{B,\,R}
\Big)
[m+h_1]_{h_1+h_2-h_3-k+1}[n+h_2+1]_k \nonu \\
&&-
\Big(
S^{h_1+1,h_2+2,h_3+1,k}_{F,\,L}
-S^{h_1+1,h_2+2,h_3+1,k}_{B,\,L}\Big)
[m+h_1]_{k}[n+h_2+1]_{h_1+h_2-h_3-k+1}
\Bigg)(\tilde{\Phi}^{(h_3),ij}_1)_{m+n}
\Bigg],
\nonu \\
&&
\acomm{(\tilde{\Phi}^{(h_2),i}_{\frac{3}{2}})_r}{(\tilde{\Phi}^{(h_2),j}_{\frac{3}{2}})_s}
=
\delta^{ij}\,C_{\frac{3}{2},\frac{3}{2}}^{h_1,h_2}(\lambda)\,
      [r+h_1+\tfrac{1}{2}]_{h_1+h_2+2}\,\delta_{r+s}
      \nonu
\\
&& +
\delta^{ij}
\sum_{h_3=1}^{h_1+h_2+2\,\,}\sum_{k=0}^{h_1+h_2-h_3+2}
(-1)^{h_1+h_2}\,4^{2(h_1+h_2-h_3)-2}\,
(2h_3-1)!
\nonu \\
&&
\times\Bigg[
\Big(U^{\,\,h_1+1,h_2+1,h_3,k}_{F}-U^{\,\,h_2+1,h_1+1,h_3,k}_{B}\Big)
[r+h_1+\tfrac{1}{2}]_{h_1+h_2-h_3-k+2}
[s+h_2+\tfrac{1}{2}]_{k}
\nonu \\
&&
+\Big(U^{\,\,h_2+1,h_1+1,h_3,k}_{F}-U^{\,\,h_1+1,h_2+1,h_3,k}_{B}\Big)
[r+h_1+\tfrac{1}{2}]_{k}
[s+h_2+\tfrac{1}{2}]_{h_1+h_2-h_3-k+2}
\,\Bigg](\Phi^{(h_3)}_{0})_{r+s}
\nonu \\
&&
+\delta^{ij}
\sum_{h_3=-1}^{h_1+h_2\,\,}\sum_{k=0}^{h_1+h_2-h_3}
2\,(-1)^{h_1+h_2+1}\,4^{2(h_1+h_2-h_3)-4}\,
(2h_3+2)!
\nonu \\
&& \times\Bigg[
\Big((h_3+1+2\lambda)U^{\,\,h_1+1,h_2+1,h_3+2,k}_{F}
+(h_3+2-2\lambda)U^{\,\,h_2+1,h_1+1,h_3+2,k}_{B}\Big)
\nonu \\
&& \times
[r+h_1+\tfrac{1}{2}]_{h_1+h_2-h_3-k}
[s+h_2+\tfrac{1}{2}]_{k} 
\nonu
\\
&&+
\Big(
(h_3+1+2\lambda)U^{\,\,h_2+1,h_1+1,h_3+2,k}_{F}
+(h_3+2-2\lambda)U^{\,\,h_1+1,h_2+1,h_3+2,k}_{B}
\Big)
\nonu \\
&& \times
[r+h_1+\tfrac{1}{2}]_{k}
[s+h_2+\tfrac{1}{2}]_{h_1+h_2-h_3-k} 
\,\Bigg](\tilde{\Phi}^{(h_3)}_{2})_{r+s}
\nonu \\
&& +
\sum_{h_3=0}^{h_1+h_2+1\,\,}\sum_{k=0}^{h_1+h_2-h_3+1}
(-1)^{h_1+h_2+1}\,4^{2(h_1+h_2-h_3)-3}\,(2h_3+1)!\,
\Bigg[
\nonu \\
&&
\times\Bigg(
\Big(U^{\,\,h_1+1,h_2+1,h_3+1,k}_{F}+U^{\,\,h_2+1,h_1+1,h_3+1,k}_{B}\Big)
[r+h_1+\tfrac{1}{2}]_{h_1+h_2-h_3-k+1}
[s+h_2+\tfrac{1}{2}]_{k}
\nonu \\
&&
-\Big(U^{\,\,h_2+1,h_1+1,h_3+1,k}_{F}+U^{\,\,h_1+1,h_2+1,h_3+1,k}_{B}\Big)
[r+h_1+\tfrac{1}{2}]_{k}
[s+h_2+\tfrac{1}{2}]_{h_1+h_2-h_3-k+1}
\Bigg)(\Phi^{(h_3),ij}_{1})_{r+s}
\nonu \\
&&
+\Bigg(
\Big(U^{\,\,h_1+1,h_2+1,h_3+1,k}_{F}-U^{\,\,h_2+1,h_1+1,h_3+1,k}_{B}\Big)
[r+h_1+\tfrac{1}{2}]_{h_1+h_2-h_3-k+1}
[s+h_2+\tfrac{1}{2}]_{k}
\nonu \\
&&
-\Big(U^{h_2+1,h_1+1,h_3+1,k}_{F}-U^{h_1+1,h_2+1,h_3+1,k}_{B}\Big)
[r+h_1+\tfrac{1}{2}]_{k}
[s+h_2+\tfrac{1}{2}]_{h_1+h_2-h_3-k+1}
 \Bigg) (\tilde{\Phi}^{(h_3),ij}_{1})_{r+s}\Bigg],
\nonu \\
&&
\comm{(\tilde{\Phi}^{(h_2),i}_{\frac{3}{2}})_r}{(\tilde{\Phi}^{(h_2)}_{2})_m}
=
\sum_{h_3=0}^{h_1+h_2+2\,\,}\sum_{k=0}^{h_1+h_2-h_3+2}
(-1)^{h_1+h_2}\,4^{2(h_1+h_2-h_3)-2}\,(2h_3)!
\nonu \\
&&
\times\Bigg[
\Big(T^{\,\,h_2+2,h_1+1,h_3,k}_{F}+\bar{T}^{\,\,h_2+2,h_1+1,h_3,k}_{B}\Big)
[r+h_1+\tfrac{1}{2}]_{k}
[m+h_2+1]_{h_1+h_2-h_3-k+2}
\nonu \\
&&
-\Big(\bar{T}^{\,\,h_2+2,h_1+1,h_3,k}_{F}+T^{\,\,h_2+2,h_1+1,h_3,k}_{B}\Big)
[r+h_1+\tfrac{1}{2}]_{h_1+h_2-h_3-k+2}
[m+h_2+1]_{k}
 \Bigg] (\Phi^{(h_3),i}_{\frac{1}{2}})_{r+m}
\nonu \\
&& +
\sum_{h_3=-1}^{h_1+h_2+1\,\,}\sum_{k=0}^{h_1+h_2-h_3+1}
(-1)^{h_1+h_2}\,4^{2(h_1+h_2-h_3)-3}\,(2h_3+2)!
\nonu \\
&&
\times\Bigg[
\Big(T^{\,\,h_2+2,h_1+1,h_3+1,k}_{F}-\bar{T}^{\,\,h_2+2,h_1+1,h_3+1,k}_{B}\Big)
[r+h_1+\tfrac{1}{2}]_{k}
[m+h_2+1]_{h_1+h_2-h_3-k+1}
\nonu \\
&&
+\Big(\bar{T}^{\,\,h_2+2,h_1+1,h_3+1,k}_{F}-T^{\,\,h_2+2,h_1+1,h_3+1,k}_{B}\Big)
[r+h_1+\tfrac{1}{2}]_{h_1+h_2-h_3-k+1}
[m+h_2+1]_{k}
 \Bigg] (\tilde{\Phi}^{(h_3),i}_{\frac{3}{2}})_{r+m},
\nonu \\
&&\comm{(\tilde{\Phi}^{(h_1)}_{2})_m}{(\tilde{\Phi}^{(h_2)}_{2})_n}
=C_{2,2}^{h_1,h_2}(\lambda)\,[m+h_1+1]_{h_1+h_2+3}\,\delta_{m+n}
\nonu \\
&& +\sum_{h_3=1}^{h_1+h_2+3\,\,}\sum_{k=0}^{h_1+h_2-h_3+3}
(-1)^{h_1+h_2+1}\,4^{2(h_1+h_2-h_3)-1}\,
(2h_3-1)! \nonu \\
&& \times \Bigg[
\Big(
S^{\,\,h_1+2,h_2+2,h_3,k}_{F,\,R}
-S^{\,\,h_1+2,h_2+2,h_3,k}_{B,\,R}
\Big)
[m+h_1+1]_{h_1+h_2-h_3-k+3}[n+h_2+1]_k
\nonu \\
&& -
\Big(
S^{\,\,h_1+2,h_2+2,h_3,k}_{F,\,L}
-S^{\,\,h_1+2,h_2+2,h_3,k}_{B,\,L}
\Big)
[m+h_1+1]_k [n+h_2+1]_{h_1+h_2-h_3-k+3}\,\,
\Bigg](\Phi^{(h_3)}_0)_{m+n}
\nonu 
\\
&& +\sum_{h_3=-1}^{h_1+h_2+1\,\,}\sum_{k=0}^{h_1+h_2-h_3+1}
(-1)^{h_1+h_2}\,2^{-1}\,4^{2(h_1+h_2-h_3)-2}
\,(2h_3+2)! \nonu \\
&& \times \Bigg[
\bigg(
(h_3+1+2\lambda)
S^{\,\,h_1+2,h_2+2,h_3+2,k}_{F,\,R}
+(h_3+2-2\lambda)
S^{\,\,h_1+2,h_2+2,h_3+2,k}_{B,\,R}\bigg)
\nonu \\
&& \times
[m+h_1+1]_{h_1+h_2-h_3-k+1}[n+h_2+1]_k
\nonu
\\
&&-
\bigg(
(h_3+1+2\lambda)
S^{\,\,h_1+2,h_2+2,h_3+2,k}_{F,\,L}
+(h_3+2-2\lambda)S^{\,\,h_1+2,h_2+2,h_3+2,k}_{B,\,L}
\bigg) \nonu \\
&&
\times[m+h_1+1]_k [n+h_2+1]_{h_1+h_2-h_3-k+1}
\,\,\Bigg](\tilde{\Phi}^{(h_3)}_2)_{m+n}.
\label{n4relations}
\eea
Note that
we introduce the fields $
\tilde{\Phi}_1^{ (h), i j } \equiv
\frac{1}{2} \epsilon^{i j k l }   {\Phi}_1^{ (h), k l }$.
In the sixth commutator, we
use a simplified notation
$[j \leftrightarrow k]$ which stands for
the substitution $j \leftrightarrow k$
in the previous expressions.
We do not write down the arguments for the structure constants
appearing in (\ref{STRUCT}) in (\ref{n4relations})
\footnote{The corresponding OPEs can be obtained from the
analysis of the footnote \ref{commtoope}.} and 
the structure constants are given by (\ref{STRUCT}).

The $\la$ dependent coefficients appearing in (\ref{n4relations}) are
\bea
C_{0,2}^{h_1,h_2}(\lambda) & \equiv &
\frac{N\,(-1)^{h_1}\,4^{2(h_1+h_2)-6}\,}{(2h_1-1)}\,
\sum_{i_1=0}^{h_1-1}\sum_{i_2=0}^{h_2+1}
\frac{i_1 !\,i_2 !}{(i_1+i_2+1)!}
\nonu \\
&
\times & \Bigg[(h_1-2\lambda)
a^{i_1}(h_1,\lambda+\tfrac{1}{2})
a^{i_2}(h_2+2,\lambda+\tfrac{1}{2})\nonu \\
& + &
(h_1-1+2\lambda)
a^{i_1}\big(h_1,\lambda\big)
a^{i_2}\big(h_2+2,\lambda\big)
\Bigg],
\nonu \\
C_{\frac{1}{2},\frac{1}{2}}^{h_1,h_2}(\lambda)
& \equiv &
2\,N\,(-1)^{h_1}\,4^{2(h_1+h_2-4)}\,
\sum_{i_1=0}^{h_1}\sum_{i_2=0}^{h_2}
\frac{i_1 !\,i_2 !}{(i_1+i_2+1)!}
 \nonu \\
&
\times & \Bigg[
\beta^{i_1}(h_1+1,\lambda)
\alpha^{i_2}(h_2+1,\lambda)+
\alpha^{i_1}(h_1+1,\lambda)
\beta^{i_2}(h_2+1,\lambda)
\Bigg],
\nonu \\
C_{\frac{1}{2},\frac{3}{2}}^{h_1,h_2}(\lambda)& \equiv&
2\,N\,(-1)^{h_1+1}\,4^{2(h_1+h_2)-7}\,
\sum_{i_1=0}^{h_1}\sum_{i_2=0}^{h_2+1}
\frac{i_1 !\,i_2 !}{(i_1+i_2+1)!}
\nonu \\
&
\times & \Bigg(
\beta^{i_1}(h_1+1,\lambda)
\alpha^{i_2}(h_2+2,\lambda)-
\alpha^{i_1}(h_1+1,\lambda)
\beta^{i_2}(h_2+2,\lambda)
\Bigg),
\nonu \\
C_{1,1,\,\pm}^{h_1,h_2}(\lambda)& \equiv &
2\,N\,(-1)^{h_1+1}\,4^{2(h_1+h_2)-7}\,
\sum_{i_1=0}^{h_1}\sum_{i_2=0}^{h_2}
\frac{i_1 !\,i_2 !}{(i_1+i_2+1)!}
\nonu \\
&
\times & \Bigg[
a^{i_1}(h_1+1,\lambda+\tfrac{1}{2})
a^{i_2}(h_2+1,\lambda+\tfrac{1}{2})
\pm
a^{i_1}(h_1+1,\lambda)
a^{i_2}(h_2+1,\lambda)
\Bigg],
\nonu \\
C_{\frac{3}{2},\frac{3}{2}}^{h_1,h_2}(\lambda)& \equiv&
2\,N\,(-1)^{h_1}\,4^{2(h_1+h_2)-6}\,
\sum_{i_1=0}^{h_1+1\,\,}\sum_{i_2=0}^{h_2+1}
\frac{i_1 !\,i_2 !}{(i_1+i_2+1)!}
\nonu \\
&
\times& \Bigg[
\beta^{i_1}\big(h_1+2,\lambda\big)
\alpha^{i_2}\big(h_2+2,\lambda\big)+
\alpha^{i_1}\big(h_1+2,\lambda\big)
\beta^{i_2}\big(h_2+2,\lambda\big)
\Bigg],
\nonu \\
C_{2,2}^{h_1,h_2}(\lambda) & \equiv &
2\,N\,(-1)^{h_1+1}\,4^{2(h_1+h_2)-5}
\sum_{i_1=0}^{h_1+1\,\,}\sum_{i_2=0}^{h_2+1}
\frac{i_1 !\,i_2 !}{(i_1+i_2+1)!}
\nonu \\
&
\times & \Bigg[
a^{i_1}(h_1+2,\lambda+\tfrac{1}{2})
a^{i_2}(h_2+2,\lambda+\tfrac{1}{2})
-
a^{i_1}(h_1+2,\lambda)
a^{i_2}(h_2+2,\lambda)
\Bigg],
\label{bigC}
\eea
where the previous relations (\ref{coeff}) can be applied.
Note that
the first and the third in (\ref{bigC})
have nonzero values although the components
of the ${\cal N}=4$ multiplet on the left hand side
are different from each other because of (\ref{3half})
and (\ref{lastcomp}).

\section{  The basic component OPEs  for ${\cal N}=4$ supersymmetric
  OPE  }

The fundamental five OPEs
for the ${\cal N}=4$ superspace description (\ref{SUPER})
are enough to write the component OPEs in terms of
${\cal N}=4$ super OPE because
other corresponding OPEs to (\ref{n4relations})
can be obtained
by taking super derivatives and putting the fermionic coordinates
to zero in ${\cal N}=4$ superspace
\footnote{ \label{Relations}
The exact relations between the components and its
superfields in (\ref{bigPhi}) can be summarized by
\cite{Schoutens,AK1509}
\bea
\Phi_{0}^{(h)}  & \leftrightarrow &  
    {\bf \Phi}^{(h)},
    \qquad
\Phi_{\frac{1}{2}}^{(h),i}  \leftrightarrow   D^i
    {\bf \Phi}^{(h)},  
\qquad
\Phi_{1}^{(h),ij}   \leftrightarrow   -\frac{1}{2!} \,
 \varepsilon^{ijkl} \, D^k D^l {\bf \Phi}^{(h)},
\nonu \\
\Phi_{\frac{3}{2}}^{(h),i}  & \leftrightarrow &  \frac{1}{3!} \,
\varepsilon^{ijkl} \, D^j D^k D^l {\bf \Phi}^{(h)},
\qquad
\Phi_{2}^{(h)}    \leftrightarrow   \frac{1}{ 4!} \,
\varepsilon^{ijkl} \, D^{i} D^j D^k D^l {\bf \Phi}^{(h)}.
\nonu
\eea}.
They are given by
\bea
\Phi^{(h_1)}_0(z)\,\Phi^{(h_2)}_0(w)
    &= & \frac{1}{(z-w)^{h_1+h_2}}\,C_{0,0}^{h_1,h_2}(\lambda)\,(h_1+h_2-1)!
\nonu    \\
&+&
\sum_{p=1}^{h_1+h_2-1\,\,}\sum_{h_3=1}^{h_1+h_2-p}\frac{1}{(z-w)^p}\,\Bigg[f^{h_1,h_2,h_3,p}_1(\lambda)\,\partial_w^{h_1+h_2-h_3-p}\,\Phi^{(h_3)}_0
\nonu     \\
    &+ & f^{h_1,h_2,h_3,p}_2(\lambda)\,\partial_w^{h_1+h_2-h_3-p}
    \Big(\Phi^{(h_3-2)}_2\!-\frac{(1-4\lambda)}{(2h_3-3)}\partial_w^2 \Phi^{(h_3-2)}_0\Big)\,\Bigg](w)+\cdots,
 \nonu \\
 \Phi^{(h_1),i}_\frac{1}{2}(z)\,\Phi^{(h_2)}_0(w)
 &= & \sum_{p=1}^{h_1+h_2\,\,}\frac{1}{(z-w)^p}\,\Bigg[
\sum_{h_3=0}^{h_1+h_2-p}
   f^{h_1,h_2,h_3,p}_3(\lambda)\,\partial_w^{h_1+h_2-h_3-p}\,
  \Phi^{(h_3),i}_\frac{1}{2}
\nonu     \\
&- &\sum_{h_3=0}^{h_1+h_2-p-1}
f^{h_1,h_2,h_3,p}_4(\lambda)\,\partial_w^{h_1+h_2-h_3-p-1}
\Big(\Phi^{(h_3),i}_\frac{3}{2}\!-\frac{(1-4\lambda)}{(2h_3+1)}\partial_w \Phi^{(h_3),i}_\frac{1}{2}\Big)\,\Bigg](w)
 \nonu \\
 & + & \cdots,
\nonu \\
\Phi^{(h_1),ij}_1(z)\,\Phi^{(h_2)}_0(w)
&=&
\sum_{p=1}^{h_1+h_2\,\,}\sum_{h_3=0}^{h_1+h_2-p}\frac{1}{(z-w)^p}\,
    \Bigg[-f^{h_1,h_2,h_3,p}_5(\lambda)\,\partial_w^{h_1+h_2-h_3-p}\,\Phi^{(h_3),ij}_1
\nonu     \\
    &- & f^{h_1,h_2,h_3,p}_6(\lambda)\,\partial_w^{h_1+h_2-h_3-p}\,
    \tilde{\Phi}^{(h_3),ij}_1\,\Bigg](w)+\cdots,
    \nonu \\   
    \Phi^{(h_1),i}_\frac{3}{2}(z)\,\Phi^{(h_2)}_0(w)
    &= & \sum_{p=1}^{h_1+h_2+1}\,\,\frac{1}{(z-w)^p}\,
    \Bigg[-\sum_{h_3=0}^{h_1+h_2-p} f^{h_1,h_2,h_3,p}_7(\lambda)\,
    \partial_w^{h_1+h_2-h_3-p}\,\Phi^{(h_3),i}_\frac{3}{2}
    \nonu \\
    &+ & \sum_{h_3=0}^{h_1+h_2-p+1}
    f^{h_1,h_2,h_3,p}_8(\lambda)\,\partial_w^{h_1+h_2-h_3-p+1}
    \Phi^{(h_3),i}_\frac{1}{2}\,\Bigg](w)+\cdots,
    \nonu \\    
    \Phi^{(h_1)}_2(z)\,\Phi^{(h_2)}_0(w)
&= &
    \frac{(h_1+h_2+1)!}{(z-w)^{h_1+h_2+2}}\,
    \bigg[(-1)^{h_1+h_2}C_{0,2}^{h_2,h_1}(\lambda)+\frac{(1-4\lambda)}{(2h_1+1)}C_{0,0}^{h_1,h_2}(\lambda)\bigg]
\nonu    \\
&+ &
\sum_{p=1}^{h_1+h_2+1\,\,}\sum_{h_3=1}^{h_1+h_2-p+2}\frac{1}{(z-w)^p}\,\Bigg[
      f^{h_1,h_2,h_3,p}_9(\lambda)\,\partial_w^{h_1+h_2-h_3-p+2}
      \nonu \\
      & \times &
\Big(\Phi^{(h_3-2)}_2\!-\frac{(1-4\lambda)}{(2h_3-3)}\partial_w^2 \Phi^{(h_3-2)}_0\Big)
\nonu    \\
    &+ & f^{h_1,h_2,h_3,p}_{10}(\lambda)\,
    \partial_w^{h_1+h_2-h_3-p+2}\,\Phi^{(h_3)}_0\,\Bigg](w)+\cdots.
\label{n4sope}
\eea
For fixed $\Phi^{(h_2)}_0(w)$, all the five components with $z$
dependence are acted in (\ref{n4sope}) and some of the coefficients
in (\ref{C00}) and (\ref{bigC}) are used. 
Based on (\ref{n4sope}) together with the footnote \ref{Relations},
the ${\cal N}=4$ super OPE (\ref{SUPER}) can be determined completely.

The $\la$ dependent structure constants appearing in 
(\ref{n4sope}) are given by
\bea
f^{\,\,h_1,h_2,h_3,p}_1(\lambda)
&\equiv & \sum_{k=0}^{h_1+h_2-h_3-1}
\frac{4^{2(h_1+h_2-h_3)-4}(2h_3-1)!(p-1)!}{(2h_1-1)(2h_2-1)}
\nonu \\
&\times & \Bigg[
(-1)^{h_1+h_2+k+1}\binom{k}{h_1+h_2-h_3-p}
\Big(
(h_1-2\lambda)(h_2-2\lambda)S^{\,\,h_1,h_2,h_3,k}_{F,R}
\nonu \\
&
- & (h_1-1+2\lambda)(h_2-1+2\lambda)S^{\,\,h_1,h_2,h_3,k}_{B,R}\,
\Big)
\nonu \\
& + & (-1)^{h_3+k+1}\binom{h_1+h_2-h_3-k-1}{h_1+h_2-h_3-p}
\Big(
(h_1-2\lambda)(h_2-2\lambda)S^{\,\,h_1,h_2,h_3,k}_{F,L}
\nonu \\
&
-& (h_1-1+2\lambda)(h_2-1+2\lambda)S^{\,\,h_1,h_2,h_3,k}_{B,L}\,
\Big)
\Bigg],
\nonu \\
f^{\,\,h_1,h_2,h_3,p}_2(\lambda)
& \equiv &
\sum_{k=0}^{h_1+h_2-h_3-1}
\frac{4^{2(h_1+h_2-h_3)-1}(2h_3-2)!(p-1)!}{2(2h_1-1)(2h_2-1)}
\Bigg[
  (-1)^{h_1+h_2+k} \nonu \\
  & \times & \binom{k}{h_1+h_2-h_3-p}
\Big(
(h_1-2\lambda)(h_2-2\lambda)(h_3-1+2\lambda)S^{\,\,h_1,h_2,h_3,k}_{F,R}
\nonu \\
&
+ & (h_1-1+2\lambda)(h_2-1+2\lambda)(h_3-2\lambda)
S^{\,\,h_1,h_2,h_3,k}_{B,R}\,
\Big)
\nonu \\
&+ & (-1)^{h_3+k}\binom{h_1+h_2-h_3-k-1}{h_1+h_2-h_3-p}
\Big(
(h_1-2\lambda)(h_2-2\lambda)(h_3-1+2\lambda)
\nonu \\
& \times & S^{\,\,h_1,h_2,h_3,k}_{F,L}
+ (h_1-1+2\lambda)(h_2-1+2\lambda)(h_3-2\lambda)
S^{\,\,h_1,h_2,h_3,k}_{B,L}\,
\Big)
\Bigg],
\nonu \\
f^{\,\,h_1,h_2,h_3,p}_3(\lambda)
& \equiv &
\sum_{k=0}^{h_1+h_2-h_3-1}
\frac{2\times4^{2(h_1+h_2-h_3)-5}(2h_3)!(p-1)!}{(2h_2-1)}
\nonu \\
&\times &
\Bigg[
(-1)^{h_3+k+1}\binom{h_1+h_2-h_3-k-1}{h_1+h_2-h_3-p}
\nonu \\
& \times &
\Big(
(h_2-2\lambda)T^{\,\,h_2,h_1,h_3,k}_{F}
+(h_2-1+2\lambda)\bar{T}^{\,\,h_2,h_1,h_3,k}_{B}
\Big)
\nonu \\
&+ &
(-1)^{h_1+h_2+k}\binom{k}{h_1+h_2-h_3-p}
\nonu \\
& \times &
\Big(
(h_2-2\lambda)\bar{T}^{\,\,h_2,h_1,h_3,k}_{F,L}
+(h_2-1+2\lambda)T^{\,\,h_2,h_1,h_3,k}_{B}
\Big)
\Bigg],
\nonu \\
f^{\,\,h_1,h_2,h_3,p}_4(\lambda)
& \equiv & \sum_{k=0}^{h_1+h_2-h_3-2}
\frac{2\times4^{2(h_1+h_2-h_3)-6}(2h_3+2)!(p-1)!}{(2h_2-1)}
\nonu \\
&\times&
\Bigg[
(-1)^{h_3+k+1}\binom{h_1+h_2-h_3-k-2}{h_1+h_2-h_3-p-1}
\nonu \\
& \times &
\Big(
(h_2-2\lambda)T^{\,\,h_2,h_1,h_3+1,k}_{F}
-(h_2-1+2\lambda)\bar{T}^{\,\,h_2,h_1,h_3+1,k}_{B}
\Big)
\nonu \\
&+&
(-1)^{h_1+h_2+k}\binom{k}{h_1+h_2-h_3-p-1}
\nonu \\
& \times &
\Big(
(h_2-2\lambda)\bar{T}^{\,\,h_2,h_1,h_3+1,k}_{F,L}
-(h_2-1+2\lambda)T^{\,\,h_2,h_1,h_3+1,k}_{B}
\Big)
\Bigg],
\nonu \\
f^{\,\,h_1,h_2,h_3,p}_5(\lambda)
& \equiv &
\sum_{k=0}^{h_1+h_2-h_3-1}
\frac{2\times4^{2(h_1+h_2-h_3)-5}(2h_3+1)!(p-1)!}{(2h_2-1)}
\nonu \\
&\times&
\Bigg[
(-1)^{h_3+k}\binom{h_1+h_2-h_3-k-1}{h_1+h_2-h_3-p}
\nonu \\
& \times &
\Big(
-(h_2-2\lambda)S^{\,\,h_2,h_1+1,h_3+1,k}_{F,R}
+(h_2-1+2\lambda)S^{\,\,h_2,h_1+1,h_3+1,k}_{B,R}
\Big)
\nonu \\
&+&
(-1)^{h_1+h_2+k}\binom{k}{h_1+h_2-h_3-p}
\nonu \\
& \times &
\Big(
-(h_2-2\lambda)S^{\,\,h_2,h_1+1,h_3+1,k}_{F,L}
+(h_2-1+2\lambda)S^{\,\,h_2,h_1+1,h_3+1,k}_{B,L}
\Big)
\Bigg],
\nonu \\
f^{\,\,h_1,h_2,h_3,p}_6(\lambda)
& \equiv &
\sum_{k=0}^{h_1+h_2-h_3-1}
\frac{2\times4^{2(h_1+h_2-h_3)-5}(2h_3+1)!(p-1)!}{(2h_2-1)}
\nonu \\
&\times&
\Bigg[
(-1)^{h_3+k}\binom{h_1+h_2-h_3-k-1}{h_1+h_2-h_3-p}
\nonu \\
& \times &
\Big(
-(h_2-2\lambda)S^{\,\,h_2,h_1+1,h_3+1,k}_{F,R}
-(h_2-1+2\lambda)S^{\,\,h_2,h_1+1,h_3+1,k}_{B,R}
\Big)
\nonu \\
&+ &
(-1)^{h_1+h_2+k}\binom{k}{h_1+h_2-h_3-p}
\nonu \\
& \times &
\Big(
-(h_2-2\lambda)S^{\,\,h_2,h_1+1,h_3+1,k}_{F,L}
-(h_2-1+2\lambda)S^{\,\,h_2,h_1+1,h_3+1,k}_{B,L}
\Big)
\Bigg],
\nonu \\
f^{\,\,h_1,h_2,h_3,p}_7(\lambda)
& \equiv & -f^{\,\,h_1+1,h_2,h_3+1,p}_3(\lambda)+
\sum_{k=0}^{h_1+h_2-h_3-1}
\frac{4^{2(h_1+h_2-h_3)-5}(2h_3+2)!(p-1)! (1-4\lambda)}{2(2h_1+1)(2h_2-1)}
\nonu \\
&\times & \Bigg[
(-1)^{h_3+k}\binom{h_1+h_2-h_3-k-2}{h_1+h_2-h_3-p}
\nonu \\
& \times &
\Big(
(h_2-2\lambda)T^{\,\,h_2,h_1,h_3+1,k}_{F}
-(h_2-1+2\lambda)\bar{T}^{\,\,h_2,h_1,h_3+1,k}_{B}
\Big) \nonu
\\
&+ &
(-1)^{h_1+h_2+k+1}\binom{k}{h_1+h_2-h_3-p}
\nonu \\
& \times &
\Big(
(h_2-2\lambda)\bar{T}^{\,\,h_2,h_1,h_3+1,k}_{F}
-(h_2-1+2\lambda)T^{\,\,h_2,h_1,h_3+1,k}_{B}
\Big)
\Bigg],
\nonu \\
f^{\,\,h_1,h_2,h_3,p}_8(\lambda)
& \equiv & -4^{-2}\,f^{\,\,h_1+1,h_2,h_3-1,p}_4(\lambda)
\nonu \\
&
+&
\sum_{k=0}^{h_1+h_2-h_3}
\frac{ 4^{2(h_1+h_2-h_3)-4}(2h_3)!(p-1)! (1-4\lambda)}{2(2h_1+1)(2h_2-1)}
\nonu \\
&\times & \Bigg[
(-1)^{h_3+k}\binom{h_1+h_2-h_3-k-1}{h_1+h_2-h_3-p+1}
\nonu \\
&\times&
\Big(
(h_2-2\lambda)T^{\,\,h_2,h_1,h_3,k}_{F}
+(h_2-1+2\lambda)\bar{T}^{\,\,h_2,h_1,h_3,k}_{B}
\Big)
\nonu \\
&
+&
(-1)^{h_1+h_2+k+1}\binom{k}{h_1+h_2-h_3-p+1}
\nonu \\
& \times &
\Big(
(h_2-2\lambda)\bar{T}^{\,\,h_2,h_1,h_3,k}_{F}
+(h_2-1+2\lambda)T^{\,\,h_2,h_1,h_3,k}_{B}
\Big)
\Bigg],
\nonu \\
f^{\,\,h_1,h_2,h_3,p}_{9}(\lambda)
  & \equiv & \sum_{k=0}^{h_1+h_2-h_3+1}
\frac{4^{2(h_1+h_2-h_3)}(2h_3-2)!(p-1)!}{(2h_2-1)}\Bigg[
\nonu \\
&\times&
\Bigg(
(-1)^{h_3+k+1}\binom{h_1+h_2-h_3-k+1}{h_1+h_2-h_3-p+2}
\nonu \\
& \times &
\Big(
(h_2-2\lambda)(h_3-1+2\lambda)S^{\,\,h_2,h_1+2,h_3,k}_{F,R}
\nonu \\
& - & (h_2-1+2\lambda)(h_3-2\lambda)S^{\,\,h_2,h_1+2,h_3,k}_{B,R}
\Big)
\nonu \\
&+&
(-1)^{h_1+h_2+k+1}\binom{k}{h_1+h_2-h_3-p+2}
\nonu \\
& \times &
\Big(
(h_2-2\lambda)(h_3-1+2\lambda)S^{\,\,h_2,h_1+2,h_3,k}_{F,L}
\nonu \\
& - & (h_2-1+2\lambda)(h_3-2\lambda)S^{\,\,h_2,h_1+2,h_3,k}_{B,L}
\Big)
\Bigg)
\nonu \\
&+&
\frac{(1-4\lambda)}{2\times 4(2h_1+1)(2h_1-1)}\Bigg(
(-1)^{h_1+h_2+k}\binom{k}{h_1+h_2-h_3-p+2}
\nonu \\
& \times &
\Big(
(h_1-2\lambda)(h_2-2\lambda)(h_3-1+2\lambda)
S^{\,\,h_1,h_2,h_3,k}_{F,R}
\nonu \\
& 
+&
(h_1-1+2\lambda)(h_2-1+2\lambda)(h_3-2\lambda)
S^{\,\,h_1,h_2,h_3,k}_{B,R}
\Big)
\nonu \\
&+ &
(-1)^{h_3+k}\binom{h_1+h_2-h_3-k-1}{h_1+h_2-h_3-p+2}
\nonu \\
& \times &
\Big(
(h_1-2\lambda)(h_2-2\lambda)(h_3-1+2\lambda)
S^{\,\,h_1,h_2,h_3,k}_{F,L}
\nonu \\
& 
+&
(h_1-1+2\lambda)(h_2-1+2\lambda)(h_3-2\lambda)
S^{\,\,h_1,h_2,h_3,k}_{B,L}
\Big)
\Bigg)
\Bigg],
\nonu  \\
f^{\,\,h_1,h_2,h_3,p}_{10}(\lambda)
  &\equiv & \sum_{k=0}^{h_1+h_2-h_3+1}
\frac{4^{2(h_1+h_2-h_3)-3}(2h_3-1)!(p-1)!}{(2h_2-1)}\Bigg[
\nonu \\
&\times&
2\Bigg(
(-1)^{h_3+k}\binom{h_1+h_2-h_3-k+1}{h_1+h_2-h_3-p+2}
\nonu \\
& \times &
\Big(
(h_2-2\lambda)S^{\,\,h_2,h_1+2,h_3,k}_{F,R}
+(h_2-1+2\lambda)S^{\,\,h_2,h_1+2,h_3,k}_{B,R}
\Big)
\nonu \\
&+&
(-1)^{h_1+h_2+k}\binom{k}{h_1+h_2-h_3-p+2}
\nonu \\
& \times &
\Big(
(h_2-2\lambda)S^{\,\,h_2,h_1+2,h_3,k}_{F,L}
+(h_2-1+2\lambda)S^{\,\,h_2,h_1+2,h_3,k}_{B,L}
\Big)
\Bigg)
\nonu \\
&+&
\frac{(1-4\lambda)}{4(2h_1+1)(2h_1-1)}\Bigg(
(-1)^{h_1+h_2+k+1}\binom{k}{h_1+h_2-h_3-p+2}
\nonu \\
& \times &
\Big(
(h_1-2\lambda)(h_2-2\lambda)S^{\,\,h_1,h_2,h_3,k}_{F,R}
\nonu \\
& - & (h_1-1+2\lambda)(h_2-1+2\lambda)S^{\,\,h_1,h_2,h_3,k}_{B,R}
\Big)
\nonu \\
&-&
(-1)^{h_3+k}\binom{h_1+h_2-h_3-k-1}{h_1+h_2-h_3-p+2}
\nonu \\
& \times &
\Big(
(h_1-2\lambda)(h_2-2\lambda)S^{\,\,h_1,h_2,h_3,k}_{F,L}
\nonu \\
& - & (h_1-1+2\lambda)(h_2-1+2\lambda)S^{\,\,h_1,h_2,h_3,k}_{B,L}
\Big)
\Bigg)
\Bigg].
\label{10f}
\eea
Again, the arguments $\la$ in (\ref{STRUCT}) are ignored in (\ref{10f})
for convenience.
The summation over $k$ is put to these structure constants
with other $k$ dependent factors.

The ${\cal N}=4$ super OPE between the ${\cal N}=4$
stress energy tensor and the ${\cal N}=4$ multiplet
can be summarized by
\bea
& & \bold{J}(Z_1)\,\bold{\Phi}^{(h)}(Z_2)
=
\frac{1}{z_{12}^{h}}\,C_{1}^{h}(\lambda)
+\frac{\theta^{4-0}_{12}}{z_{12}^{h+2}}\,C_{2}^{h}(\lambda)
\nonu \\
&&
+\sum_{p=1}^{h\,\,}\frac{1}{z_{12}^p}\,
\Bigg[\,
\sum_{h_3=1}^{h-p}
g^{h,h_3,p}_1(\lambda)\,\partial^{h-h_3-p} {\bf \Phi}^{(h_3)}
+\sum_{h_3=2}^{h-p}\,
g^{h,h_3,p}_2(\lambda)\,\partial^{h-h_3-p} D^{4-0} {\bf \Phi}^{(h_3-2)}
\nonu \\
&& +v_1^{h,p}(\lambda)\,\partial^{h-p}{\bf J}
\,\Bigg](Z_2)
\nonu \\
& &
+\sum_{p=1}^{h\,\,}\frac{\theta^i_{12}}{z_{12}^p}\,
\Bigg[\,
\sum_{h_3=0}^{h-p}
g^{h,h_3,p}_3(\lambda)\,\partial^{h-h_3-p} D^i {\bf \Phi}^{(h_3)}
+\sum_{h_3=0}^{h-p-1}\,
g^{h,h_3,p}_4(\lambda)\,\partial^{h-h_3-p-1} D^{4-i} {\bf \Phi}^{(h_3)}
\,\Bigg](Z_2)
\nonu \\
&&
+\sum_{p=1}^{h\,\,}\sum_{h_3=0}^{h-p}\frac{\theta^{4-ij}_{12}}{z_{12}^p}\,
\Bigg[\,
g^{h,h_3,p}_5(\lambda)\,D^{4-ij}
+
g^{h,h_3,p}_6(\lambda)\,D^{ij}
\,\Bigg] \partial^{h-h_3-p} {\bf \Phi}^{(h_3)}(Z_2)
\nonu \\
&&
+\sum_{p=1}^{h+1\,\,}\frac{\theta^{4-i}_{12}}{z_{12}^p}\,
\Bigg[\,
\sum_{h_3=0}^{h-p}
g^{h,h_3,p}_7(\lambda)\,\partial^{h-h_3-p} D^{4-i} {\bf \Phi}^{(h_3)}
+
\sum_{h_3=0}^{h-p+1}
g^{h,h_3,p}_8(\lambda)\,\partial^{h-h_3-p+1} D^i {\bf \Phi}^{(h_3)}
\,\Bigg](Z_2)
\nonu \\
&&
+\sum_{p=1}^{h+1\,\,}\frac{\theta^{4-0}_{12}}{z_{12}^p}\,
\Bigg[\,
\sum_{h_3=2}^{h-p+2}
g^{h,h_3,p}_9(\lambda)\,\partial^{h-h_3-p+2} D^{4-0} {\bf \Phi}^{(h_3-2)}
+\sum_{h_3=1}^{h-p+2}\,
g^{h,h_3,p}_{10}(\lambda)\,\partial^{h-h_3-p+2}  {\bf \Phi}^{(h_3)}
\nonu \\
&& +v^{h,p}_2(\lambda)\,\partial^{h-p+2}{\bf J}
\,\Bigg](Z_2) + \cdots.
\label{jphi}
\eea
All the terms in (\ref{jphi}) should have the spin $h$.
On the right hand side, there exists the ${\cal N}=4$ stress energy
tensor ${\bf J}$.

The central terms and the structure constants
appearing in (\ref{jphi}) are given by
\bea
C^{h}_1(\lambda) & \equiv & -\frac{N\,4^{2(h-2)}(h-1)!}{(2h-1)}
\sum_{i=0}^{h-1}\frac{i!}{(i+1)!}
\Bigg[(h-2\lambda)a^i(h,\lambda+\tfrac{1}{2})
+(h-1+2\lambda)a^i(h,\lambda) \Bigg]\,,
\nonu \\
C^{h}_2(\lambda) & \equiv & -\frac{N\,4^{2(h-2)}(h+1)!}{(2h-1)}
\sum_{i=0}^{h-1}\frac{1}{(i+1)(i+2)}
\Bigg[(h-2\lambda)(2+3i)a^i(h,\lambda+\tfrac{1}{2})
  \nonu \\
  & + & (h-1+2\lambda)(i-2)a^i(h,\lambda) \Bigg]\,,
\nonu \\
g^{h,h_3,p}_1(\lambda) & \equiv &
\frac{256}{p}
\Bigg[
f^{-1,h,h_3,p+1}_{10}(\lambda)-\frac{(1-4\lambda)}{(2h_3+1)}\,f^{-1,h,h_3+2,p+1}_9(\lambda)
\Bigg]\,,
\nonu \\
g^{h,h_3,p}_2(\lambda) & \equiv &
\frac{256}{p}
\,
f^{-1,h,h_3,p+1}_9(\lambda)\,,
\nonu \\
g^{h,h_3,p}_3(\lambda) & \equiv &
128
\Bigg[
f^{0,h,h_3,p}_3(\lambda)+\frac{(1-4\lambda)}{(2h_3+1)}\,f^{0,h,h_3,p}_4(\lambda)
\Bigg]\,,
\nonu \\
g^{h,h_3,p}_4(\lambda) & \equiv &
128\,f^{0,h,h_3,p}_4(\lambda)\,,
\nonu \\
g^{h,h_3,p}_5(\lambda) & \equiv &
64\,f^{0,h,h_3,p}_5(\lambda)\,,
\nonu \\
g^{h,h_3,p}_6(\lambda) & \equiv &
64\,f^{0,h,h_3,p}_6(\lambda)\,,
\nonu \\
g^{h,h_3,p}_7(\lambda) & \equiv &
\frac{4^{2(h-h_3)-1}}{2(2h-1)}(2h_3+2)!(p-1)!
\,
\sum_{k=0}^{h-h_3-1}
\Bigg[
(-1)^{h_3+k+1}
\binom{h-h_3-k-1}{h-h_3-p}
\nonu \\
& \times & \bigg(
(h-2\lambda)\,T_{F}^{\,\,h,1,h_3+1,k}(\lambda)
+(h-1+2\lambda)\,\bar{T}_{B}^{\,\,h,1,h_3+1,k}(\lambda)
\bigg)
\nonu \\
&
+& (-1)^{h_2+k}
\binom{k}{h-h_3-p}
\bigg(
(h-1+2\lambda)\,T_{B}^{\,\,h,1,h_3+1,k}(\lambda)
+(h-2\lambda)\,\bar{T}_{F}^{\,\,h,1,h_3+1,k}(\lambda)
\bigg)
\nonu \\
&
- & \frac{(1-4\lambda)}{2}
\bigg(
(-1)^{h+k}\binom{k}{h-h_3-p}
(h-2\lambda)\,\bar{T}_{F}^{\,\,h,0,h_3+1,k}(\lambda)
\nonu \\
&
+&
(-1)^{h_3+k}\binom{h-h_3-k-2}{h-h_3-p}
(h-1+2\lambda)\,\bar{T}_{B}^{\,\,h,0,h_3+1,k}(\lambda)
\bigg)
\Bigg],
\nonu \\
g^{h,h_3,p}_8(\lambda) & \equiv &
-\frac{4^{2(h-h_3)}}{2(2h-1)}(2h_3)!(p-1)!
\,
\sum_{k=0}^{h-h_3}
\Bigg[
(-1)^{h_3+k}
\binom{h-h_3-k}{h-h_3-p+1}
\nonu \\
& \times & 
\bigg(
(h-2\lambda)\,T_{F}^{\,\,h,1,h_3,k}(\lambda)
-(h-1+2\lambda)\,\bar{T}_{B}^{\,\,h,1,h_3,k}(\lambda)
\bigg)
\nonu \\
&
+& (-1)^{h_2+k}
\binom{k}{h-h_3-p+1}
\bigg(
(h-1+2\lambda)\,T_{B}^{\,\,h,1,h_3,k}(\lambda)
-(h-2\lambda)\,\bar{T}_{F}^{\,\,h,1,h_3,k}(\lambda)
\bigg)
\nonu \\
&
+& \frac{(1-4\lambda)}{2}
\bigg(
(-1)^{h+k}\binom{k}{h-h_3-p+1}
(h-2\lambda)\,\bar{T}_{F}^{\,\,h,0,h_3,k}(\lambda)
\nonu \\
&
-&
(-1)^{h_3+k}\binom{h-h_3-k-1}{h-h_3-p+1}
(h-1+2\lambda)\,\bar{T}_{B}^{\,\,h,0,h_3,k}(\lambda)
\bigg)
\Bigg]
+\frac{(1-4\lambda)}{(2h_3+1)}\,g^{h,h_3,p}_7(\lambda),
\nonu \\
g^{h,h_3,p}_9(\lambda) & \equiv &
\frac{4^{2(h-h_3)+3}}{(2h-1)}(2h_3-2)!(p-1)!
\,
\sum_{k=0}^{h-h_3+1}
\Bigg[
(-1)^{h_2+k+1}
\binom{k}{h-h_3-p+2}
\nonu \\
& \times & \bigg(
(h-2\lambda)(h_3-1+2\lambda)\,S_{F,R}^{\,\,2,h,h_3,k}(\lambda)
-(h-1+2\lambda)(h_3-2\lambda)\,S_{B,R}^{\,\,2,h,h_3,k}(\lambda)
\bigg)
\nonu \\
&
-& (-1)^{h_3+k}
\binom{h-h_3-k+1}{h-h_3-p+2}
\bigg(
(h-2\lambda)(h_3-1+2\lambda)\,S_{F,L}^{\,\,2,h,h_3,k}(\lambda)
\nonu \\
&- & (h-1+2\lambda)(h_3-2\lambda)\,S_{B,L}^{\,\,2,h,h_3,k}(\lambda)
\bigg)
\nonu \\
&
+& \frac{(1-4\lambda)}{4}
(-1)^{h+k}\binom{k}{h-h_3-p+2}
\bigg(
(h-2\lambda)(h_3-1+2\lambda)\,S_{F,R}^{\,\,1,h,h_3,k}(\lambda)
\nonu \\
&
-& (h-1+2\lambda)(h_3-2\lambda)\,S_{B,R}^{\,\,1,h,h_3,k}(\lambda)
\bigg)
\nonu \\
&
-& \frac{(1-4\lambda)}{4}(-1)^{h_3+k}\binom{h-h_3-k}{h-h_3-p+2}
\bigg(
(h-2\lambda)(h_3-1+2\lambda)\,S_{F,L}^{\,\,1,h,h_3,k}(\lambda)
\nonu \\
&-&
(h-1+2\lambda)(h_3-2\lambda)\,S_{B,L}^{\,\,1,h,h_3,k}(\lambda)
\bigg)
\Bigg],
\nonu \\
g^{h,h_3,p}_{10}(\lambda) & \equiv &
\frac{4^{2(h-h_3)+1}}{2(2h-1)}(2h_3-1)!(p-1)!
\,
\sum_{k=0}^{h-h_3+1}
\Bigg[
(-1)^{h_2+k}
\binom{k}{h-h_3-p+2}
\nonu \\
& \times & \bigg(
(h-2\lambda)\,S_{F,R}^{\,\,2,h,h_3,k}(\lambda)
+(h-1+2\lambda)\,S_{B,R}^{\,\,2,h,h_3,k}(\lambda)
\bigg)
\nonu \\
&
+& (-1)^{h_3+k}
\binom{h-h_3-k+1}{h-h_3-p+2}
\bigg(
(h-2\lambda)\,S_{F,L}^{\,\,2,h,h_3,k}(\lambda)
+(h-1+2\lambda)\,S_{B,L}^{\,\,2,h,h_3,k}(\lambda)
\bigg)
\nonu \\
&
-& \frac{(1-4\lambda)}{4}
(-1)^{h+k}\binom{k}{h-h_3-p+2}
\bigg(
(h-2\lambda)\,S_{F,R}^{\,\,1,h,h_3,k}(\lambda)
\nonu \\
&
+& (h-1+2\lambda)\,S_{B,R}^{\,\,1,h,h_3,k}(\lambda)
\bigg)
\nonu \\
&
+& \frac{(1-4\lambda)}{4}(-1)^{h_3+k}\binom{h-h_3-k}{h-h_3-p+2}
\bigg(
(h-2\lambda)\,S_{F,L}^{\,\,1,h,h_3,k}(\lambda)
\nonu \\
&
+& (h-1+2\lambda)\,S_{B,L}^{\,\,1,h,h_3,k}(\lambda)
\bigg)
\Bigg]
-\frac{(1-4\lambda)}{(2h_3+1)}\,g^{h,h_3+2,p}_9(\lambda),
\nonu \\
v^{h,p}_1(\lambda)& \equiv & -\frac{1}{p}\Bigg[
f^{-1,h,1,p+1}_9(\lambda)+4(1-4\lambda)\,f^{-1,h,2,p+1}_9(\lambda)
\Bigg],
\nonu \\
v^{h,p}_2(\lambda) & \equiv &
-\frac{1}{256}\Bigg[
\,g^{h,1,p}_9(\lambda)+4(1-4\lambda)\,g^{h,2,p}_9(\lambda)\,
\Bigg],
\label{jphiquantity}
\eea
where the previous relations (\ref{STRUCT}) and (\ref{10f})
are used in (\ref{jphiquantity}).


\section{ The  ${\cal N}=2$ supersymmetric linear
  $W_{\infty}^{K,K}[\la]$ algebra }

The whole $26$ (anti)commutators are summarized by
\bea
\comm{(W_{F,h_1}^{\lambda})_m}{(W_{F,h_2}^{\lambda})_n}
&=&
K\,c_{W_F}^{h_1,h_2}(\la)\,q^{h_1+h_2-4}\,[m+h_1-1]_{h_1+h_2-1}\,\delta_{m+n}
\nonu \\
&+ & \sum_{h_3=1}^{h_1+h_2-1\,\, }\sum_{k=0}^{h_1+h_2-h_3-1}
\,
(-1)^{h_3}\,(4q)^{h_1+h_2-h_3-2}\,(2h_3-1)!\nonu \\
&\times&
\bigg(\,\,S^{\,\,h_1,h_2,h_3,k}_{F,\,R}\, [m+h_1-1]_{h_1+h_2-h_3-k-1}\,[n+h_2-1]_{k}
\nonu \\
&-&
S^{\,\,h_1,h_2,h_3,k}_{F,\,L}\,  [m+h_1-1]_{k}\,[n+h_2-1]_{h_1+h_2-h_3-k-1}\,\,\bigg)\,(W_{F,h_3}^{\lambda})_{m+n},    
\nonu \\
\comm{(W_{F,h_1}^{\lambda})_m}{(W_{F,h_2}^{\lambda,\hat{A}})_n}
&= & \sum_{h_3=1}^{h_1+h_2-1\,\, }\sum_{k=0}^{h_1+h_2-h_3-1} 
\,
(-1)^{h_3}\,(4q)^{h_1+h_2-h_3-2}\,(2h_3-1)!
\nonu \\
&\times &
\bigg(\,\,S^{\,\,h_1,h_2,h_3,k}_{F,\,R}\, [m+h_1-1]_{h_1+h_2-h_3-k-1}\,[n+h_2-1]_{k}
\nonu \\
&- &
S^{\,\,h_1,h_2,h_3,k}_{F,\,L}\,  [m+h_1-1]_{k}\,[n+h_2-1]_{h_1+h_2-h_3-k-1}\,\,\bigg)\,(W_{F,h_3}^{\lambda,\hat{A}})_{m+n},    
\nonu \\
\comm{(W_{F,h_1}^{\lambda,\hat{A}})_m}{(W_{F,h_2}^{\lambda,\hat{B}})_n}
&=&
\delta^{\hat{A}\hat{B}}\,
c_{W_F}^{h_1,h_2}(\la)\,q^{h_1+h_2-4}\,[m+h_1-1]_{h_1+h_2-1}\,\delta_{m+n} \nonu \\
&+ & \frac{1}{2}\sum_{h_3=1}^{h_1+h_2-1\,\, }\sum_{k=0}^{h_1+h_2-h_3-1}
\,
(-1)^{h_3}\,(4q)^{h_1+h_2-h_3-2}\,(2h_3-1)! \nonu \\
&\times&
\Bigg[ \bigg(\,\,S^{\,\,h_1,h_2,h_3,k}_{F,\,R}\, [m+h_1-1]_{h_1+h_2-h_3-k-1}\,[n+h_2-1]_{k}
\nonu \\
&-&
S^{\,\,h_1,h_2,h_3,k}_{F,\,L}\,  [m+h_1-1]_{k}\,[n+h_2-1]_{h_1+h_2-h_3-k-1}\,\,\bigg)
\nonu \\
& \times & \bigg(
\frac{2}{K}\delta^{\hat{A}\hat{B}}\,\,(W_{F,h_3}^{\lambda})_{m+n}
+d^{\hat{A}\hat{B}\hat{C}}\,(W_{F,h_3}^{\lambda,\hat{C}})_{m+n}
\bigg)
\nonu \\
&+&
\bigg(\,\,S^{\,\,h_1,h_2,h_3,k}_{F,\,R}\, [m+h_1-1]_{h_1+h_2-h_3-k-1}\,[n+h_2-1]_{k}
\nonu \\
&+ &
S^{\,\,h_1,h_2,h_3,k}_{F,\,L}\,  [m+h_1-1]_{k}\,[n+h_2-1]_{h_1+h_2-h_3-k-1}\,\,\bigg)\nonu \\
& \times &
i\,f^{\hat{A}\hat{B}\hat{C}}\,(W_{F,h_3}^{\lambda,\hat{C}})_{m+n} \,\Bigg],
\nonu \\
\comm{(W_{B,h_1}^{\lambda})_m}{(W_{B,h_2}^{\lambda})_n}
&=&
K\,c_{W_B}^{h_1,h_2}(\la)\,q^{h_1+h_2-4}\,[m+h_1-1]_{h_1+h_2-1}\,\delta_{m+n}
\nonu \\
&+ & \sum_{h_3=1}^{h_1+h_2-1\,\, }\sum_{k=0}^{h_1+h_2-h_3-1}
\,
(-1)^{h_3}\,(4q)^{h_1+h_2-h_3-2}\,(2h_3-1)!
\nonu \\
&\times &
\bigg(\,\,\,S^{\,\,h_1,h_2,h_3,k}_{B,\,R}\,
     [m+h_1-1]_{h_1+h_2-h_3-k-1}\,[n+h_2-1]_{k}
\nonu \\
&-&
\,S^{\,\,h_1,h_2,h_3,k}_{B,\,L}\,  [m+h_1-1]_{k}\,[n+h_2-1]_{h_1+h_2-h_3-k-1}\,\,\bigg)\,(W_{B,h_3}^{\lambda})_{m+n},  
\nonu \\
\comm{(W_{B,h_1}^{\lambda})_m}{(W_{B,h_2}^{\lambda,\hat{A}})_n}
&=&
\sum_{h_3=1}^{h_1+h_2-1\,\, }\sum_{k=0}^{h_1+h_2-h_3-1}
\,
(-1)^{h_3}\,(4q)^{h_1+h_2-h_3-2}\,(2h_3-1)!\nonu \\
&\times &
\bigg(\,\,\,S^{\,\,h_1,h_2,h_3,k}_{B,\,R}\, [m+h_1-1]_{h_1+h_2-h_3-k-1}\,[n+h_2-1]_{k}
\nonu \\
&- & S^{\,\,h_1,h_2,h_3,k}_{B,\,L}\,  [m+h_1-1]_{k}\,[n+h_2-1]_{h_1+h_2-h_3-k-1}\,\,\bigg)\,(W_{B,h_3}^{\lambda,\hat{A}})_{m+n},    
\nonu \\
\comm{(W_{B,h_1}^{\lambda,\hat{A}})_m}{(W_{B,h_2}^{\lambda,\hat{B}})_n}
&=&
\delta^{\hat{A}\hat{B}}\,
c_{W_B}^{h_1,h_2}(\la)
\,q^{h_1+h_2-4}\,[m+h_1-1]_{h_1+h_2-1}\,\delta_{m+n} \nonu \\
&+ & \frac{1}{2}\sum_{h_3=1}^{h_1+h_2-1\,\, }\sum_{k=0}^{h_1+h_2-h_3-1}
\,
(-1)^{h_3}\,(4q)^{h_1+h_2-h_3-2}\,(2h_3-1)! \nonu \\
&\times&
\Bigg[\bigg(\,\,\,S^{\,\,h_1,h_2,h_3,k}_{B,\,R}\, [m+h_1-1]_{h_1+h_2-h_3-k-1}\,[n+h_2-1]_{k}
\nonu \\
&- &
S^{\,\,h_1,h_2,h_3,k}_{B,\,L}\,  [m+h_1-1]_{k}\,[n+h_2-1]_{h_1+h_2-h_3-k-1}\,\,\bigg)\nonu \\
& \times & \bigg(
\frac{2}{K}\delta^{\hat{A}\hat{B}}\,\,(W_{B,h_3}^{\lambda})_{m+n}
+d^{\hat{A}\hat{B}\hat{C}}\,(W_{B,h_3}^{\lambda,\hat{C}})_{m+n}\bigg)
\nonu \\
&+ &
\bigg(\,\,S^{\,\,h_1,h_2,h_3,k}_{B,\,R}\, [m+h_1-1]_{h_1+h_2-h_3-k-1}\,[n+h_2-1]_{k}
\nonu \\
&+ & S^{\,\,h_1,h_2,h_3,k}_{B,\,L}\,  [m+h_1-1]_{k}\,[n+h_2-1]_{h_1+h_2-h_3-k-1}\,\,\bigg)\nonu \\
& \times &
i\,f^{\hat{A}\hat{B}\hat{C}}\,(W_{B,h_3}^{\lambda,\hat{C}})_{m+n} \,\Bigg],
\nonu \\
\comm{(W_{F,h_1}^{\lambda})_m}{(Q_{h_2+\frac{1}{2}}^{\lambda})_r}
&= & \sum_{h_3=1}^{h_1+h_2-1\,\, }\sum_{k=0}^{h_1+h_2-h_3-1}
\,
(-1)^{h_3}\,(4q)^{h_1+h_2-h_3-2}\,(2h_3)! \nonu \\
&\times & 
T_F^{\,\,h_1,h_2,h_3,k}\,\,[m+h_1-1]_{h_1+h_2-h_3-k-1}\,[r+h_2-\tfrac{1}{2}]_k\,
(Q_{h_3+\frac{1}{2}}^{\lambda})_{m+r},    
\nonu \\
\comm{(W_{F,h_1}^{\lambda})_m}{(Q_{h_2+\frac{1}{2}}^{\lambda,\hat{A}})_r}
& = & \comm{(W_{F,h_1}^{\lambda,\hat{A}})_m}{(Q_{h_2+\frac{1}{2}}^{\lambda})_r}
\nonu \\
&=& \sum_{h_3=1}^{h_1+h_2-1\,\, }\sum_{k=0}^{h_1+h_2-h_3-1}
\,
(-1)^{h_3}\,(4q)^{h_1+h_2-h_3-2}\,(2h_3)! \nonu \\
&\times& 
T_F^{\,\,h_1,h_2,h_3,k}\,\,[m+h_1-1]_{h_1+h_2-h_3-k-1}\,[r+h_2-\tfrac{1}{2}]_k\,
(Q_{h_3+\frac{1}{2}}^{\lambda,\hat{A}})_{m+r},  
\nonu \\
\comm{(W_{F,h_1}^{\lambda,\hat{A}})_m}{(Q_{h_2+\frac{1}{2}}^{\lambda,\hat{B}})_r}
&=&
\frac{1}{2}\sum_{h_3=1}^{h_1+h_2-1\,\, }\sum_{k=0}^{h_1+h_2-h_3-1}
\,
(-1)^{h_3}\,(4q)^{h_1+h_2-h_3-2}\,(2h_3)!
\nonu \\
& \times & T_F^{\,\,h_1,h_2,h_3,k}
\,
[m+h_1-1]_{h_1+h_2-h_3-k-1}\,
[r+h_2-\tfrac{1}{2}]_k\,
\nonu \\
&\times &
\Bigg(\frac{2}{K}\delta^{\hat{A}\hat{B}}\,(Q_{h_3+\frac{1}{2}}^{\lambda})_{m+r} +
(i\,f^{\hat{A}\hat{B}\hat{C}}+d^{\hat{A}\hat{B}\hat{C}})\,(Q_{h_3+\frac{1}{2}}^{\lambda, \hat{C}})_{m+r} 
\,
\Bigg),
\nonu \\
\comm{(W_{F,h_1}^{\lambda})_m}{(\bar{Q}_{h_2+\frac{1}{2}}^{\lambda})_r}
&=& \sum_{h_3=0}^{h_1+h_2-1\,\, }\sum_{k=0}^{h_1+h_2-h_3-1}
\,
(-1)^{h_3}\,(4q)^{h_1+h_2-h_3-2}\,(2h_3)! \nonu \\
& \times & 
\bar{T}_F^{\,\,h_1,h_2,h_3,k}\,\,[m+h_1-1]_{k}\,[r+h_2-\tfrac{1}{2}]_{h_1+h_2-h_3-k-1}\,
(\bar{Q}_{h_3+\frac{1}{2}}^{\lambda})_{m+r},    
\nonu \\
\comm{(W_{F,h_1}^{\lambda})_m}{(\bar{Q}_{h_2+\frac{1}{2}}^{\lambda,\hat{A}})_r}
& = &
\comm{(W_{F,h_1}^{\lambda,\hat{A}})_m}{(\bar{Q}_{h_2+\frac{1}{2}}^{\lambda})_r}
\nonu \\
&=& \sum_{h_3=0}^{h_1+h_2-1\,\, }\sum_{k=0}^{h_1+h_2-h_3-1}
\,
(-1)^{h_3}\,(4q)^{h_1+h_2-h_3-2}\,(2h_3)! \nonu \\
&\times& 
\bar{T}_F^{\,\,h_1,h_2,h_3,k}\,\,[m+h_1-1]_{k}\,[r+h_2-\tfrac{1}{2}]_{h_1+h_2-h_3-k-1}\,
(\bar{Q}_{h_3+\frac{1}{2}}^{\lambda,\hat{A}})_{m+r},  
\nonu \\
\comm{(W_{F,h_1}^{\lambda,\hat{A}})_m}{(\bar{Q}_{h_2+\frac{1}{2}}^{\lambda,\hat{B}})_r}
&=&
\frac{1}{2}\sum_{h_3=0}^{h_1+h_2-1\,\, }\sum_{k=0}^{h_1+h_2-h_3-1}
\,
(-1)^{h_3}\,(4q)^{h_1+h_2-h_3-2}\,(2h_3)!
\nonu \\
& \times & \bar{T}_F^{\,\,h_1,h_2,h_3,k}
\, [m+h_1-1]_{k}\,[r+h_2-\tfrac{1}{2}]_{h_1+h_2-h_3-k-1}\,
\nonu \\
& \times &
\Bigg(\frac{2}{K}\delta^{\hat{A}\hat{B}}\,(Q_{h_3+\frac{1}{2}}^{\lambda})_{m+r} 
-(i\,f^{\hat{A}\hat{B}\hat{C}}-d^{\hat{A}\hat{B}\hat{C}})\,(Q_{h_3+\frac{1}{2}}^{\lambda, \hat{C}})_{m+r} 
\,
\Bigg),
\nonu \\
\comm{(W_{B,h_1}^{\lambda})_m}{(Q_{h_2+\frac{1}{2}}^{\lambda})_r}
&=& \sum_{h_3=1}^{h_1+h_2-1\,\, }\sum_{k=0}^{h_1+h_2-h_3-1}
\,
(-1)^{h_3+1}\,(4q)^{h_1+h_2-h_3-2}\,(2h_3)! \nonu \\
&\times & 
T_B^{\,\,h_1,h_2,h_3,k}\,\,[m+h_1-1]_k\,[r+h_2-\tfrac{1}{2}]_{h_1+h_2-h_3-k-1}\,
(Q_{h_3+\frac{1}{2}}^{\lambda})_{m+r},   
\nonu \\
\comm{(W_{B,h_1}^{\lambda})_m}{(Q_{F,h_2+\frac{1}{2}}^{\lambda,\hat{A}})_r}
& = & \comm{(W_{B,h_1}^{\lambda,\hat{A}})_m}{(Q_{h_2+\frac{1}{2}}^{\lambda})_r}
\nonu \\
&=& \sum_{h_3=1}^{h_1+h_2-1\,\, }\sum_{k=0}^{h_1+h_2-h_3-1}
\,
(-1)^{h_3+1}\,(4q)^{h_1+h_2-h_3-2}\,(2h_3)! \nonu \\
&\times& 
T_B^{\,\,h_1,h_2,h_3,k}\,\,[m+h_1-1]_k\,[r+h_2-\tfrac{1}{2}]_{h_1+h_2-h_3-k-1}\,
(Q_{h_3+\frac{1}{2}}^{\lambda,\hat{A}})_{m+r}, 
\nonu \\
\comm{(W_{B,h_1}^{\lambda,\hat{A}})_m}{(Q_{h_2+\frac{1}{2}}^{\lambda,\hat{B}})_r}
&=&
\frac{1}{2}\sum_{h_3=1}^{h_1+h_2-1\,\, }\sum_{k=0}^{h_1+h_2-h_3-1}
\,
(-1)^{h_3+1}\,(4q)^{h_1+h_2-h_3-2}\,(2h_3)!
\nonu \\
& \times & T_B^{\,\,h_1,h_2,h_3,k}
\,
      [m+h_1-1]_k\,[r+h_2-\tfrac{1}{2}]_{h_1+h_2-h_3-k-1}\,
      \nonu \\
      & \times & 
\Bigg(\frac{2}{K}\delta^{\hat{A}\hat{B}}\,(Q_{h_3+\frac{1}{2}}^{\lambda})_{m+r} 
-(i\,f^{\hat{A}\hat{B}\hat{C}}-d^{\hat{A}\hat{B}\hat{C}})\,
(Q_{h_3+\frac{1}{2}}^{\lambda, \hat{C}})_{m+r} 
\,
\Bigg),
\nonu \\
\comm{(W_{B,h_1}^{\lambda})_m}{(\bar{Q}_{h_2+\frac{1}{2}}^{\lambda})_r}
&=& \sum_{h_3=0}^{h_1+h_2-1\,\, }\sum_{k=0}^{h_1+h_2-h_3-1}
\,
(-1)^{h_3+1}\,(4q)^{h_1+h_2-h_3-2}\,(2h_3)! \nonu \\
&\times& 
\bar{T}_B^{\,\,h_1,h_2,h_3,k}\,\,[m+h_1-1]_{h_1+h_2-h_3-k-1}\,[r+h_2-\tfrac{1}{2}]_k\,
(\bar{Q}_{h_3+\frac{1}{2}}^{\lambda})_{m+r},
\nonu \\
\comm{(W_{B,h_1}^{\lambda})_m}{(\bar{Q}_{h_2+\frac{1}{2}}^{\lambda,\hat{A}})_r}
& = & \comm{(W_{B,h_1}^{\lambda,\hat{A}})_m}{(\bar{Q}_{h_2+\frac{1}{2}}^{\lambda})_r}
\nonu \\
&=& \sum_{h_3=0}^{h_1+h_2-1\,\, }\sum_{k=0}^{h_1+h_2-h_3-1}
\,
(-1)^{h_3+1}\,(4q)^{h_1+h_2-h_3-2}\,(2h_3)! \nonu \\
&\times& 
\bar{T}_B^{\,\,h_1,h_2,h_3,k}\,\,[m+h_1-1]_{h_1+h_2-h_3-k-1}\,[r+h_2-\tfrac{1}{2}]_k\,
(\bar{Q}_{h_3+\frac{1}{2}}^{\lambda,\hat{A}})_{m+r},   
\nonu \\
\comm{(W_{B,h_1}^{\lambda,\hat{A}})_m}{(\bar{Q}_{h_2+\frac{1}{2}}^{\lambda,\hat{B}})_r}
&=&
\sum_{h_3=0}^{h_1+h_2-1\,\, }\sum_{k=0}^{h_1+h_2-h_3-1}
\,
(-1)^{h_3+1}\,(4q)^{h_1+h_2-h_3-2}\,(2h_3)!
\nonu \\
& \times & \bar{T}_B^{\,\,h_1,h_2,h_3,k}(\lambda)
\,
[m+h_1-1]_{h_1+h_2-h_3-k-1}\,[r+h_2-\tfrac{1}{2}]_k\,
\nonu \\
.& \times &
\Bigg(\frac{2}{K}\delta^{\hat{A}\hat{B}}\,(Q_{h_3+\frac{1}{2}}^{\lambda})_{m+r} 
+(i\,f^{\hat{A}\hat{B}\hat{C}}+d^{\hat{A}\hat{B}\hat{C}})\,
(Q_{h_3+\frac{1}{2}}^{\lambda, \hat{C}})_{m+r} 
\,
\Bigg),
\nonu \\
\acomm{(Q_{h_1+\frac{1}{2}}^{\lambda})_r}{(\bar{Q}_{h_2+\frac{1}{2}}^{\lambda})_s}
&=& K\,c_{Q\bar{Q}}^{h_1,h_2}(\la)
\,q^{h_1+h_2-2}\,
[r+h_1-\tfrac{1}{2}]_{h_1+h_2}\,\delta_{r+s} \nonu \\
&+ & \sum_{h_3=1}^{h_1+h_2\,\, }\sum_{k=0}^{h_1+h_2-h_3}
\,
2\,(-1)^{h_3+1}\,(4q)^{h_1+h_2-h_3}\,(2h_3-1)! \nonu \\
&\times&
\Bigg(U_F^{\,\,h_1,h_2,h_3,k}\, 
[r+h_1-\tfrac{1}{2}]_{h1+h_2-h_3-k}\,[s+h_2-\tfrac{1}{2}]_{k}\,(W_{F,h_3}^{\lambda})_{r+s}
\nonu \\
&+ & U_B^{\,\,h_1,h_2,h_3,k}\, 
[r+h_1-\tfrac{1}{2}]_{k}\,[s+h_2-\tfrac{1}{2}]_{h1+h_2-h_3-k}\, (W_{B,h_3}^{\lambda})_{r+s}\Bigg),
\nonu \\
\acomm{(Q_{h_1+\frac{1}{2}}^{\lambda,\hat{A}})_r}{(\bar{Q}_{h_2+\frac{1}{2}}^{\lambda})_s} & = &
\acomm{(Q_{h_2+\frac{1}{2}}^{\lambda})_r}{(\bar{Q}_{h_1+\frac{1}{2}}^{\lambda,\hat{A}})_s}
\nonu \\
&=& \sum_{h_3=1}^{h_1+h_2 \,\, }\sum_{k=0}^{h_1+h_2-h_3}
\,
2\,(-1)^{h_3+1}\,(4q)^{h_1+h_2-h_3}\,(2h_3-1)! \nonu \\
&\times&
\Bigg(U_F^{\,\,h_1,h_2,h_3,k}\, 
[r+h_1-\tfrac{1}{2}]_{h1+h_2-h_3-k}\,[s+h_2-\tfrac{1}{2}]_{k}\,(W_{F,h_3}^{\lambda,\hat{A}})_{r+s}
\nonu \\
&+ & U_B^{\,\,h_1,h_2,h_3,k}\, 
[r+h_1-\tfrac{1}{2}]_{k}\,[s+h_2-\tfrac{1}{2}]_{h1+h_2-h_3-k}\, (W_{B,h_3}^{\lambda,\hat{A}})_{r+s}\Bigg),
\nonu \\
\acomm{(Q_{h_1+\frac{1}{2}}^{\lambda,\hat{A}})_r}{(\bar{Q}_{h_2+\frac{1}{2}}^{\lambda,\hat{B}})_s}
&=& \delta^{\hat{A}\hat{B}}\,c_{Q\bar{Q}}^{h_1,h_2}(\la)
\,q^{h_1+h_2-2}\,
   [r+h_1-\tfrac{1}{2}]_{h_1+h_2}\,\delta_{r+s}
   \nonu \\
&+ & \sum_{h_3=1}^{h_1+h_2 \,\, }\sum_{k=0}^{h_1+h_2-h_3}
\,
(-1)^{h_3+1}\,(4q)^{h_1+h_2-h_3}\,(2h_3-1)! \nonu \\
&\times&
\Bigg[\bigg(\, U_F^{\,\,h_1,h_2,h_3,k}\, 
[r+h_1-\tfrac{1}{2}]_{h1+h_2-h_3-k}\,[s+h_2-\tfrac{1}{2}]_{k} \nonu \\
&+ & U_B^{\,\,h_1,h_2,h_3,k}\, 
[r+h_1-\tfrac{1}{2}]_{k}\,[s+h_2-\tfrac{1}{2}]_{h1+h_2-h_3-k}\,\bigg)
\nonu \\
& \times & \bigg(
\frac{2}{K}\delta^{\hat{A}\hat{B}}(W_{F,h_3}^{\lambda})_{r+s}
+d^{\hat{A}\hat{B}\hat{C}}\,(W_{F,h_3}^{\lambda,\hat{C}})_{r+s}\bigg)
\nonu \\
&+ &
\bigg(\, U_F^{\,\,h_1,h_2,h_3,k}\, 
            [r+h_1-\tfrac{1}{2}]_{h1+h_2-h_3-k}\,[s+h_2-\tfrac{1}{2}]_{k}
            \nonu \\
&- & U_B^{\,\,h_1,h_2,h_3,k}\, 
[r+h_1-\tfrac{1}{2}]_{k}\,[s+h_2-\tfrac{1}{2}]_{h1+h_2-h_3-k}\,\bigg)
\nonu \\
& \times &
i\,f^{\hat{A}\hat{B}\hat{C}}\,(W_{B,h_3}^{\lambda,\hat{C}})_{r+s}\,\Bigg].
\label{26}
\eea
The central terms are given by (\ref{cwf}) and (\ref{remcentral}).
For $K \geq 3$, there are nonzero $d$ symbols.
In (\ref{26}),
we do not write down
the $\la$ dependences appearing in (\ref{STRUCT})
explicitly in order to save the space
\footnote{The corresponding OPEs can be obtained from the
  description of the footnote \ref{commtoope}.} and
again, the structure constants are given by (\ref{STRUCT}). 

We continue to present the description in (\ref{six})
for other cases having the adjoint indices and we obtain
the following results for several $h_1$ and $h_2$
\bea
W_{F,1}^{\lambda}(z)\,W_{F,1}^{\lambda,\hat{A}}(w)
&= & 0 +\cdots,
\nonu \\
W_{F,1}^{\lambda}(z)\,W_{F,2}^{\lambda,\hat{A}}(w)
&= &
\frac{1}{(z-w)^2}\,W_{F,1}^{\lambda,\hat{A}}(w)+\cdots,
\nonu \\
W_{F,1}^{\lambda}(z)\,W_{F,3}^{\lambda,\hat{A}}(w)
&=&
\frac{1}{(z-w)^3}\,8\, q\,\lambda\,W_{F,1}^{\lambda,\hat{A}}(w)
+\frac{1}{(z-w)^2}\,\Bigg[
2\,W_{F,2}^{\lambda,\hat{A}}
-4\, q\,\lambda\,\partial_w W_{F,1}^{\lambda,\hat{A}}
\Bigg](w)+\cdots,
\nonu 
\\
W_{F,2}^{\lambda}(z)\,W_{F,2}^{\lambda,\hat{A}}(w)
&=&
\frac{1}{(z-w)^2}\,2\,W_{F,2}^{\lambda,\hat{A}}(w)
+\frac{1}{(z-w)}\,\partial_w W_{F,2}^{\lambda,\hat{A}}(w)+\cdots,
\nonu \\
W_{F,2}^{\lambda}(z)\,W_{F,3}^{\lambda,\hat{A}}(w)
&=&
\frac{1}{(z-w)^4}\,16\, q^2\,(1-2\lambda)(1+2\lambda)\,
W_{F,1}^{\lambda,\hat{A}}(w)
+\frac{1}{(z-w)^2}\,3\,W_{F,3}^{\lambda,\hat{A}}(w)
\nonu \\
& + & \frac{1}{(z-w)}\,\partial_w W_{F,3}^{\lambda,\hat{A}}(w)+\cdots,
\nonu 
\\
W_{F,3}^{\lambda}(z)\,W_{F,3}^{\lambda,\hat{A}}(w)
&=&
\frac{1}{(z-w)^4}\,64\, q^2\,(1-\lambda)(1+\lambda)\,W_{F,2}^{\lambda,\hat{A}}(w)
\nonu \\
& + &
\frac{1}{(z-w)^3}\, 32\, q^2\,(1-\lambda)(1+\lambda)\,\partial_w
W_{F,2}^{\lambda,\hat{A}}(w)
\nonu \\
&+&
\frac{1}{(z-w)^2}
\Bigg[
4\,W_{F,4}^{\lambda,\hat{A}}
+
\frac{48 \, q^2}{5}\,(1-\lambda)(1+\lambda)\,\partial_w^2 W_{F,2}^{\lambda,\hat{A}}
\Bigg](w)
\nonu \\
&+&
\frac{1}{(z-w)}
\Bigg[
2\,\partial_w W_{F,4}^{\lambda,\hat{A}}
+\frac{32\, q^2}{15}\,
(1-\lambda)(1+\lambda)\,\partial_w^3 W_{F,2}^{\lambda,\hat{A}}
\Bigg](w)+\cdots,
\nonu
\\
    W_{F,1}^{\lambda,\hat{A}}(z)\,W_{F,1}^{\lambda,\hat{B}}(w)
&=& \frac{1}{(z-w)^2}\,\delta^{\hat{A}\hat{B}}\,\frac{N}{16\, q^2}
    -\frac{1}{(z-w)}\,
    \frac{i}{4\, q}\,f^{\hat{A}\hat{B}\hat{C}}\,W_{F,1}^{\lambda,\hat{C}}(w) +\cdots,
\nonu \\
W_{F,1}^{\lambda,\hat{A}}(z)\,W_{F,2}^{\lambda,\hat{B}}(w)
&=& \frac{1}{(z-w)^3}\,\delta^{\hat{A}\hat{B}}\,\frac{N\,\lambda}{2\, q}
\nonu \\
&+& \frac{1}{(z-w)^2}
\Bigg[
\delta^{\hat{A}\hat{B}}\,\frac{1}{ K}\,W_{F,1}^{\lambda}
+\frac{1}{2}\,d^{\hat{A}\hat{B}\hat{C}}\,W_{F,1}^{\lambda,\hat{C}}
-i\,\lambda\,f^{\hat{A}\hat{B}\hat{C}}\,W_{F,1}^{\lambda,\hat{C}}
\Bigg](w)
\nonu \\
&
-& \frac{1}{(z-w)}
\,\frac{i}{4\, q}\,
f^{\hat{A}\hat{B}\hat{C}}\,W_{F,2}^{\lambda,\hat{C}}(w)
+ \cdots,
\nonu \\
W_{F,1}^{\lambda,\hat{A}}(z)\,W_{F,3}^{\lambda,\hat{B}}(w)
&=& \frac{1}{(z-w)^4}\,2\,N\,\lambda^2\, \delta^{\hat{A}\hat{B}}
\nonu \\
&+&\frac{1}{(z-w)^3}
\Bigg[
\delta^{\hat{A}\hat{B}}\,\frac{8 \, q\,\lambda}{K}\,W_{F,1}^{\lambda}
+4 \, q\, \lambda\,d^{\hat{A}\hat{B}\hat{C}}\,W_{F,1}^{\lambda,\hat{C}}\nonu \\
& - &
\frac{4\, q}{3}(1+2\lambda^2)\,i\,
f^{\hat{A}\hat{B}\hat{C}}\,W_{F,1}^{\lambda,\hat{C}}
\Bigg](w)
\nonu \\
&
+& \frac{1}{(z-w)^2}
\Bigg[
\delta^{\hat{A}\hat{B}}\Big(\frac{2}{K}\,W_{F,2}^{\lambda}
-q\,\frac{4\,\lambda}{K}\,\partial_w W_{F,1}^{\lambda}\Big)
+d^{\hat{A}\hat{B}\hat{C}}\Big(
W_{F,2}^{\lambda,\hat{C}}-2\,
q\,\lambda\,\partial_w W_{F,1}^{\lambda,\hat{C}}\Big)
\nonu \\
&
-& i\,f^{\hat{A}\hat{B}\hat{C}}\Big(
\lambda\,W_{F,2}^{\lambda,\hat{C}}-\frac{2\, q}{
  3}(1+2\lambda^2)\,\partial_w W_{F,1}^{\lambda,\hat{C}}\Big)
\Bigg](w)
\nonu \\
&
-& \frac{1}{(z-w)}\,\frac{i}{4\, q}\,
f^{\hat{A}\hat{B}\hat{C}}\,W_{F,3}^{\lambda,\hat{C}}(w)+\cdots,
\nonu \\
W_{F,2}^{\lambda,\hat{A}}(z)\,W_{F,2}^{\lambda,\hat{B}}(w)
&=& \frac{1}{(z-w)^4}\,\frac{N\,(1-12\lambda^2)}{2}\,
\delta^{\hat{A}\hat{B}}
\nonu \\
&-& \frac{1}{(z-w)^3}\,
2\, q\,(1-2\lambda)(1+2\lambda)\,i\, f^{\hat{A}\hat{B}\hat{C}}\,
W_{F,1}^{\lambda,\hat{C}}(w)
\nonu \\
&
+& \frac{1}{(z-w)^2}
\Bigg[
\delta^{\hat{A}\hat{B}}\,\frac{2}{K}\,W_{F,2}^{\lambda}
+d^{\hat{A}\hat{B}\hat{C}}\,W_{F,2}^{\lambda,\hat{C}}
\nonu \\
& - &
q\,(1-2\lambda)(1+2\lambda)\,i\,
f^{\hat{A}\hat{B}\hat{C}}\,\partial W_{F,1}^{\lambda,\hat{C}}
\Bigg](w)
\nonu \\
&
+& \frac{1}{(z-w)}
\Bigg[
\delta^{\hat{A}\hat{B}}\,
\frac{1}{K}\,\partial_w W_{F,2}^{\lambda}
+\frac{1}{2}\,d^{\hat{A}\hat{B}\hat{C}}\,\partial_w W_{F,2}^{\lambda,\hat{C}}
\nonu \\
&
-& i\,f^{\hat{A}\hat{B}\hat{C}}
\Big(
\frac{1}{4\, q}\,W_{F,3}^{\lambda,\hat{C}}
+\frac{q}{3}(1-2\lambda)(1+2\lambda)\,\partial_w^2 W_{F,1}^{\lambda,\hat{C}}
\,\Big)
\Bigg](w) + \cdots,
\nonu \\
W_{F,2}^{\lambda,\hat{A}}(z)\,W_{F,3}^{\lambda,\hat{B}}(w)
&=&
\frac{1}{(z-w)^5}\,\delta^{\hat{A}\hat{B}}\,8 \,q
\,N\,\lambda(1-2\lambda)(1+2\lambda)
\nonu \\
&
+& \frac{1}{(z-w)^4}
\Bigg[
  \delta^{\hat{A}\hat{B}}\,\frac{16 \, q^2\,
    (1-2\lambda)(1+2\lambda)}{K}W_{F,1}^{\lambda}
\nonu \\
& + &
8\,
q^2\,(1-2\lambda)(1+2\lambda)\,d^{\hat{A}\hat{B}\hat{C}}\,W_{F,1}^{\lambda,\hat{C}}
\nonu \\
&
- & 8\, q^2\,\lambda(1-2\lambda)(1+2\lambda)\,i\,
f^{\hat{A}\hat{B}\hat{C}}\,W_{F,1}^{\lambda,\hat{C}}
\Bigg](w)
\nonu \\
&
-& \frac{1}{(z-w)^3}
\,\frac{16 \, q}{3}(1-\lambda)(1+\lambda)\,i\, f^{\hat{A}\hat{B}\hat{C}}\,
W_{F,2}^{\lambda,\hat{C}}(w)
\nonu \\
&
+& \frac{1}{(z-w)^2}
\Bigg[
\delta^{\hat{A}\hat{B}}\,\frac{3}{K}\, W_{F,3}^{\lambda}
+\frac{3}{2}\,d^{\hat{A}\hat{B}\hat{C}}\,W_{F,3}^{\lambda,\hat{C}}
\nonu \\
& - &
\frac{4\, q}{3}(1-\lambda)(1+\lambda)\,i\,
f^{\hat{A}\hat{B}\hat{C}}\,\partial_w W_{F,2}^{\lambda,\hat{C}}
\Bigg](w)
\nonu \\
&
+& \frac{1}{(z-w)}
\Bigg[
\delta^{\hat{A}\hat{B}}\,\frac{1}{K}\,\partial_w W_{F,3}^{\lambda}
+\frac{1}{2}\,d^{\hat{A}\hat{B}\hat{C}}\,\partial_w W_{F,3}^{\lambda,\hat{C}}
\nonu \\
&
-& i\,f^{\hat{A}\hat{B}\hat{C}}
\Big(
\frac{1}{4\, q}\,W_{F,4}^{\lambda,\hat{C}}
+\frac{4\, q}{15}
(1-\lambda)(1+\lambda)\,\partial_w^2 W_{F,2}^{\lambda,\hat{C}}
\Big)
\Bigg](w)+\cdots,
\nonu \\
W_{F,3}^{\lambda,\hat{A}}(z)\,W_{F,3}^{\lambda,\hat{B}}(w)
&=&
\frac{1}{(z-w)^6}\,\frac{32\,q^2\, N\,(1-15\lambda^2+20\lambda^4)}{3}\,
\delta^{\hat{A}\hat{B}}
\nonu \\
&
-&
\frac{1}{(z-w)^5}\,\frac{128\, q^3}{3}(1-\lambda)(1+\lambda)(1-2\lambda)(1+2\lambda)
\,i\, f^{\hat{A}\hat{B}\hat{C}}\,W_{F,1}^{\lambda,\hat{C}}(w)
\nonu \\
&
+&
\frac{1}{(z-w)^4}
\Bigg[
  \delta^{\hat{A}\hat{B}}\,\frac{64  q^2
    (1-\lambda)(1+\lambda)}{K} W_{F,2}^{\lambda}
  +32  q^2
  (1-\lambda)(1+\lambda)\,d^{\hat{A}\hat{B}\hat{C}} W_{F,2}^{\lambda,\hat{C}}
\nonu \\
& 
-& \frac{64\, q^3}{3}(1-\lambda)(1+\lambda)(1-2\lambda)(1+2\lambda)\,
i\, f^{\hat{A}\hat{B}\hat{C}}
\,\partial_w W_{F,1}^{\lambda,\hat{C}}
\Bigg](w)
\nonu \\
&
+&
\frac{1}{(z-w)^3}
\Bigg[
  \delta^{\hat{A}\hat{B}}\,\frac{32\, q^2\,
    (1-\lambda)(1+\lambda)}{K}\,\partial_w W_{F,2}^{\lambda}
\nonu \\
& + &
16 \, q^2\,
(1-\lambda)(1+\lambda)\,d^{\hat{A}\hat{B}\hat{C}}\,\partial_w W_{F,2}^{\lambda,\hat{C}} 
- i\,f^{\hat{A}\hat{B}\hat{C}}
\Big(\frac{8\, q}{3}(2-\lambda)(2+\lambda)\,W_{F,3}^{\lambda,\hat{C}}
\nonu \\
& 
+& \frac{64\, q^3}{3}(1-\lambda)(1+\lambda)(1-2\lambda)(1+2\lambda)
\,\partial_w^2 W_{F,1}^{\lambda,\hat{C}}\Big)
\Bigg](w)
\nonu \\
&
+ &
\frac{1}{(z-w)^2}
\Bigg[
\delta^{\hat{A}\hat{B}}\Big(
\frac{4}{K}\,W_{F,4}^{\lambda}
+\frac{48\, q^2\,
  (1-\lambda)(1+\lambda)}{5\,K}\,\partial_w^2 W_{F,2}^{\lambda}
\Big)
\nonu \\
& 
+&
d^{\hat{A}\hat{B}\hat{C}}\Big(
2\,W_{F,4}^{\lambda,\hat{C}}
+\frac{24\, q^2}{5}(1-\lambda)(1+\lambda)\,\partial_w^2 W_{F,2}^{\lambda,\hat{C}}
\Big)
\nonu \\
& 
-& i\,f^{\hat{A}\hat{B}\hat{C}}
\Big(\frac{4\, q}{
  3}(2-\lambda)(2+\lambda)\,\partial_w W_{F,3}^{\lambda,\hat{C}}
\nonu \\
&
+& \frac{16\, q^3}{9}(1-\lambda)(1+\lambda)(1-2\lambda)(1+2\lambda)
\,\partial_w^3 W_{F,1}^{\lambda,\hat{C}}\Big)
\Bigg](w)
\nonu \\
&
+&
\frac{1}{(z-w)}
\Bigg[
\delta^{\hat{A}\hat{B}}\Big(
\frac{2}{K}\,\partial_w W_{F,4}^{\lambda}
+\frac{32\, q^2\,
  (1-\lambda)(1+\lambda)}{15\,K}\,\partial_w^3 W_{F,2}^{\lambda}
\Big)
\nonu \\
& 
+&
d^{\hat{A}\hat{B}\hat{C}}\Big(
\partial_w W_{F,4}^{\lambda,\hat{C}}
+
\frac{16\, q^2}{15}(1-\lambda)(1+\lambda)\,\partial_w^3 W_{F,2}^{\lambda,\hat{C}}
\Big)
\nonu \\
& 
-& i\,f^{\hat{A}\hat{B}\hat{C}}
\Big(
\frac{1}{4\, q}\,W_{F,5}^{\lambda,\hat{C}}
+\frac{8\, q}{21}(2-\lambda)(2+\lambda)\,\partial_w^2 W_{F,3}^{\lambda,\hat{C}}
\nonu \\
& 
+& \frac{16\, q^3}{45}(1-\lambda)(1+\lambda)(1-2\lambda)(1+2\lambda)
\,\partial_w^4 W_{F,1}^{\lambda,\hat{C}}\Big)
\Bigg](w)+\cdots.
\label{OSother}
\eea
At $\la=0$ with $N=1$,
the six OPEs (\ref{six}) and (\ref{OSother})
can be seen from the previous work \cite{OS}.
By redefining the fields having the adjoint index
with the extra factor $q$ and ignoring the central terms,
we observe that
there are the second and first order poles
in the first half of (\ref{OSother}) (for $h_1=1$,
there is only the second order pole)
while there exists
the first order pole in the
last half of (\ref{OSother}) under the $q \rightarrow 0$ limit with
$\la=0$ \cite{Ahn2111}.

We present some structure constants we are using above
for each $h_1, h_2, h_3$ and $k$,
from (\ref{WFWFstructR}),
as follows:
\bea
    S_{F,R}^{\,\,1,1,1,0}(\lambda)&=& 1 \,, \qquad
    S_{F,R}^{\,\,1,2,1,0}(\lambda)=\frac{1}{2}(2\lambda -1) \,,  \qquad
    S_{F,R}^{\,\,1,2,1,1}(\lambda)=0  \,, \nonu
    \\
       S_{F,R}^{\,\,1,2,2,0}(\lambda)&= & -\frac{1}{6} \,, \qquad
       S_{F,R}^{\,\,1,3,1,0}(\lambda)=
       \frac{1}{3}(\lambda -1)(2\lambda -1)\,, \qquad
       S_{F,R}^{\,\,1,3,1,1}(\lambda)=
       \frac{1}{6}(\lambda -1)(2\lambda -1)\,,
       \nonu \\
    S_{F,R}^{\,\,1,3,1,2}(\lambda)&=& 0\,, \qquad
    S_{F,R}^{\,\,1,3,2,0}(\lambda)= -\frac{1}{6}(\lambda-1)\,, \qquad
    S_{F,R}^{\,\,1,3,2,1}(\lambda)= 0\,, \nonu
    \\
     S_{F,R}^{\,\,1,3,3,0}(\lambda)&=& \frac{1}{120}\,, \qquad
     S_{F,R}^{\,\,2,1,1,0}(\lambda)= 0\,, \qquad 
     S_{F,R}^{\,\,2,1,1,1}(\lambda)= \frac{1}{2}(2\lambda+1)\,, \nonu
     \\  
     S_{F,R}^{\,\,2,1,2,0}(\lambda)&=& -\frac{1}{6}\,,   
     S_{F,R}^{\,\,2,2,1,0}(\lambda)=
     -\frac{1}{12}(2\lambda-1)(2\lambda+1) \,,   
     S_{F,R}^{\,\,2,2,1,1}(\lambda)=
     \frac{1}{12}(2\lambda-1)(2\lambda+1)\,, \nonu \\  
     S_{F,R}^{\,\,2,2,1,2}(\lambda)&=&
     -\frac{1}{12}(2\lambda-1)(2\lambda+1) \,, \qquad  
     S_{F,R}^{\,\,2,2,2,0}(\lambda)= \frac{1}{12}\,, \qquad  
     S_{F,R}^{\,\,2,2,2,1}(\lambda)=-\frac{1}{12}\,, \nonu
     \\
     S_{F,R}^{\,\,2,2,3,0}(\lambda)&= & \frac{1}{120}\,, \qquad  
     S_{F,R}^{\,\,2,3,1,0}(\lambda)=
     -\frac{1}{12}(\lambda-1)(2\lambda-1)(2\lambda+1)\,, \qquad  
     S_{F,R}^{\,\,2,3,1,1}(\lambda)=0\,, \nonu \\  
     S_{F,R}^{\,\,2,3,1,2}(\lambda)&=& 0\,, \qquad  
     S_{F,R}^{\,\,2,3,1,3}(\lambda)=0\,, \qquad  
     S_{F,R}^{\,\,2,3,2,0}(\lambda)=
     \frac{1}{15}(\lambda-1)(\lambda+1)\,, \nonu \\  
     S_{F,R}^{\,\,2,3,2,1}(\lambda)&=&
     -\frac{1}{30}(\lambda-1)(\lambda+1)\,,   
     S_{F,R}^{\,\,2,3,2,2}(\lambda)=
     \frac{1}{90}(\lambda-1)(\lambda+1)\,,   
     S_{F,R}^{\,\,2,3,3,0}(\lambda)=-\frac{1}{120}\,,
     \nonu \\
     S_{F,R}^{\,\,2,3,3,1}(\lambda)&= & \frac{1}{240}\,, \qquad  
       S_{F,R}^{\,\,2,3,4,0}(\lambda)=-\frac{1}{5040}\,, \qquad
      S_{F,R}^{\,\,3,1,1,0}(\lambda)=0\,, \nonu \\
      S_{F,R}^{\,\,3,1,1,1}(\lambda)&=&
      \frac{1}{6}(\lambda+1)(2\lambda+1)\,, \qquad
      S_{F,R}^{\,\,3,1,1,2}(\lambda)=
      \frac{1}{3}(\lambda+1)(2\lambda+1)\,, \qquad
         S_{F,R}^{\,\,3,1,2,0}(\lambda)=0\,, \nonu \\
        S_{F,R}^{\,\,3,1,2,1}(\lambda)&=& -\frac{1}{6}(\lambda+1)\,, \qquad
        S_{F,R}^{\,\,3,1,3,0}(\lambda)=\frac{1}{120}\,, \qquad
        S_{F,R}^{\,\,3,2,1,0}(\lambda)=0\,, \nonu
        \\
        S_{F,R}^{\,\,3,2,1,1}(\lambda)&= & 0\,, \qquad
        S_{F,R}^{\,\,3,2,1,2}(\lambda)=0\,, \qquad
        S_{F,R}^{\,\,3,2,1,3}(\lambda)=-\frac{1}{12}(\lambda+1)
        (2\lambda-1)(2\lambda+1)\,, \nonu \\
        S_{F,R}^{\,\,3,2,2,0}(\lambda)&=&
        \frac{1}{90}(\lambda-1)(\lambda+1)\,, \qquad
        S_{F,R}^{\,\,3,2,2,1}(\lambda)=
        -\frac{1}{30}(\lambda-1)(\lambda+1)\,,
        \nonu \\
        S_{F,R}^{\,\,3,2,2,2}(\lambda) & = &
        \frac{1}{15}(\lambda-1)(\lambda+1)\,, \qquad  
       S_{F,R}^{\,\,3,2,3,0}(\lambda)= -\frac{1}{240}\,, \qquad
       S_{F,R}^{\,\,3,2,3,1}(\lambda)=\frac{1}{120}\,, \nonu \\   
       S_{F,R}^{\,\,3,2,4,0}(\lambda)& = & -\frac{1}{5040}\,, \qquad
    S_{F,R}^{\,\,3,3,1,0}(\lambda)=
    \frac{1}{180}(\lambda-1)(\lambda+1)(2\lambda-1)(2\lambda+1)\,,
    \nonu \\ 
    S_{F,R}^{\,\,3,3,1,1}(\lambda)&=&
    -\frac{1}{180}(\lambda-1)(\lambda+1)(2\lambda-1)(2\lambda+1)\,,
    \nonu \\
    S_{F,R}^{\,\,3,3,1,2}(\lambda) & = &
    \frac{1}{180}(\lambda-1)(\lambda+1)(2\lambda-1)(2\lambda+1)\,,
    \nonu \\
    S_{F,R}^{\,\,3,3,1,3}(\lambda)&=&
    -\frac{1}{180}(\lambda-1)(\lambda+1)(2\lambda-1)(2\lambda+1)\,,
    \nonu \\ 
    S_{F,R}^{\,\,3,3,1,4}(\lambda)&=&
    \frac{1}{180}(\lambda-1)(\lambda+1)(2\lambda-1)(2\lambda+1)\,,
    \nonu \\ 
    S_{F,R}^{\,\,3,3,2,0}(\lambda)&=&
    -\frac{1}{90}(\lambda-1)(\lambda+1)\,, \qquad 
    S_{F,R}^{\,\,3,3,2,1}(\lambda)=
    \frac{1}{60}(\lambda-1)(\lambda+1)\,, \nonu
    \\
    S_{F,R}^{\,\,3,3,2,2}(\lambda)&=&
    -\frac{1}{60}(\lambda-1)(\lambda+1)\,, \qquad 
    S_{F,R}^{\,\,3,3,2,3}(\lambda)=
    \frac{1}{90}(\lambda-1)(\lambda+1)\,,
    \nonu
    \\
    S_{F,R}^{\,\,3,3,3,0}(\lambda)&=&
    -\frac{1}{1260}(\lambda-2)(\lambda+2)\,, \qquad 
    S_{F,R}^{\,\,3,3,3,1}(\lambda)=
    \frac{1}{840}(\lambda-2)(\lambda+2)\,, \nonu \\ 
    S_{F,R}^{\,\,3,3,3,2}(\lambda)&=&
    -\frac{1}{1260}(\lambda-2)(\lambda+2)\,, \qquad 
     S_{F,R}^{\,\,3,3,4,0}(\lambda)=\frac{1}{5040}\,, \qquad 
    S_{F,R}^{\,\,3,3,4,1}(\lambda)=-\frac{1}{5040}\,, \nonu \\ 
    S_{F,R}^{\,\,3,3,5,0}(\lambda)&=& \frac{1}{362880}\,,
    \qquad
   S_{F,L}^{\,\,h_2,h_1,h_3,k}(\lambda)= S_{F,R}^{\,\,h_1,h_2,h_3,k}(\lambda),
\label{lastequation}
    \eea
    where
    the corresponding $ S_{F,L}^{\,\,h_2,h_1,h_3,k}(\lambda)$
    can be obtained by using the last relation of (\ref{lastequation}).
    Note that the $\la$ dependences above do not
    have a $\la$ factor and therefore, they do not vanish
    at $\la=0$.
    Some of the structure constants do not depend on
    the $\la$.
    For other structure constants, the similar analsis
    can be done.
    It is rather nontrivial to observe the previous relations
    (\ref{zerostruct}) because
    these structure constants at $\la=0$ have their nontrivial numerical
    values.
    We have checked that
    all the $\la$ dependent structure constants for nonzero $\la$
    have the factor $\la$ explicitly after
    subtracted by those evaluated at $\la=0$.

By introducing a new variable    
\bea
h_3= h_1+h_2-h-2\,,
\label{h3}
  \eea
  we can sum over the variable $k$ for the first commutator
  in (\ref{26}) as follows:
\bea
P_F^{h_1,h_2,h}(m,n,\la) & \equiv &
\sum_{k=0}^{h+1}
(-1)^{h_1+h_2-h-2}\,4^{h}\,(2(h_1+h_2-h)-5)! \nonu \\
& \times &
\bigg(\,\,S^{\,\,h_1,h_2,h_1+h_2-h-2,k}_{F,\,R}(\lambda)\, [m+h_1-1]_{h-k+1}\,[n+h_2-1]_{k}
\nonu \\
&- & S^{\,\,h_1,h_2,h_1+h_2-h-2,k}_{F,\,L}(\lambda)\,  [m+h_1-1]_{k}\,[n+h_2-1]_{h-k+1}\,\,\bigg).
\label{largeP}
\eea
Then we obtain the following structure constants by using
(\ref{lastequation}), (\ref{h3}) and (\ref{largeP}),
\bea
P_F^{1,1,-1}(m,n,\la) & = & 0\,,
\qquad
P_F^{1,2,-1}(m,n,\la)  =  0\,,
\qquad
P_F^{1,2,0}(m,n,\la)  =  m\,,
\nonu \\
P_F^{1,3,-1}(m,n,\la) & = & 0\,,
\qquad
P_F^{1,3,0}(m,n,\la)=2m\,,
\qquad
P_F^{1,3,1}(m,n,\la)=4m(2m+n)\lambda\,,
\nonu \\
P_F^{2,1,-1}(m,n,\la) & = & 0\,,
\qquad
P_F^{2,1,0}(m,n,\la)=-n\,,
\qquad
P_F^{2,2,-1}(m,n,\la)=0\,,
\nonu \\
P_F^{2,2,0}(m,n,\la)& = & (m-n)\,,
\qquad
P_F^{2,2,1}(m,n,\la)=0\,,
\qquad
P_F^{2,3,-1}(m,n,\la)=0\,,
\nonu \\
P_F^{2,3,0}(m,n,\la)& = & (2m-n)\,,
\qquad
P_F^{2,3,1}(m,n,\la)=0\,,
\nonu \\
P_F^{2,3,2}(m,n,\la) & = & -\frac{8}{3}(m-1)m(m+1)(2\lambda-1)(2\lambda+1)\,,
\qquad
P_F^{3,1,-1}(m,n,\la) =  0\,,
\nonu \\
P_F^{3,1,0}(m,n,\la) & = & -2n\,,
\qquad
P_F^{3,1,1}(m,n,\la)  =  -4n(m+2n)\lambda\,,
\nonu \\
P_F^{3,2,-1}(m,n,\la) & = & 0\,,
\qquad
P_F^{3,2,0}(m,n,\la)=(m-2n)\,,
\qquad
P_F^{3,2,1}(m,n,\la)=0\,,
\nonu \\
P_F^{3,2,2}(m,n,\la) & = &
\frac{8}{3}(n-1)n(n+1)(2\lambda-1)(2\lambda+1)\,,
\qquad
P_F^{3,3,-1}(m,n,\la)=0\,,
\nonu \\
P_F^{3,3,0}(m,n,\la) & = & 2(m-n)\,,
\qquad
P_F^{3,3,1}(m,n,\la)  =  0\,,
\nonu \\
P_F^{3,3,2}(m,n,\la) & = &
-\frac{16}{15}(m-n)(2m^2-m\,n+2n^2-8)(\lambda-1)(\lambda+1)\,,
\nonu \\
P_F^{3,3,3}(m,n,\la) & = & 0\,.
\label{Pexp}
\eea

On the other hand, we write down the previous structure constants
appearing in \cite{Ahn2203} as follows:
\bea
p_F^{1,2,0}(m,n,\la) & = & m\,,
\qquad
p_F^{1,3,0}(m,n,\la)=2m\,,
\qquad
p_F^{2,1,0}(m,n,\la) =  -n\,,
\nonu \\
p_F^{2,2,0}(m,n,\la)& = & (m-n)\,,
\qquad
p_F^{2,3,0}(m,n,\la)  =  (2m-n)\,,
\nonu \\
p_F^{2,3,2}(m,n,\la)& = &
-\frac{8}{3}(m-1)m(m+1)(2\lambda-1)(2\lambda+1)\,,
\qquad
\nonu \\
p_F^{3,1,0}(m,n,\la) & = & -2n\,,
\qquad
p_F^{3,2,0}(m,n,\la)  =   (m-2n)\,,
\nonu \\
p_F^{3,2,2}(m,n,\la) & = &
\frac{8}{3}(n-1)n(n+1)(2\lambda-1)(2\lambda+1)\,,
\qquad
\nonu \\
p_F^{3,3,0}(m,n,\la)& = & 2(m-n)\,,
\qquad
\nonu \\
p_F^{3,3,2}(m,n,\la) & = &
-\frac{16}{15}(m-n)(2m^2-m\,n+2n^2-8)(\lambda-1)(\lambda+1)\,.
\label{pexp}
\eea
Comparing (\ref{Pexp}) with (\ref{pexp}),
the structure constants $P_F^{1,3,1}(m,n,\la)$ and $P_F^{3,1,1}(m,n,\la) $
do not appear in (\ref{pexp}).
This indicates that the structure constants in \cite{AK2009}
are not appropriate for the present work. 

For convenience, we present the following OPEs
corresponding to (\ref{six})
\bea
W_{B,1}^{\lambda}(z)\,W_{B,1}^{\lambda}(w)
&=& -\frac{1}{(z-w)^2}\,\frac{N\,K}{16\, q^2}+\cdots\,,
\nonu \\
W_{B,1}^{\lambda}(z)\,W_{B,2}^{\lambda}(w)
&= & -\frac{1}{(z-w)^3}\,\frac{N\,K\,(2\lambda-1)}{4\,q}
+\frac{1}{(z-w)^2}\,W_{B,1}^{\lambda}(w)+\cdots\,,
\nonu \\
W_{B,1}^{\lambda}(z)\,W_{B,3}^{\lambda}(w)
&=&
-\frac{1}{(z-w)^4}\,\frac{N\,K\,(2\lambda-1)^2}{2}
+\frac{1}{(z-w)^3}\,4 q\,(2\lambda-1) \,W_{B,1}^{\lambda}(w)
\nonu \\
&+& \frac{1}{(z-w)^2}\,\Bigg[
2\,W_{B,2}^{\lambda}
-2 q\, (2\lambda-1)\,\partial_w W_{B,1}^{\lambda}
\Bigg](w)+\cdots,
\nonu \\
W_{B,2}^{\lambda}(z)\,W_{B,2}^{\lambda}(w)
&=&
\frac{1}{(z-w)^4}\,N\,K\,(1-6\lambda+6\lambda^2)
+\frac{1}{(z-w)^2}\,2\,W_{B,2}^{\lambda}(w)
+\frac{1}{(z-w)}\,\partial_w W_{B,2}^{\lambda}(w)
\nonu \\
& + & \cdots,
\nonu \\
W_{B,2}^{\lambda}(z)\,W_{B,3}^{\lambda}(w)
&=&
\frac{1}{(z-w)^5}\,16 q\, N\, K\, (\lambda - 1)\lambda (2\lambda - 1)
+\frac{1}{(z-w)^4}\,64 q^2\,\lambda(\lambda-1)\,W_{B,1}^{\lambda}(w)
\nonu \\
&
+& \frac{1}{(z-w)^2}\,3\, W_{B,3}^{\lambda}(w)
+\frac{1}{(z-w)}\,\partial_w W_{B,3}^{\lambda}(w)+\cdots,
\nonu \\
W_{B,2}^{\lambda}(z)\,W_{B,4}^{\lambda}(w)
&=&
\frac{1}{(z-w)^6}\,32 q^2\,N\,K \,\lambda(\lambda-1)(2\lambda-1)^2
\nonu \\
&
-& \frac{1}{(z-w)^5}\,256 q^3\,\lambda(\lambda-1)(2\lambda-1)\,
W_{B,1}^{\lambda}(w)
\nonu \\
&
+& \frac{1}{(z-w)^4}
\Bigg[\!\!
- \frac{192 q^2}{5}(3\lambda^2-3\lambda-1)\,W_{B,2}^{\lambda}
\nonu \\
& + & 128 q^3
\,\lambda(\lambda-1)(2\lambda-1)\,\partial_w W_{B,1}^{\lambda}
\Bigg](w)
\nonu \\
&
+& \frac{1}{(z-w)^2}\,4\,W_{B,4}^{\lambda}(w)
+\frac{1}{(z-w)}\,\partial_w W_{B,4}^{\lambda}(w)
+\cdots,
\nonu \\
W_{B,3}^{\lambda}(z)\,W_{B,3}^{\lambda}(w)
&=&
-\frac{1}{(z-w)^6}\,\frac{16 q^2\,N\,K}{3}
(-3+10\lambda+30\lambda^2-80\lambda^3+40\lambda^4)
\nonu \\
&
-& \frac{1}{(z-w)^4}\,16q^2\,(2\lambda-3)(2\lambda+1)\,
W_{B,2}^{\lambda}(w)
\nonu \\
& - &
\frac{1}{(z-w)^3}\,8q^2\,(2\lambda-3)(2\lambda+1)\,
\partial_w W_{B,2}^{\lambda}(w)
\nonu \\
&
+& \frac{1}{(z-w)^2}
\Bigg[
  4\,W_{B,4}^{\lambda}-
  \frac{12 q^2}{5}(2\lambda-3)(2\lambda+1)\,\partial_w^2 W_{B,2}^{\lambda}
\Bigg](w)
\nonu \\
&
+& \frac{1}{(z-w)}
\Bigg[
  2\,\partial_w W_{B,4}^{\lambda}-
  \frac{8 q^2}{15}(2\lambda-3)(2\lambda+1)\,\partial_w^3 W_{B,2}^{\lambda}
\Bigg](w)
+\cdots.
\label{LAST}
\eea
Note that the structure constants appearing in (\ref{LAST})
can be reproduced by changing $\la \rightarrow \la -\frac{1}{2}$
in (\ref{six}) with opposite signs for the central terms.




\end{document}